\def\draftversion{false}
\def\showall{true}  % Set to false to hide figures

\RequirePackage{ifthen}
\ifthenelse{\equal{\draftversion}{true}}{
  \documentclass[aps,pra,10pt,galley,amsmath,amssymb,
                 longbibliography,superscriptaddress,nofootinbib]{revtex4}
}{
  \documentclass[aps,pra,10pt,twocolumn,amsmath,amssymb,floatfix,
    longbibliography,superscriptaddress,nofootinbib]{revtex4-1}
}

\usepackage{amsmath,amssymb,amsfonts,mathtools}
\usepackage{amsthm}
\usepackage{bm,bbm} % bold math
\usepackage{nicefrac}
\usepackage{graphicx}
\usepackage{hhline}
\usepackage[utf8]{inputenc}
\usepackage{subfigure}
\usepackage{xcolor}

\usepackage{soul}  % \st    for strike-out
\usepackage{hyperref}
\hypersetup{
    colorlinks=true,       
    linkcolor=blue,          
    citecolor=blue,       
    filecolor=magenta,      
    urlcolor=blue          
}

\ifthenelse{\equal{\draftversion}{true}}{
  \marginparwidth 2.7in
  \marginparsep 0.5in
  \newcounter{comm} % counter for commentaries
  \def\commnext{\stepcounter{comm}}
  \def\commtext{{\bf\color{blue}[\arabic{comm}]}}
  \def\commmar{{\bf\color{blue}[\arabic{comm}]}}
  \def\trm#1{\commnext\marginpar{\small TR\commmar: #1}\commtext}
  \def\tom#1{\commnext\marginpar{\small TO\commmar: #1}\commtext}
  \def\ism#1{\commnext\marginpar{\small IS\commmar: #1}\commtext}
  \def\dvm#1{\commnext\marginpar{\small DV\commmar: #1}\commtext}
  
  \def\parsedate #1:20#2#3#4#5#6#7#8\empty{#4#5/#6#7/20#2#3}
  \def\moddate{\expandafter\parsedate\pdffilemoddate{\jobname.tex}\empty}
  
}{
  \def\trm#1{}
  \def\tom#1{}
  \def\ism#1{}
  \def\dvm#1{}
  
}

\ifthenelse{\equal{\showall}{false}}{
  \renewcommand\includegraphics[2][]{{\bf Figure not shown.}}
  \renewcommand\input[1][]{{\bf Not shown\ }}
}{}

\newcommand{\code}[1]{\textsc{#1}}

\newcommand{\beq}{\begin{equation}}
\newcommand{\eeq}{\end{equation}}
\newcommand{\nn}{\nonumber\\}

\newcommand{\eq}[1]{Eq.~(\ref{eq:#1})}
\newcommand{\Eq}[1]{Equation~(\ref{eq:#1})}
\newcommand{\eqs}[2]{Eqs.~(\ref{eq:#1}) and (\ref{eq:#2})}
\newcommand{\Eqs}[2]{Equations~(\ref{eq:#1}) and (\ref{eq:#2})}

\newcommand{\eqr}[2]{Eqs.~(\ref{eq:#1}-\ref{eq:#2})}
\newcommand{\Eqr}[2]{Equations~(\ref{eq:#1}-\ref{eq:#2})}

\newcommand{\fref}[1]{Fig.~\ref{fig:#1}}

\newcommand{\fsref}[1]{Figs.~\ref{fig:#1}}
\newcommand{\Fref}[1]{Figure~\ref{fig:#1}}

\newcommand{\sref}[1]{Sec.~\ref{sec:#1}}

\newcommand{\aref}[1]{Appendix~\ref{sec:#1}}

\newcommand{\tref}[1]{Table~\ref{tab:#1}} 

\ifthenelse{\equal{\draftversion}{true}}{
  \newcommand{\eql}[1]{\label{eq:#1}\hbox{\Red{\small\;\;[#1]}}}
  \newcommand{\figl}[1]{\label{fig:#1}\Red{\small\;\;[Fig:~#1]}}
  \newcommand{\secl}[1]{\label{sec:#1}\Red{\small\;\;[Sec:~#1]}}
  
}{}
\ifthenelse{\equal{\draftversion}{false}}{
  \newcommand{\eql}[1]{\label{eq:#1}}
  \newcommand{\figl}[1]{\label{fig:#1}}
  \newcommand{\secl}[1]{\label{sec:#1}}
  
}{}

\newcommand{\ket}[1]{\vert#1\rangle}
\newcommand{\bra}[1]{\langle#1\vert}
\newcommand{\ip}[2]{\langle#1\vert#2\rangle}
\newcommand{\me}[3]{\langle#1\vert#2\vert#3\rangle}

\newcommand{\wt}[1]{\widetilde{#1}}

\newcommand{\Imag}{{\rm Im\,}}

\def\k{{\bf k}}
\def\kk{{\boldsymbol \kappa}}
\def\R{{\bf R}}
\def\A{{\bf A}}

\def\rr{{\bf r}} % (cannot use \r, messes up with \AA=\r{A})

\def\a{{\bf a}}
\def\b{{\bf b}}
\def\xhat{\hat{\bf x}}
\def\yhat{\hat{\bf y}}
\def\zhat{\hat{\bf z}}

\def\bnk{_{n\k}}
\def\bmk{_{m\k}}

\def\0{\mathbf{0}}

\def\half{\tfrac{1}{2}}

\def\Naf{\overline N_{\rm A}}
\def\Nafp{\overline N_{{\rm A}^+}}
\def\Nafm{\overline N_{{\rm A}^-}}
\def\Nafpm{\overline N_{{\rm A}^\pm}}

\def\Nad{\wt N_{\rm A}}
\def\Wa{W_{\rm A}}
\def\Caf{\overline C_{\rm A}}
\def\Cafp{\overline C_{{\rm A}^+}}
\def\Cafm{\overline C_{{\rm A}^-}}
\def\Cafpm{\overline C_{{\rm A}^\pm}}
\def\Cad{\wt C_{\rm A}}
\def\Cadp{\wt C_{{\rm A}^+}}
\def\Cadm{\wt C_{{\rm A}^-}}
\def\Cadpm{\wt C_{{\rm A}^\pm}}
\def\pA{p_{\rm A}}

\def\Nbf{\overline N_{\rm B}}

\def\Nbfpm{\overline N_{{\rm B}^\pm}}
\def\Nbfmp{\overline N_{{\rm B}^\mp}}
\def\Nbd{\wt N_{\rm B}}
\def\Wb{W_{\rm B}}
\def\Cbf{\overline C_{\rm B}}
\def\Cbfp{\overline C_{{\rm B}^+}}
\def\Cbfm{\overline C_{{\rm B}^-}}
\def\Cbfpm{\overline C_{{\rm B}^\pm}}
\def\Cbd{\wt C_{\rm B}}

\def\Cbdpm{\wt C_{{\rm B}^\pm}}
\def\pB{p_{\rm B}}

\def\uc{_{\rm UC}}

\def\Nuc{\wt N\uc}
\def\Cuc{\wt C\uc}
\def\Cucp{\wt C_{{\rm UC}^+}}
\def\Cucm{\wt C_{{\rm UC}^-}}

\def\G{_{\rm G}}
\def\X{_{\rm X}}

\def\z{\mathbbm{Z}}
\def\zt{\mathbbm{Z}_2}

\def\T{\mathcal{T}}
\def\I{\mathcal{I}}

\begin{document}

\title{Mirror Chern numbers in the hybrid Wannier representation}

\author{Tom\'{a}\v{s} Rauch}
\affiliation{Friedrich-Schiller-University Jena, 07743 Jena, Germany}

\author{Thomas Olsen}
\affiliation{Computational Atomic-scale Materials Design,
  Department of Physics, Technical University of Denmark, 2800
  Kgs. Lyngby Denmark}

\author{David Vanderbilt} \affiliation{Department of Physics and
  Astronomy, Rutgers University, Piscataway, New Jersey 08854-8019,
  USA}

\author{Ivo Souza} \affiliation{Centro de F{\'i}sica de Materiales,
  Universidad del Pa{\'i}s Vasco, 20018 San Sebasti{\'a}n,
  Spain} \affiliation{Ikerbasque Foundation, 48013 Bilbao, Spain
  }

\begin{abstract}
The topology of electronic states in band insulators with mirror
symmetry can be classified in two different ways.  One is in terms of
the mirror Chern number, an integer that counts the number of
protected Dirac cones in the Brillouin zone of high-symmetry
surfaces. The other is via a $\zt$ index that distinguishes between
systems that have a nonzero quantized orbital magnetoelectric coupling
(``axion-odd''), and those that do not (``axion-even''); this
classification can also be induced by other symmetries in the magnetic
point group, including time reversal and inversion.  A systematic
characterization of the axion $\zt$ topology has previously been
obtained by representing the valence states in terms of hybrid Wannier
functions localized along one chosen crystallographic direction, and
inspecting the associated Wannier band structure.  Here we focus on
mirror symmetry, and extend that characterization to the mirror Chern
number. We choose the direction orthogonal to the mirror plane as the
Wannierization direction, and show that the mirror Chern number can be
determined from the winding numbers of the touching points between
Wannier bands on mirror-invariant planes, and from the Chern numbers
of flat bands pinned to those planes. In this representation, the
relation between the mirror Chern number and the axion $\zt$ index is
readily established.  The formalism is illustrated by means of {\it ab
  initio} calculations for SnTe in the monolayer and bulk forms,
complemented by tight-binding calculations for a toy model.
\end{abstract}
\maketitle

%========================
\section{Introduction}
\secl{intro}
%========================

The band theory of solids has been enriched in recent years by a
vigorous study of its topological aspects. That effort resulted in a
systematic topological classification of insulators on the basis of
symmetry, and in the identification of a large number of topological
materials.  After an initial focus on the role of time-reversal
symmetry, it was realized that crystallographic symmetries could also
protect topological behaviors, leading to the notion of ``topological
crystalline insulators.''

The assignment of an insulator to a particular topological class can
be made by evaluating the corresponding topological invariant.
Depending on the protecting symmetry, that invariant may assume one of
two values ($\zt$ classification), or it may assume any integer value
($\z$ classification).  Other types of classifications such as $\z_4$
also occur, but they do not concern us here. When the invariant
vanishes the system is classified as trivial, and otherwise it is
classified as nontrivial or topological. Topological insulators
typically display robust gapless states at the boundary, which provide
an experimental signature of topological behavior.

In some cases, the same symmetry may induce two different topological
classifications. This happens for example with mirror symmetry, where
a $\z$ classification in terms of the mirror Chern number
(MCN)~\cite{teo-prb08,ando-arcmp15}
coexists with a $\zt$ classification based on the quantized axion
angle.  The two classifications are not independent, and elucidating
the relation between them is one goal of the present work.

The axion $\zt$ classification of three-dimensional (3D) insulators is
based on the orbital magnetoelectric effect.  In brief, the isotropic
part of the linear orbital magnetoelectric tensor is conveniently
expressed in terms of the axion angle~$\theta$, which is only defined
modulo $2\pi$ as a bulk property.  In the presence of ``axion-odd''
symmetries that flip its sign, the axion angle can only assume two
values: $\theta=0$ (trivial), and $\theta=\pi$
(topological)~\cite{qi-prb08,essin-prl09,vanderbilt-book18,armitage-scipost19,nenno-nrp20,sekine-jap21}.

The axion $\zt$ index was originally introduced for time-reversal
invariant insulators, where it was shown to be equivalent to the
``strong'' $\zt$ index $\nu_0=0$ or $1$, that is, $\theta=\pi\nu_0$.
More generally, axion-odd symmetries can be classified as proper
rotations combined with time reversal (including time reversal
itself), and improper rotations (including inversion and reflection)
not combined with time reversal; in both cases, the associated
symmetry operation in the magnetic space group may include a
fractional translation. This results in a large number of magnetic
space groups that can host axion-odd topological insulators. A recent
realization is the MnBi$_2$Te$_4$ family of antiferromagnetic
materials~\cite{otrokov-nat19,nenno-nrp20,sekine-jap21}, whose axion
topology is protected by the time reversal operation combined with a
half-lattice translation as envisioned in Ref.~\cite{mong-prb10}.

To aid the computational search for axionic topological insulators, it
is useful to devise simple procedures for determining the (quantized)
axion angle $\theta$. Unfortunately, subtle gauge issues make its
direct evaluation from the valence Bloch states rather challenging in
general~\cite{vanderbilt-book18}. Notable exceptions are
centrosymmetric insulators, both nonmagnetic and magnetic. For such
systems, the axion $\zt$ index can be
obtained by counting the number of odd-parity states at high-symmetry
points in the Brillouin zone (BZ)~\cite{fu-prb07,turner-prb12}.

Recently, an alternative procedure was introduced based on
representing the valence states in terms of hybrid Wannier (HW)
functions that are maximally localized along a chosen crystallographic
direction $z$. The HW centers along $z$, also known as ``Wilson-loop
eigenvalues,'' form a band structure when plotted as a function of
$k_x$ and $k_y$; in the presence of one or more axion-odd symmetries,
the quantized $\theta$ value can be determined from this ``Wannier
band structure,'' often by mere visual
inspection~\cite{varnava-prb20}.

In the HW representation, axion-odd symmetries are naturally
classified as ``$z$-preserving'' or ``$z$-reversing,'' and the rules
for deducing the axion $\zt$ index are different in each case (they
also depend on whether or not the symmetry operation involves a
fractional translation along~$z$)~\cite{varnava-prb20}.  Time reversal
is an example of a $z$-preserving operation, while inversion is $z$
reversing. Mirror operations may be placed in one group or the other,
depending on whether the Wannierization direction $z$ lies in the
reflection plane (vertical mirror) or is orthogonal to it (horizontal
mirror). In this work we make the latter choice, so that the mirror
operation of interest becomes
\beq
M_z:z\rightarrow -z\,,
\eql{Mz-def}
\eeq
which is manifestly $z$ reversing.

A simple mirror symmetry without a glide component protects not only
the axion $\zt$ classification, but also a $\z$ or $\z\times\z$
classification based on one or two MCNs, depending on the type of
mirror.  This raises the question of whether the HW representation
might also be convenient for determining the MCNs, and for
illuminating their relationship to the quantized axion angle.

In this work, we address the above questions by investigating in
detail the Wannier bands in the presence of $M_z$ symmetry. We clarify
the generic behaviors that are expected, and discuss the rules for
deducing the MCNs.  By comparing those rules with the ones
obtained in Ref.~\cite{varnava-prb20} for the
axion $\zt$ index, we establish the relation between the two
classifications.

The paper is organized as follows. In \sref{prelim} we first
distinguish between ``type-1'' and ``type-2'' crystallographic mirror
operations; we then review the definitions of Chern invariants and
MCNs in terms of the Bloch states in the filled bands; finally, we
introduce maximally localized HW functions spanning the valence
states, and assign Chern numbers to isolated groups of Wannier bands.
This background material sets the stage for the developments in the
remainder of the paper.  In \sref{mirror-wannier} we discuss the
generic features of the Wannier band structure in the presence of
$M_z$ symmetry, and obtain a relation between Chern numbers and
winding numbers in groups of bands touching on a mirror plane.  The
rules for deducing the MCNs from the Chern numbers and winding numbers
on the mirror planes are given in \sref{MCN}, where their relation to
the quantized axion angle is also established. In \sref{methods} we
describe the numerical methods that are used in \sref{results} to
apply the formalism to several prototypical systems. We summarize and
conclude in \sref{summary}, and present in three Appendices some
derivations that were left out of the main text.

% =======================
\section{Preliminaries}
\secl{prelim}
%=======================

%------------------------------------------------
\subsection{Two types of crystallographic mirrors}
\secl{types}
% -----------------------------------------------

\begin{figure*}
\centering
\includegraphics[width=2.5in]{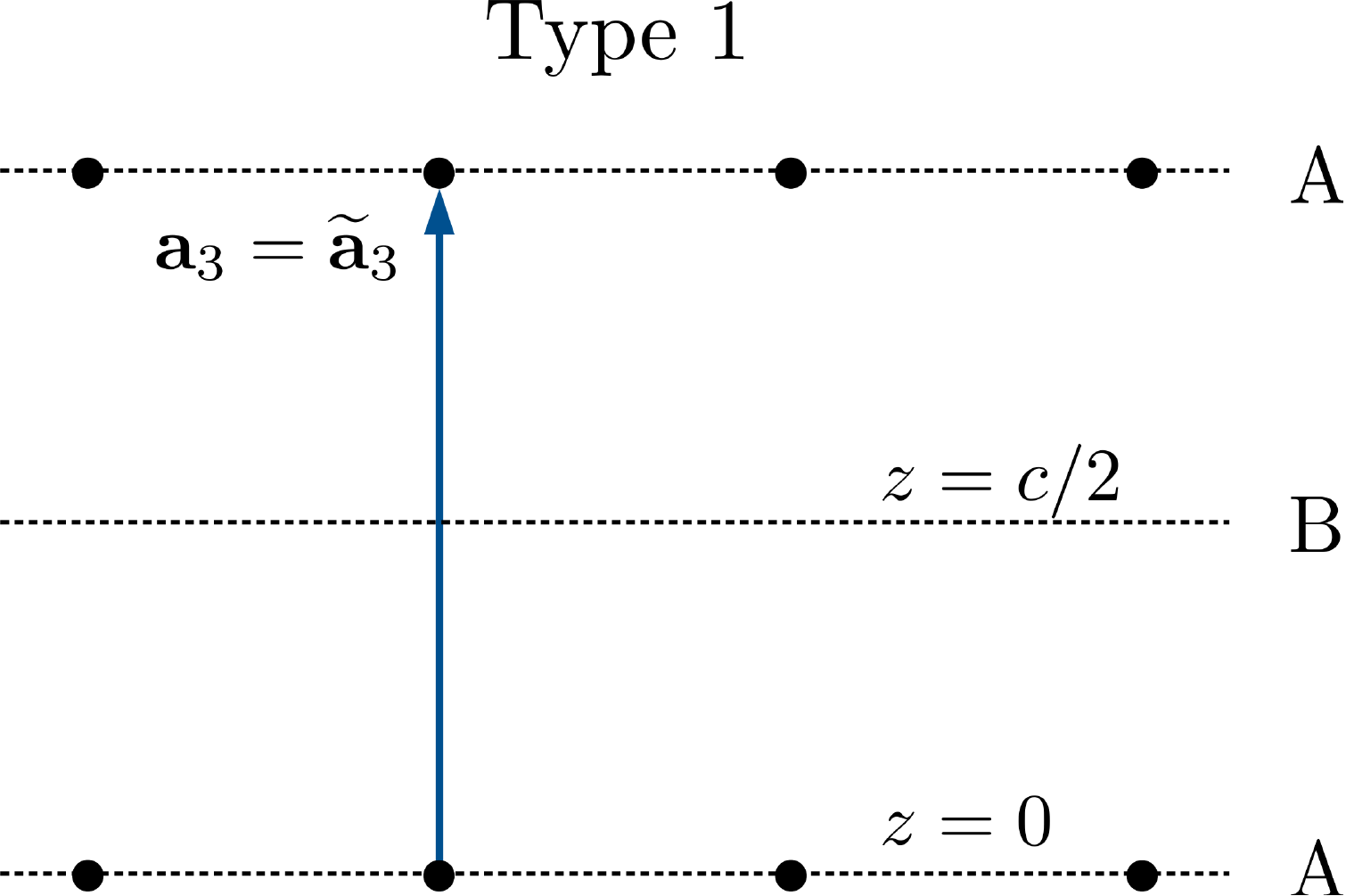}
\hspace{0.05cm}
\includegraphics[width=2.325in]{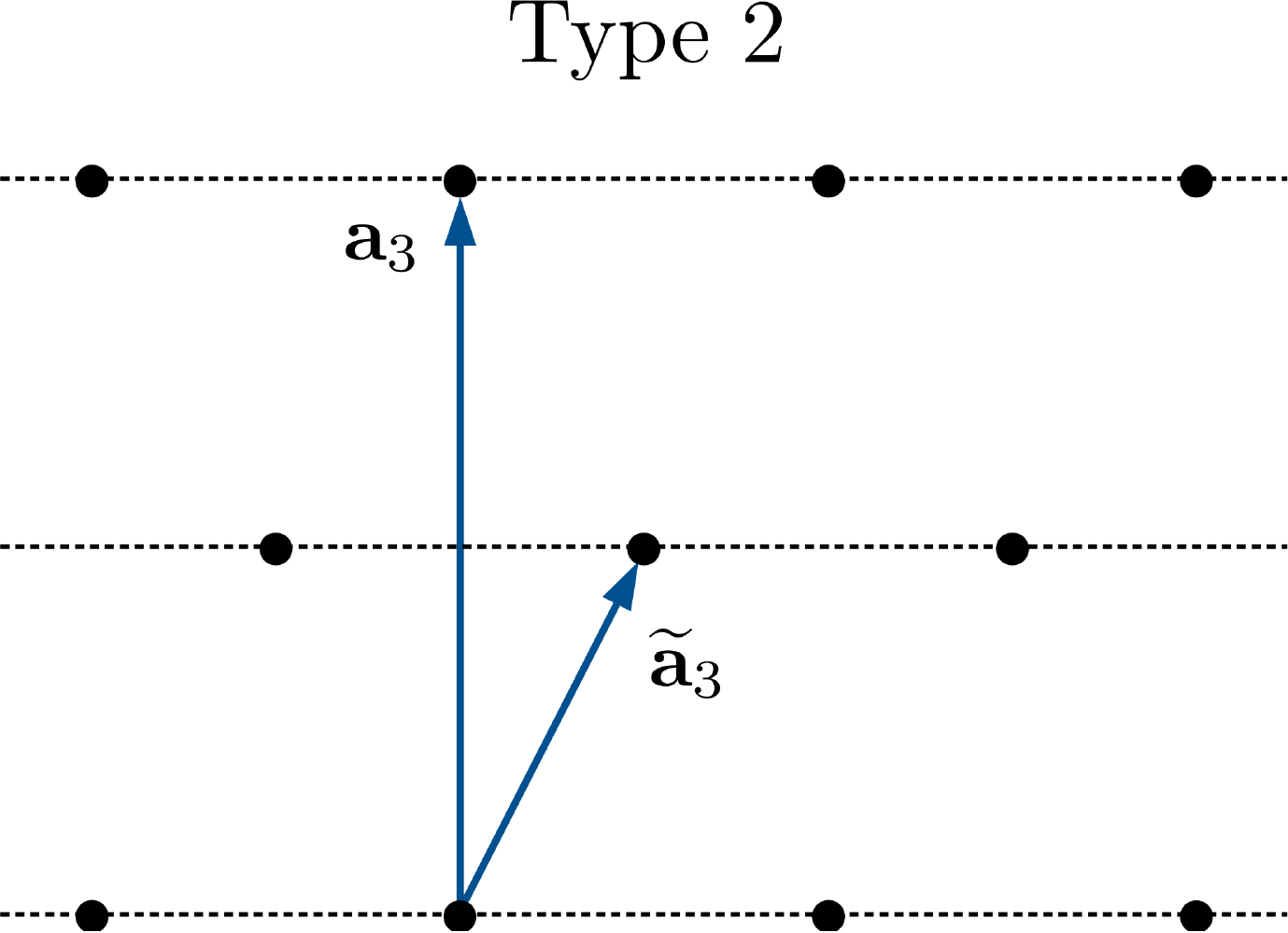}
\vskip 1.0cm
\includegraphics[width=2.5in]{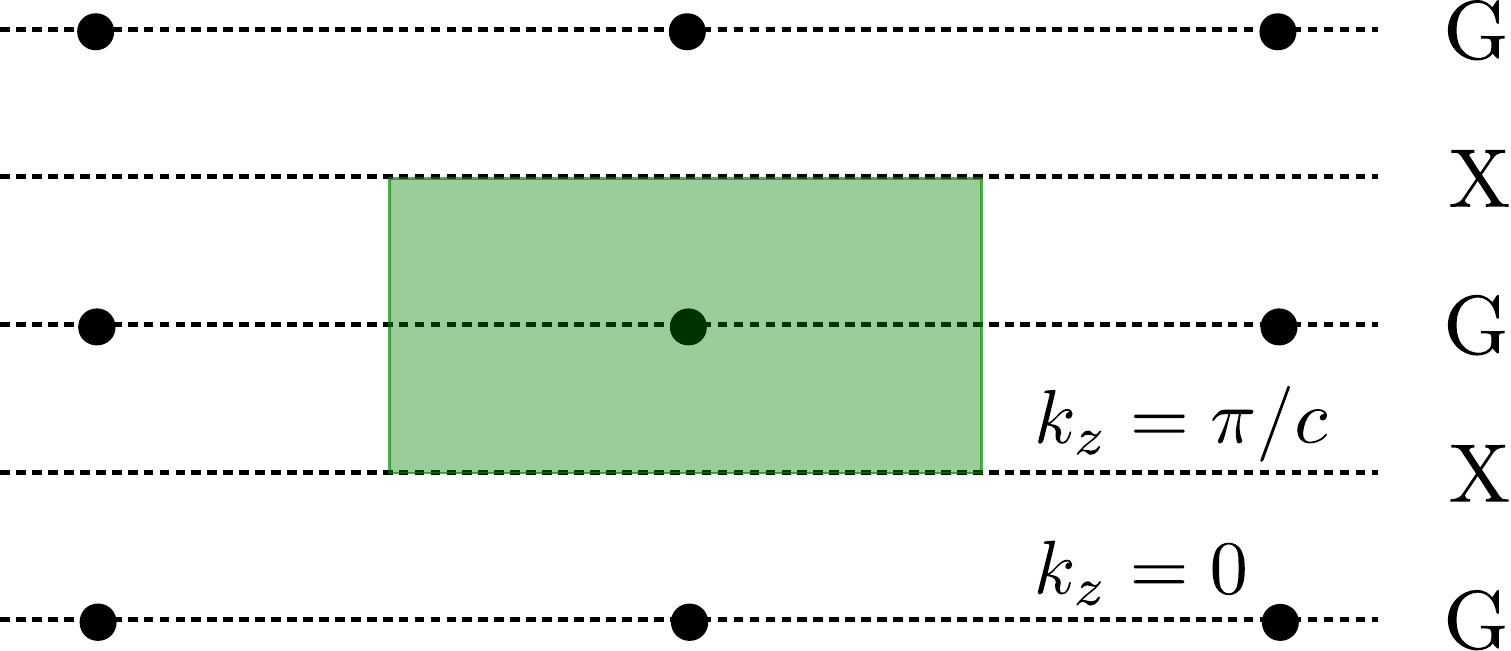}
\hspace{0.05cm}
\includegraphics[width=2.33in]{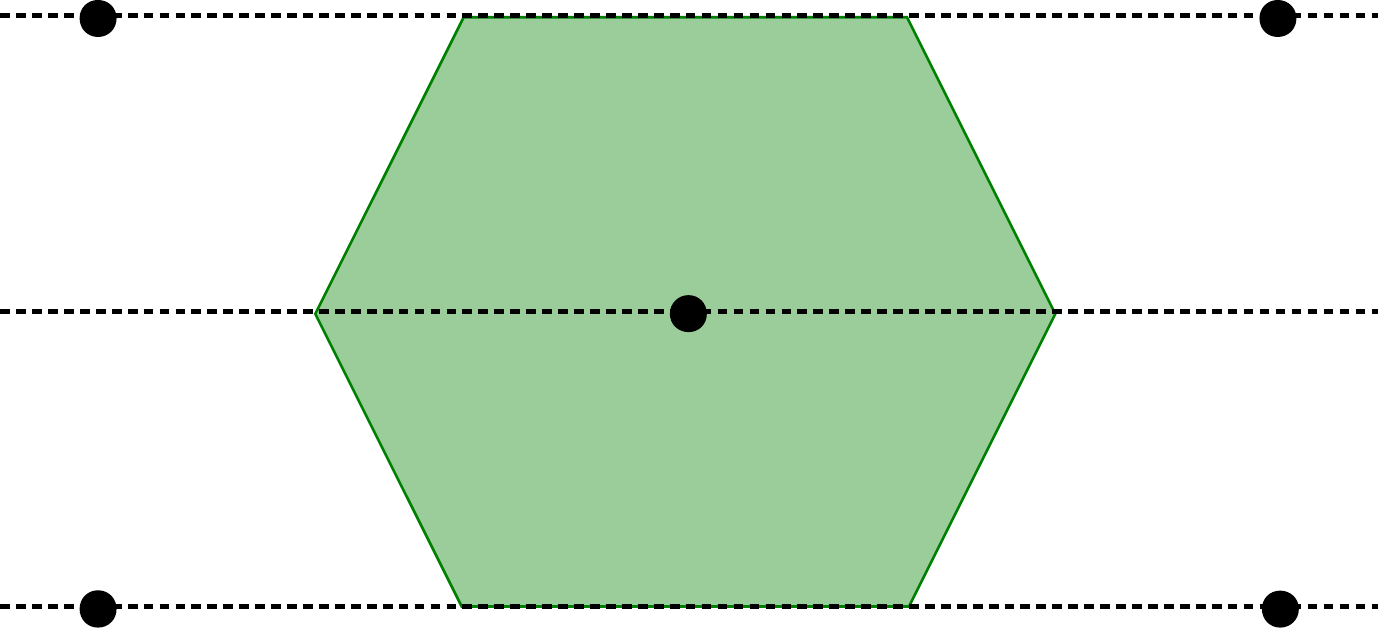}
\caption{The upper panel shows schematically a pair of 2D crystals
  lying on the $(x,z$) plane; each has one atom per primitive cell
  (black dots), and lattice constant $c$ along $z$. The crystal on the
  left has a rectangular lattice and a type-1 horizontal mirror, with
  inequivalent mirror lines $z=0\text{ mod $c$}$ (A) and
  $z=c/2\text{ mod $c$}$ (B), shown as dashed lines; the one on the
  right has a centered rectangular lattice and a type-2 mirror, with
  equivalent mirror lines A and B. The lattice vectors $\a_3$ and
  $\wt\a_3$ are defined in the main text.  The lower panel shows the
  reciprocal lattices, with a separation of $2\pi/c$ between
  horizontal lattice lines G. On the left the periodicity along $k_z$
  is $2\pi/c$, and hence both $k_z=0\text{ mod $2\pi/c$}$ (G) and
  $k_z=\pi/c\text{ mod $2\pi/c$}$ (X) are pointwise-invariant mirror
  lines, as indicated by the dashed lines. On the right, where the
  periodicity along $k_z$ is $4\pi/c$, G is a mirror-invariant line
  but X is not. The associated Brillouin zones are indicated by the
  shaded green areas.}
\figl{1}
\end{figure*}

We begin by observing that if a crystal is left invariant under an
$M_z$ reflection operation, then its Bravais lattice must contain
vectors pointing along $z$. To construct the shortest such vector
$\a_3=c\zhat$, we pick the shortest vector $\wt\a_3$ connecting
lattice points on adjacent horizontal lattice planes. If $\wt\a_3$
points along $z$ then we take it as $\a_3$, and we say that the mirror
is of type 1. Otherwise we choose the vector $\a_3=\wt\a_3-M_z\wt\a_3$
connecting second-neighbor lattice planes, and the mirror is of
type~2.

The two types of crystallographic mirrors are exemplified in 2D in
\fref{1}, where the mirror lines $z=0$ and $c/2$ are labeled A and B,
and the reciprocal-space lines $k_z=0$ and $k_z=\pi/c$ are labeled G
and X.  The same notation will be used in 3D, where A and B (G and X)
become planes in real (reciprocal) space.

The distinction between mirror operations that leave pointwise
invariant two inequivalent planes in the BZ, and those that leave
invariant only one BZ plane, was made in
Refs.~\cite{varjas-prb15,fulga-prb16}.  Since MCNs are defined on such
planes~\cite{teo-prb08,ando-arcmp15}, a 3D insulator with a type-1
mirror is characterized by two separate MCNs $\mu\G$ and $\mu\X$,
while for a type-2 mirror there is a single MCN~$\mu\G$. If the
crystallographic space group contains additional mirror operations,
there will be additional MCNs associated with them.

%----------------------------------------------
\subsection{Chern invariants in band insulators}
%----------------------------------------------

%................................
\subsubsection{Generic insulators}
%................................

Before introducing MCNs for insulators with reflection symmetry, let
us define Chern invariants for generic 2D and 3D band insulators in
terms of the $\k$-space Berry curvature of the valence
states~\cite{vanderbilt-book18}.

In 2D, the Berry curvature of a Bloch state $\ket{\psi\bnk}$ with
cell-periodic part $\ket{u\bnk}$ is a scalar defined as
\beq
\Omega\bnk=-2\Imag\ip{\partial_{k_x} u\bnk}{\partial_{k_y} u\bnk}
\eql{Omega-n-2D}
\eeq
where $\k=(k_x,k_y)$, and the Chern number is given by
\beq
C=\frac{1}{2\pi}\int_{\rm 2D BZ}\sum_{n=1}^J\,\Omega\bnk\,d^2k
\eql{C}
\eeq
where the summation is over the $J$ filled energy bands. Since
the Berry curvature has units of length squared, $C$ is a
dimensionless number, and for topological reasons it must be an
integer. The Chern number is a global property of the manifold of
occupied states, remaining invariant under multiband gauge
transformations described by $J\times J$ unitary matrices at each
$\k$, and it vanishes when the crystal has time-reversal symmetry. If
a 2D magnetic crystal has a nonzero Chern number $C$, when that
crystal is terminated at an edge there will be $|C|$ edge modes
crossing the bulk gap, whose chirality will depend on the sign of~$C$.

3D insulators are characterized by a Chern vector
\beq
{\bf K}=\frac{1}{2\pi}\int_{\rm 3D BZ}\sum_{n=1}^J\,
{\boldsymbol\Omega}\bnk\,d^3k\,,
\eeq
where now $\k=(k_x,k_y,k_z)$ and the Berry curvature has become a
vector field,
${\boldsymbol\Omega}\bnk=-\Imag\bra{\partial_\k
  u\bnk}\times\ket{\partial_\k u\bnk}$.  The Chern vector has units of
inverse length, and is quantized to be a reciprocal-lattice vector.
Like the Chern number in 2D, the Chern vector always vanishes in
nonmagnetic crystals.

Given a set of lattice vectors $\a_j$ and dual reciprocal-lattice
vectors $\b_j$, the expansion ${\bf K}=\sum_j\, C_j\b_j$ defines a
triad of integer Chern indices~$C_j$.  Let us orient the Cartesian
axes such that $\a_3=c\zhat$.  The vectors $\b_1$ and $\b_2$ then lie
on the $(x,y)$ plane, and the third Chern index can be expressed as
\beq
C_3=\frac{c}{2\pi}\int_0^{2\pi/c}\,C(k_z)\,dk_z\,,
\eql{C3}
\eeq
where
\beq
C(k_z)=\frac{1}{2\pi}\int_{\rm 2D BZ}\sum_{n=1}^J\,
\Omega^z_n(k_x,k_y,k_z)\,dk_xdk_y\,.
\eql{C-kz}
\eeq
The integral in \eq{C-kz} is over a slice of the 3D BZ spanned by
$\b_1$ and $\b_2$ at fixed $k_z$. By viewing it as an effective 2D BZ
and comparing with \eq{C}, it becomes clear that $C(k_z)$ is a Chern
number; and since in a gapped system its integer value cannot change
with the continuous parameter $k_z$, \eq{C3} reduces to $C_3=C(k_z)$
evaluated at any $k_z$. The Chern indices of 3D insulators can
therefore be evaluated as Chern numbers defined over individual BZ
slices.

%........................................
\subsubsection{Mirror-symmetric insulators}
%........................................

We now consider a 3D crystalline insulator with mirror symmetry $M_z$,
and assume that its Chern vector ${\bf K}$ vanishes.  A new
integer-valued topological index, the MCN, can be defined for such a
system as follows~\cite{teo-prb08,ando-arcmp15}.

On the mirror-invariant BZ planes, G and possibly X, the energy
eigenstates are also eigenstates of $M_z$.  The eigenvalues are
$i^Fp$, where $p=\pm 1$ is the ``mirror parity'' and $F=0$ or $1$ when
the electrons are treated as spinless or spinful particles,
respectively.  The occupied Bloch states on those planes can therefore
be grouped into ``even'' ($p=+1$) and ``odd'' ($p=-1$) sectors under
reflection about the A plane $z=0$, each carrying its own Chern
number.  The Chern numbers of the two sectors on the G plane $k_z=0$
are given by
\beq
C\G^\pm=\frac{1}{2\pi}\int_{\rm 2D BZ}
\sum_{n=1}^J\,f\bnk^\pm
\Omega^z_n(k_x,k_y,k_z=0)\,dk_xdk_y\,,
\eql{C-G}
\eeq
where $f\bnk^+=1-f\bnk^-$ equals one or zero for a state with
$p=\pm 1$, respectively. The MCN is defined as
\beq
\mu\G=\frac{1}{2}\left( C\G^+ - C\G^- \right)\,,
\eql{muG}
\eeq
and it is guaranteed to be an integer since $C\G^+ + C\G^-=C_3$
vanishes by assumption. If the mirror is of type 1, the plane X
carries a second MCN
\beq
\mu\X=\frac{1}{2}\left( C\X^+ - C\X^- \right)\,,
\eql{muX}
\eeq
where $C\X^\pm$ is obtained by replacing $k_z=0$ with $k_z=\pi/c$ in
\eq{C-G}. The MCNs remain invariant under multiband gauge
transformations that do not mix the two mirror-parity sectors. When
they are nonzero, protected gapless modes appear on surfaces that
retain the mirror symmetry $M_z$, with $|\mu\G|$ and $|\mu\X|$
counting the number of Dirac cones on the two $M_z$-invariant lines in
the surface BZ~\cite{hsieh-nc12}.

In the case of a 2D or quasi-2D insulator with reflection symmetry
$M_z$ about its own plane, the entire 2D BZ is left invariant under
$M_z$. Such a system has a unique MCN
\beq
\mu_{\rm 2D}=\frac{1}{2}\left(C_+-C_-\right)\,,
\eql{mu-2D}
\eeq
where $C_+$ and $C_-$ are obtained by inserting the 2D Berry curvature
of \eq{Omega-n-2D} in \eq{C-G}. When the net Chern number $C=C_++C_-$
vanishes, $|\mu_{\rm 2D}|$ becomes an integer that counts the number
of pairs of counterpropagating chiral edge modes~\cite{liu-nmat14}.

We note in passing that spin-orbit coupling is required to obtain
non-vanishing MCNs in systems that are either non-magnetic or whose
magnetic order is collinear.

%---------------------------------------------
\subsection{The hybrid Wannier representation}
%---------------------------------------------

%.........................................................
\subsubsection{Hybrid Wannier functions and Wannier bands}
\secl{hwf}
%.........................................................

HW functions are obtained from the valence Bloch states of a 2D or 3D
crystalline insulator by carrying out the Wannier construction along a
chosen reciprocal-lattice direction. They are therefore localized
along one direction only, in contrast to ordinary Wannier functions
which are localized in all spatial directions.

Let us momentarily return to a generic 3D insulating crystal, not
necessarily mirror-symmetric. We denote by $z$ the chosen localization
direction and let $\kk=(k_x,k_y)$, so that the wavevector in the 3D BZ
becomes $\k=(\kk,k_z$).  Given a gauge for the Bloch states that is
periodic in $k_z$,
$\ket{\psi_{n\kk,k_z+2\pi/c}}=\ket{\psi_{n\kk k_z}}$, the
corresponding HW functions are defined as
\beq
\ket{h_{ln\kk}}=\frac{1}{2\pi}\int_{-\pi/c}^{\pi/c}\,
e^{-ik_zlc}e^{-i\kk\cdot\rr}\ket{\psi_{n\kk k_z}}\,dk_z,
\eql{hwf}
\eeq
where the index $l$ runs over unit cells along $z$, and $n$ runs over
the $J$ HW functions in one unit cell.  By factoring out
$e^{-i\kk\cdot\rr}$, we have made the HW functions cell periodic in
the in-plane directions, $h_{ln\kk}(\rr+\R)=h_{ln\kk}(\rr)$ for any
in-plane lattice vector $\R$.  This will be convenient later on when
we define Berry curvatures and Chern numbers in the HW representation.

For each $\kk$ in the projected 2D BZ, we choose the multiband gauge
for the Bloch states in such a way that the HW functions have the
smallest possible quadratic spread along $z$. Such maximally-localized
HW functions satisfy the eigenvalue equation~\cite{marzari-prb97}
\beq
P_\kk z P_\kk\ket{h_{ln\kk}}=z_{ln\kk}\ket{h_{ln\kk}},
\eql{PzP}
\eeq
where
$P_\kk$ is the projection operator onto the space of valence states
with in-plane wave vector $\kk$. The eigenvalues in \eq{PzP} are the
HW centers
\beq
z_{ln\kk}=\me{h_{ln\kk}}{z}{h_{ln\kk}}\,,
\eeq
which form Wannier bands. These are periodic in real space along $z$,
as well as in the in-plane reciprocal space,
\beq
z_{ln\kk}=z_{0n\kk}+lc\,,\quad  z_{ln,\kk+{\bf G}}=z_{ln\kk}\,,
\eeq
where ${\bf G}$ is an in-plane reciprocal lattice vector.

A Wannier band structure is said to be {\it gapped} if it contains at
least one Wannier band per vertical cell that is separated from the
band below by a finite gap at all $\kk$. When that is the case, we
choose the cell contents in such a way that the first band, $n=1$, has
a gap below~it.

%..............................................
\subsubsection{Chern numbers of Wannier bands}
\secl{chern-wannier}
% .............................................

The Berry curvature of a HW state is defined as
\beq
\Omega_{ln}=-2\,\Imag\ip{\partial_{k_x} h_{ln}}{\partial_{k_y} h_{ln}}\,,
\eql{Omega-n-HW}
\eeq
and periodicity along $z$ implies that $\Omega_{ln}=\Omega_{0n}$.
(Here and in the following, we will frequently drop the index~$\kk$.)
When the Wannier spectrum is gapped, it becomes possible to associate
a Chern number with each isolated group $a$ of bands within a vertical
cell,
\beq
C_{la}=\frac{1}{2\pi}\int_{\rm 2D BZ}\sum_{n\in a}\,
\Omega_{ln}\,d^2k=C_{0a}\,.
\eeq
From the HW states in a given group, one can construct Bloch-like
states at any $\k=(k_x,k_y,k_z)$ by inverting \eq{hwf}. In general
these are not energy eigenstates, and their band indices label Wannier
bands rather than energy bands.  Their Berry curvatures along $z$ are
given~by
\beq
\Omega^z_n(k_x,k_y,k_z)=
\sum_l\,e^{ik_zlc}\Omega_{0n,ln}(k_x,k_y)\,,
\eql{Omega-bloch-sum}
\eeq
where
\beq
\Omega_{0n,ln}
=i\ip{\partial_{k_x} h_{0n}}{\partial_{k_y} h_{ln}}-
i\ip{\partial_{k_y} h_{0n}}{\partial_{k_x} h_{ln}}
\eeq
is a matrix generalization of
\eq{Omega-n-HW}~\cite{taherinejad-prl15}. To evaluate the net Chern
number $C_a(k_z)$ of that group of Bloch-like states on a slice of the
3D BZ, we insert \eq{Omega-bloch-sum} in \eq{C-kz} and restrict the
summation over $n$ to $n\in a$. The contributions from the $l\not=0$
terms drop out,\footnote{The expression for $C_a(k_z)$ involves
  $\int_0^{2\pi/a}\partial_{k_x}Y_{0n,ln}(k_x)\,dk_x$ where
  $Y_{0n,ln}(k_x)=\int_0^{2\pi/b}A^y_{0n,ln}(k_x,k_y)\,dk_y$, and
  another similar integral
  $\int_0^{2\pi/b}\partial_{k_y}X_{0n,ln}(k_y)\,dk_y$.  When
  $l\not= 0$ the quantity $Y_{0n,ln}(k_x)$ becomes fully invariant
  under band-diagonal gauge transformations of the HW states.  Hence
  its value at $k_x=2\pi/a$ must be the same as at $k_x=0$, and the
  integral vanishes.}
yielding
\beq
C_a(k_z)=C_{0a}\,.
\eql{C-bloch-wannier}
\eeq
Hence the Chern numbers are the same in the Bloch-like and HW
representations, as expected since the two representations are related
by a unitary transformation.  When the group $a$ comprises all $J$
Wannier bands in one vertical cell, its Chern number becomes equal to
the Chern index $C_3$ of \eq{C3}, which vanishes by assumption.

%===========================================
\section{Mirror-symmetric Wannier bands}
%===========================================
\secl{mirror-wannier}

With the above background material in hand, we now return to our
system of interest --~a 3D insulator with $M_z$ symmetry~-- and
construct HW functions localized along the direction $z$ orthogonal to
the mirror plane.  We begin this section by discussing the generic
features of Wannier band structures with $M_z$ symmetry.

%---------------------------------------------------------
\subsection{Flat vs dispersive bands, and the uniform parity
  assumption}
%---------------------------------------------------------

If $M_z$ is a symmetry of the system, the operator $PzP$ anticommutes
with $M_z$.  It follows that if a HW function $\ket{h_{ln}}$ satisfies
\eq{PzP} with eigenvalue $z_{ln}$, $M_z\ket{h_{ln}}$ satisfies it with
eigenvalue $-z_{ln}$. Since $z_{ln}$ is only defined modulo~$c$, two
situations may occur.  (i) $\ket{h_{ln}}$ and $M_z\ket{h_{ln}}$ are
orthogonal, in which case a pair of dispersive bands appear at
$\pm z_{ln}$.  (ii) $\ket{h_{ln}}$ and $M_z\ket{h_{ln}}$ are the same
up to a phase, in which case $\ket{h_{ln}}$ is an eigenstate of $M_z$,
and a single flat band appears at either $z=0$ (A plane) or $z=c/2$ (B
plane).  The Wannier bands of the system can therefore be classified
into flat bands of even or odd mirror parity at A; flat bands of even
or odd mirror parity at B; and dispersive pairs appearing at $\pm z$.

If there are several flat bands on a given mirror plane and not all of
them have the same parity, those of opposite parity will generally
have a nonzero $PzP$ matrix element between them, and will tend to
hybridize and split to form dispersive pairs.  Thus, all flat bands
pinned at A are expected to have the same parity $\pA$, and all flat
bands pinned at B are expected to have the same parity $\pB$.
Following Ref.~\cite{varnava-prb20}, we call this the ``uniform
parity'' assumption.  As discussed in Ref.~\cite{varnava-prb20}, this
assumption is closely related to a well-known theorem on the minimum
number of zero-energy modes in bipartite
lattices~\cite{sutherland-prb86,lieb-prl89,ramachandran-prb17}.

Under the uniform parity assumption, the numbers $\Naf$ and $\Nbf$ of
flat bands at A and B can be expressed in terms of the imbalance
between even- and odd-parity valence Bloch states at the
mirror-invariant plane(s) in the BZ. For a type-1 mirror we have
\beq
\Naf=\frac{1}{2}
\left| \Delta N_{\rm G}+\Delta N_{\rm X} \right|
\eql{num-flat-1-A}
\eeq
and
\beq
\Nbf=\frac{1}{2}
\left| \Delta N_{\rm G}-\Delta N_{\rm X} \right|\,,
\eql{num-flat-1-B}
\eeq
where $\Delta N_{\rm G}$ and $\Delta N_{\rm X}$ denote the excess of
even over odd valence states at G and X, respectively. Hence if the
mirror-parity content is balanced at both G and X, flat Wannier bands
are absent from both A and B; if it is balanced only
at G but not at X or vice-versa, the same number of flat bands is
present at A and at B; and if it is unbalanced at both G and X, the
number of flat bands at B can differ from the number at A. The
corresponding relation for a type-2 mirror is
\beq
\Naf=\Nbf=\frac{1}{2}|\Delta N_{\rm G}|\,.
\eql{num-flat-2}
\eeq
\Eqr{num-flat-1-A}{num-flat-2} are derived in \aref{num-flat}.

%-----------------------------------------
\subsection{Types of generic degeneracies}
\secl{degens}
% ----------------------------------------

In this section, we consider the types of degeneracies that are
typical of the Wannier spectra of insulators with $M_z$ symmetry. We
call a degeneracy {\it generic} when it occurs without the assistance
of any symmetries other than $M_z$. If in addition the degeneracy is
codimension protected, we call it {\it accidental}.

Accidental degeneracies away from the A and B planes have codimension
three, and hence they require fine tuning.  On the mirror planes,
there are two types of generic degeneracies: multiple flat bands
pinned to the same plane, and accidental touchings, at isolated points
in the 2D BZ, between one or more pairs of dispersive bands.  Other
possibilities such as nodal lines are non-generic and will not be
considered further. In the following we focus on the A plane $z=0$,
but the discussion would be identical for the B plane $z=c/2$.

%..........................................................
\subsubsection{Point nodes between pairs of dispersive bands}
%..........................................................

If there are no flat bands pinned at $z=0$, any bands near $z=0$ must
come in dispersive pairs at $\pm z$. If there is a single pair, we
construct from the two HW functions at each $\kk$ a pair of orthogonal
states with opposite parities about $z=0$. In this basis, the $z$
operator is represented by a matrix of the form
\beq
\begin{pmatrix}
0&f_\kk\\
f^*_\kk &0
\end{pmatrix}\,,
\eql{PzP-2bands}
\eeq
with eigenvalues $z_\kk=\pm|f_\kk|$. The two bands touch at $z=0$ when
$|f_\kk|=0$, and for that to happen both the real and imaginary parts
of $f_\kk$ must vanish; this means that such degeneracies have
codimension two, and hence they occur at isolated points in the 2D
BZ. (When the bands disperse linearly close to the nodal point, the
degeneracy is called a ``Dirac node.'')  If more than one dispersive
band pair is involved, $f_\kk$ becomes a matrix. The degeneracy
condition $\det(f_\kk)=0$ again leads to point nodes on the $z=0$
plane. Generically, these are simple nodes where only two bands
meet. However, with additional symmetries or fine tuning, more than
one pair of bands may become degenerate at a given node.

In summary, pairs of dispersive Wannier bands can touch accidentally
at isolated points on a mirror plane free of flat bands.  We note that
the same happens, and for the same mathematical reasons, with the
energy bands of models with sublattice
symmetry~\cite{ramachandran-prb17}.

%.........................................
\subsubsection{Flat bands repel point nodes}
\secl{repel}
%.........................................

When one or more flat bands are present at $z=0$, they gap out the
point nodes. Let us show this for the simplest case of one flat band
surrounded by a dispersive pair. Choosing a basis of $M_z$ eigenstates
within this three-band space, the matrix representation of the $z$
operator takes the form
\beq
\label{eq:PzP-3bands}
% P_\kk zP_\kk=
\begin{pmatrix}
0&f_\kk&g_\kk\\
f^*_\kk&0&0\\
g^*_\kk&0&0
\end{pmatrix},
\eeq
where we have chosen the first basis state to have the opposite mirror
parity from the other two. The eigenvalues are $z_\kk=0$ (flat band)
and $z_\kk=\pm\sqrt{|f_\kk|^2+|g_\kk|^2}$ (dispersive pair). An
accidental degeneracy between the pair requires the real and imaginary
parts of both $f_\kk$ and $g_\kk$ to vanish (codimension four). In
general this cannot be achieved by adjusting $\kk$ alone; it also
requires fine tuning the parameters $f_\kk$ and $g_\kk$.

In conclusion, flat bands and point nodes do not generally coexist on
a mirror plane.  Although we have only shown this for the case of one
flat band plus one dispersive pair, the same result is expected to
hold when several flat bands and/or dispersive pairs are present.
That scenario has in fact been considered for the analogous problem of
energy bands in models with sublattice symmetry
~\cite{ramachandran-prb17}.

%..............................................................
\subsubsection{Spinful time-reversal symmetry excludes flat bands}
\secl{no-flat-bands}
% .............................................................

The presence of flat bands on the mirror planes can sometimes be ruled
out on the basis of symmetry. This is the case for a crystal that has
both $M_z$ symmetry and spinful time-reversal symmetry $\T$.  Since
$[P_\kk zP_\kk,\T]=0$, the standard Kramers-degeneracy argument
applies to the Wannier bands: if $\ket{h_\kk}$ is an eigenstate of
$P_\kk zP_\kk$ with eigenvalue $z_\kk$, then
$\ket{h^\prime_{-\kk}}=\T\ket{h_\kk}$ is an orthogonal eigenstate with
the same eigenvalue.  Now suppose that $\ket{h_\kk}$ is a flat-band
state at A, with $M_z$ eigenvalue $\lambda=\pm i$. Then
$\ket{h^\prime_{-\kk}}$ is also a flat-band state, and using
$[M_z,\T]=0$ we find that its mirror eigenvalue is
$\lambda^*=-\lambda$. Since the two flat bands have opposite mirror
eigenvalues, they will generally hybridize to form a dispersive pair.

Another example is a crystal that has both $M_z$ symmetry, and spinful
$\T$ combined with inversion $\I$. The combined symmetry $\I*\T$
renders the energy bands Kramers-degenerate at every $\k$, and since
$[M_z,\I*\T]=0$ and $M_z$ has purely imaginary eigenvalues, Kramers
pairs of Hamiltonian eigenstates on the invariant BZ planes have
opposite $M_z$ eigenvalues. The mirror-parity content is therefore
balanced on those planes, and from \eqr{num-flat-1-A}{num-flat-2} we
conclude that both $\Naf$ and $\Nbf$ vanish. (Note that while the
energy bands are Kramers degenerate in the presence of $\I*\T$
symmetry, the Wannier bands are not.  The difference is that $\I*\T$
commutes with the Hamiltonian, but it anticommutes with $PzP$.)
  
In summary, spinful time-reversal symmetry, either by itself or in
combination with inversion, rules out the presence of flat Wannier
bands on the mirror planes (under the uniform parity assumption).

%----------------------------------------------------
\subsection{Chern numbers in gapped band structures}
\secl{gapped}
% ---------------------------------------------------

When an $M_z$-symmetric Wannier band structure is gapped, the $J$
bands per cell can be grouped into three internally connected
collections~\cite{varnava-prb20}: one containing bands that are pinned
at A (over the entire 2D BZ or at isolated $\kk$ points), another
containing bands that are pinned at B, and a third containing
``unpinned'' bands, in the sense that they do not touch the mirror
planes anywhere in the 2D BZ. In Ref.~\cite{varnava-prb20} these three
collections were called {\it origin-centered}, {\it
  boundary-centered}, and {\it uncentered}, respectively.

Letting $\alpha=\text{A or B}$, in each vertical cell $l$ there are in
general
\begin{itemize}

\item $\overline N_{\alpha^+}$ flat bands at $\alpha$ of even parity, 

\item $\overline N_{\alpha^-}$ flat bands at $\alpha$ of odd parity,

\item $\wt N_\alpha$ dispersive bands touching at $\alpha$

\end{itemize}
in the $\alpha$-pinned collection, and $\Nuc$ dispersive bands in the
unpinned collection. (At this stage we do not yet assume uniform
parity for the flat bands, nor do we invoke the fact that flat bands
repel point nodes.)  In the home cell $l=0$, the dispersive bands in
the A-pinned collection come in pairs at $\pm z$, and those in the
B-pinned collection come in pairs at $z$ and $c-z$.  In the case of
the unpinned collection we have a choice, since the mirror-symmetric
partners never become degenerate; for definiteness, we choose the
contents of the home cell so that the bands in the unpinned collection
come in pairs at $\pm z$.

For each of the seven groups listed above, we can add up the Chern
numbers in that group to get $\overline C_{\alpha^\pm}$,
$\wt C_\alpha$, and $\Cuc$, keeping in mind that their sum $C_3$
vanishes by assumption,
\beq
C_{\rm A}+C_{\rm B}+\Cuc=0\,,
\eql{C=0}
\eeq
where
$C_\alpha=\overline C_{\alpha^+}+\overline C_{\alpha^-}+\wt C_\alpha$
is the net Chern number of the $\alpha$-pinned collection.  We further
decompose each of the three dispersive band subspaces into even and
odd sectors under reflection about their centers, and assign
separate Chern numbers to them,
\begin{subequations}
\begin{align}
\wt C_\alpha&=\wt C_{\alpha^+}+\wt C_{\alpha^-}\,,\eql{alpha}\\
\Cuc&=\Cucp + \Cucm\,.\eql{UC}
\end{align}
\eql{pairs-pm}
\end{subequations}
In \aref{winding} we
show that
\beq
\wt C_{\alpha^+}-\wt C_{\alpha^-}=W_\alpha\,,
\eql{Nw}
\eeq
where $W_\alpha$ is the sum of the winding numbers 
% \sref{winding-number-def}
of all the nodal points in the
projected 2D BZ on the $\alpha$ mirror plane.

The winding number of a nodal point $\kk_j$ is defined as
  ~\cite{asboth-book16}
\beq
W_j=\frac{1}{2\pi}\oint_{c_j}
\partial_\kk\gamma_\kk\cdot d\kk\,,
\eql{W-j}
\eeq
where the integral is over a small circle around the node. $W_j$ is an
integer, typically taking values $\pm 1$ according to how the
phase~$\gamma_\kk$ changes going around the node. In the simplest case
where a single pair of bands meet at the node, $\gamma_\kk$ is the
phase angle of the complex matrix element $f_\k$ appearing in
\eq{PzP-2bands}. If two or more pairs of bands meet at a node, $f_\k$
becomes a matrix and $\gamma_\kk$ becomes the phase angle of its
determinant (see \sref{winding-number}).  

Combining \eqs{alpha}{Nw} we obtain
\beq
W_\alpha=\wt C_\alpha-2\wt C_{\alpha^-}\,,
\eql{mod2}
\eeq
which shows that $\wt C_\alpha$ has the same even or odd
parity as $W_\alpha$.  Since band pairs in the unpinned collection do
not touch on the special planes, by applying the same argument in
\aref{winding} that leads to \eq{Nw} we obtain
\beq
\Cucp=\Cucm\,,
\eql{UC-pairs}
\eeq
which implies that their sum $\Cuc$ is always an even
number.\footnote{The fact that $\Cuc$ is even can
  also be seen as follows~\cite{varnava-prb20}.  The unpinned
  collection is formed by two disconnected groups of bands related by
  $M_z$ symmetry, which imposes the same Berry curvature at every
  $\kk$ in the two groups, and hence the same Chern number.}

%======================================================
\section{Mirror Chern numbers in the hybrid Wannier
  representation}
\secl{MCN}
%======================================================

We are finally ready to evaluate the MCNs in the HW representation,
and then relate them to the axion $\zt$ index.  In \sref{MCN-gapped}
we consider the case of a gapped Wannier spectrum, and in
\sref{MCN-gapless} we treat the gapless case.

%------------------------------------------
\subsection{Gapped Wannier band structure}
\secl{MCN-gapped}
% ------------------------------------------

To recap, a generic gapped Wannier band structure with $M_z$ symmetry
consists of seven band collections per cell. The four that are flat
have well-defined mirror parities, and the three that are dispersive
can be decomposed into even and odd sectors. This yields a total of
ten HW groups with well-defined parities, each carrying its own Chern
number.

%.............................
\subsubsection{Type-1 mirrors}
\secl{type-1}
%.............................

\begin{table}
\caption{\label{tab:parity} Parities under a type-1 mirror $M_z$ of
  Bloch-like states constructed from HW functions that are maximally
  localized along $z$. For spinful electrons, the parity is said to be
  ``even'' or ``odd'' when the $M_z$ eigenvalue is $+i$ or $-i$.}
  \begin{tabular}{l}
  \hline\hline\noalign{\smallskip}
  {\bf Bloch representation}\\\noalign{\smallskip}\noalign{\smallskip}
  \hspace{0.3cm}${\rm G}^+=\text{even about A (and even about B)}$\\
  \hspace{0.3cm}${\rm G}^-=\text{odd\,\, about A (and odd\,\, about B)}$\\ \noalign{\smallskip}\noalign{\smallskip}
  \hspace{0.3cm}${\rm X}^+\,=\text{even about A (and odd\,\, about B)}$\\
  \hspace{0.3cm}${\rm X}^-\,=\text{odd\,\, about A (and even about B)}$\\\noalign{\smallskip}\noalign{\smallskip}
  \hline\noalign{\smallskip}\noalign{\smallskip}
  {\bf Hybrid Wannier representation}\\\noalign{\smallskip}\noalign{\smallskip}
\hspace{0.3cm}${\rm A}^+$ = even about A,
generates ${\rm G}^+$ and ${\rm X}^+$\\
\hspace{0.3cm}${\rm A}^-$ = odd\,\, about A,
generates ${\rm G}^-$ and ${\rm X}^-$\\ \noalign{\smallskip}\noalign{\smallskip}
\hspace{0.3cm}${\rm B}^+$ = even about B,
generates ${\rm G}^+$ and ${\rm X}^-$\\ 
\hspace{0.3cm}${\rm B}^-$ = odd\,\, about B,
generates ${\rm G}^-$ and ${\rm X}^+$\\\noalign{\smallskip}\noalign{\smallskip}
\hspace{0.3cm}pairs C and ${\rm C}^\prime$,\,\,\,\,\,\,\,\,\,\,\,\,\,
generates ${\rm G}^+{\rm G}^-$ and ${\rm X}^+{\rm X}^-$\\
\noalign{\smallskip}\noalign{\smallskip}
\hline\hline
\end{tabular}
\end{table}

To evaluate the MCNs $\mu\G$ and $\mu\X$, we construct from each of
the ten HW groups a group of Bloch-like states by performing Bloch
sums along $z$, and recall from \eq{C-bloch-wannier} that their Chern
numbers on any constant-$k_z$ BZ slice (and, in particular, at G and
X) are the same as the Chern numbers of the parent HW groups.  The
final needed ingredient is \tref{parity}, which tells the mirror
parities at G and X of the Bloch groups coming from each of the HW
groups.  That table is valid for both spinless and spinful mirror
symmetry $M_z$, and it agrees with the parity rules for inversion
symmetry ${\cal I}$ in 1D~\cite{varnava-prb20}; this is consistent
with the fact that $M_z={\cal I}*C_2^z$ acts along $z$ in the same way
as ${\cal I}$.

To evaluate $\mu\G$, we need to split the occupied Bloch space at G
into even- and odd-parity sectors about A. According to \tref{parity},
their Chern numbers are
\beq
C\G^\pm=\left(\Cafpm+\Cadpm+\Cuc^\pm\right)+\left(\Cbfpm+\Cbdpm\right)\,,
\eeq
where the first and second groups of terms correspond to Wannier
groups that are even or odd about A and B, respectively. Inserting
this expression into \eq{muG} for $\mu\G$ and then using
\eqs{Nw}{UC-pairs}, we find
\beq
2\mu\G=\left( \Cafp - \Cafm \right) + \left(\Cbfp - \Cbfm \right)+\Wa+\Wb\,.
\eeq
Under the uniform parity assumption the first group of terms becomes
$\pA\Caf$, where $\Caf$ is the total Chern number of the flat bands at
A, all of the same parity $\pA=\pm 1$; similarly, the second group
becomes $\pB\Cbf$.  Thus we arrive at
\beq
\mu\G=\frac{1}{2}\left(\pA\Caf+\Wa\right)+
\frac{1}{2}\left(\pB\Cbf+\Wb\right)\,,
\eql{muG-w}
\eeq
and via similar steps \eq{muX} for $\mu\X$ turns into
\beq
  \mu\X=\frac{1}{2}\left(\pA\Caf+\Wa\right)-
  \frac{1}{2}\left(\pB\Cbf+\Wb\right)\,.
\eql{muX-w}
\eeq
Out of the three collections in a type-1 disconnected band structure,
the uncentered collection does not contribute to the MCNs; and the
A-centered and B-centered ones contribute as in \eqs{muG-w}{muX-w}.

\Eqs{muG-w}{muX-w} are a central result of this work, and in the
following sections we will extract several conclusions from them.  In
practical applications, those equations can often be simplified: since
flat bands and point nodes do not generically coexist on the mirror
planes, at least one of the two terms inside each parenthesis will
typically vanish.

Before proceeding, let us verify that \eq{muG-w} correctly yields an
integer value for $\mu\G$ when $C_3=0$.  First we eliminate the
winding numbers from \eq{muG-w} with the help of \eq{mod2}, and then
we take mod 2 on both sides of the resulting equation to find
\begin{align}
2\mu\G\text{ mod 2} &=
\left(\Caf+\Cad+\Cbf+\Cbd\right)\text{ mod 2}\nn&=
-\Cuc\text{ mod 2}\,,
% =0\,,
\eql{mcn-int}
\end{align}
where \eq{C=0} was used to go from the first to the second line. Given
that $\Cuc$ is an even number, we conclude that $\mu\G$ is an
integer. The proof is identical for \eq{muX-w}.

We emphasize that the separate contributions from the A- and
B-centered collection to \eqs{muG-w}{muX-w} are not always
integer-valued.  As can be seen from \eq{theta} below, those
contributions assume half-integer values when the axion angle is
quantized to $\theta=\pi$ by mirror symmetry; a concrete example where
this happens will be given in \sref{dirac}.

%----------------------------------------------------
\subsubsection{Relation to the quantized axion coupling}
\secl{theta}
%----------------------------------------------------

As mentioned in the Introduction, mirror symmetry belongs to the group
of ``axion-odd'' symmetries that reverse the sign of the axion angle
$\theta$.  When one or more such symmetries are present in a 3D
insulator with a vanishing Chern vector, $\theta$ is restricted to be
zero or $\pi$ mod $2\pi$, becoming a $\zt$ topological index.

In the case of mirror symmetry, where the band topology is already
characterized by the MCNs, there should be a relation between them and
the quantized $\theta$ value.  Below we derive that relation for an
insulator with a type-1 mirror and a gapped Wannier spectrum.  To that
end, we make use of the formalism of Ref.~\cite{varnava-prb20} for
expressing $\theta$ in the HW representation.

First we write $\mu\G+\mu\X$ by combining \eqs{muG-w}{muX-w}, and
eliminate the winding numbers using \eq{mod2}. Then we take mod 2 on
both sides to find
\beq
\left(\mu\G+\mu\X\right)\text{ mod 2}=C_{\rm A}\text{ mod 2}\,.
\eeq
Comparing with the relation
$\theta/\pi=C_{\rm A}\text{ mod 2}$~\cite{varnava-prb20}, valid for a
gapped spectrum in the presence of a $z$-reversing axion-odd symmetry
such as $M_z$, we conclude that
\beq
\frac{\theta}{\pi}=\left(\mu\G+\mu\X\right)\text{ mod 2}\,.
\eql{theta}
\eeq
Thus, the system is axion-even ($\theta=0$) or axion-odd
($\theta=\pi$) depending on whether the sum of the two MCNs associated
with $M_z$ is even or odd.  Previously, this result had been inferred
from an argument based on counting Dirac cones in the surface
BZ~\cite{varjas-prb15,fulga-prb16}.  Here, we have obtained it
directly as a formal relation between bulk quantities expressed in the
HW representation.  As we will see shortly, the same relation holds
when the Wannier spectrum is gapless.

%............................
\subsubsection{Type-2 mirrors}
\secl{type-2}
%............................

In a crystal with a type-2 mirror, where the planes A and B are
equivalent and G is the only mirror-invariant plane in reciprocal
space, the unique MCN $\mu\G$ is obtained by setting $\pB=\pA$,
$\Cbf=\Caf$, and $\Wb=\Wa$ in \eq{muG-w},
\beq
\mu\G=\pA\Caf+\Wa\,.
\eql{muG-type2}
\eeq

If flat bands are present at A, they repel the point nodes. Hence
$\Wa=0$, and therefore $|\mu\G|=|\Caf|$. Interestingly, in this case
the magnitude of the MCN does not depend on the parity of the
flat-band states; this simplifies considerably its numerical
evaluation, since one does not need to know how the basis orbitals
transform under $M_z$.  Given that only the magnitude (not the sign)
of the MCN is needed to establish the bulk-boundary correspondence,
this is a potentially useful result.

Inserting \eq{mod2} for $\Wa$ in \eq{muG-type2}, taking mod~2 on both
sides, and again comparing with $\theta/\pi=C_{\rm A}\text{ mod 2}$,
we conclude that in this case the relation between the axion $\zt$
index and the MCN reads
\beq
\frac{\theta}{\pi}=\mu\G\text{ mod 2}\,,
\eql{theta-type2}
\eeq
as stated in Ref.~\cite{fulga-prb16}.

%.............................................
\subsubsection{Weakly coupled layered crystals}
\secl{layers}
%.............................................
%
Consider a crystal composed of weakly coupled identical layers that
remain invariant under reflection about their own planes.  Following
Ref.~\cite{kim-prl15}, we assume that the layers are stacked exactly
vertically. In this case the reflection symmetry about the individual
layers becomes a type-1 mirror of the 3D structure, with two separate
MCNs $\mu\G$ and $\mu\X$. In the fully decoupled limit where there is
no $k_z$ dependence the G and X reciprocal planes become equivalent,
so that $\mu\X=\mu\G\equiv \mu_{\rm 2D}$ where $\mu_{\rm 2D}$ is the
MCN of an isolated layer [\eq{mu-2D}]. But since the MCNs are
integers, they cannot change if a weak interlayer coupling is
introduced, and from \eqs{muG-w}{muX-w} we obtain
\beq
\mu_{\rm 2D}=\frac{1}{2}\left(\pA\Caf+\Wa\right)
\eql{muG-2D}
\eeq
for the unique MCN of a weakly-coupled layered crystal.

If flat bands are present at A (the plane of a layer), then $\Wa=0$
and the net Chern number of the valence bands becomes $\Caf+\Cuc$;
since the net Chern number vanishes by assumption and $\Cuc$ is even,
$\mu_{\rm 2D}=\pA\Caf/2$ is clearly an integer.  In this case
$|\mu_{\rm 2D}|$ can be determined without knowing the parity of the
flat-band states, as in the case of a type-2 mirror with flat bands.

Let us now evaluate the axion $\zt$ index. Since
$\mu\G+\mu\X=2\mu_{\rm 2D}$ is an even number, \eq{theta} yields
\beq
\theta=0\text{ mod $2\pi$}\,.
\eeq
This is consistent with the assertion made in Ref.~~\cite{kim-prl15}
that weakly-coupled layered topological crystalline insulators are
analogous to ``weak topological insulators'' with a vanishing strong
$\zt$ invariant $\nu_0$.

%------------------------------------------
\subsection{Gapless Wannier band structure}
\secl{MCN-gapless}
%------------------------------------------

Let us now apply our formalism to a $M_z$-symmetric system with a
gapless Wannier spectrum. We start out by noting that such a spectrum
must have degeneracies at both A and B. On those special planes the
codimension is two, so point nodes are allowed. Flat bands can be
ruled out since they would repel any nodes and generate a gap, and we
assume that nodal lines are absent as well.

We are left with a scenario where there are point nodes at both A and
B, and these are connected by Wannier bands. The only way this can
happen without the assistance of other symmetries is if there are only
two Wannier bands, one in each half unit cell, since otherwise there
is generically a gap somewhere in each half cell (accidental
degeneracies away from A and B are not protected, since the
codimension is three). With the assistance of other symmetries, the
gapless spectrum may contain more than two bands per cell.

To treat the above scenario, we temporarily add a symmetric pair of
occupied orbitals at degeneracy-free planes $\pm z_0$, and initially
do not let them hop at all (completely isolated).  This will introduce
flat bands on those planes.  Now let the added orbitals hybridize with
other orbitals.  Since accidental degeneracies away from the mirror
planes are not protected, gaps will generally open up between the new
and the old Wannier bands (the only exceptions to this rule are
treated in the next paragraph). And since the added orbitals are
topologically trivial, they have no effect on the MCNs, which can now
be evaluated using the formalism of \sref{MCN-gapped} for gapped
spectra.  Setting $\Caf=\Cbf=0$ in \eqs{muG-w}{muX-w} therein, we
obtain
\beq
\mu\G=\frac{1}{2}\left(\Wa+\Wb\right)
\eql{muG-gapless}
\eeq
and
\beq
\mu\X=\frac{1}{2}\left(\Wa-\Wb\right)\,.
\eql{muX-gapless}
\eeq
But since $\Wa$ and $\Wb$ cannot be affected by orbitals inserted far
from the A and B planes, we conclude that
\eqs{muG-gapless}{muX-gapless} can be directly applied to the original
system with a gapless Wannier spectrum.

The above argument needs to be refined if the system is an axion-odd
insulator that has, in addition to $M_z$ symmetry, one or more
axion-odd symmetries that are $z$ preserving and symmorphic (e.g.,
spinful time reversal or vertical mirrors). The Wannier spectrum is
then guaranteed to be gapless, with adjacent bands touching at an odd
number of Dirac nodes~\cite{varnava-prb20}.  The solution is to weakly
break all such symmetries via some low-symmetry perturbation; the band
connectivity then becomes ``fragile,'' allowing gaps to open up once
the added orbitals hybridize with the original
ones~\cite{wieder-arxiv18,varnava-prb20}. The rest of the argument
proceeds as before, again with the conclusion that
\eqs{muG-gapless}{muX-gapless} can be directly applied to the original
system with a gapless spectrum. This scenario is illustrated in
\sref{Dirac-disconnected}, where the orbital insertion itself acts as
the symmetry-lowering perturbation.

To conclude, let us show that the relation~\eqref{eq:theta} between
the MCNs and the axion angle remains valid for gapless
spectra. \Eqs{muG-gapless}{muX-gapless} give $\mu\G+\mu\X=\Wa$, while
$\theta$ is equal to the sum of Berry phases of vanishingly small
loops around the nodes at A~\cite{varnava-prb20}. Since those Berry
phases divided by $\pi$ are equal to the node winding numbers
modulo~2~\cite{park-prb11}, \eq{theta} is immediately recovered.

%==================
\section{Methods}
\secl{methods}
%==================

%------------------------------------------------------------
\subsection{Tight-binding, {\it ab initio}, and Wannier methods}
\secl{framework}
%------------------------------------------------------------

In this work, the formalism for evaluating MCNs in the HW
representation is implemented in the tight-binding (TB) framework,
using a modified version of the \code{PythTB} code~\cite{pythtb}.
Illustrative calculations are carried out for 2D and 3D models with
mirror symmetry; some are simple toy models, while others are obtained
from {\it ab initio} calculations as described below. Each model is
specified by providing the on-site energies, the hopping amplitudes,
and the matrix elements of the position and mirror operators.

In the TB literature, it is common to assume that the position
operator is represented by a diagonal matrix in the TB basis,
\beq
\me{\varphi_{\R i}}{\rr}{\varphi_{\R' j}}=
(\R+\bm{\tau}_i)\delta_{\R,\R'}\delta_{ij}
\eql{diag-r}
\eeq
where $\bm{\tau}_i$ is the location of the $i$th basis orbital in the
home cell $\R=\0$.  This approximation is problematic for calculating
the Wannier bands of unbuckled monolayers, since it forces all bands
to lie flat on the $z=0$ plane: when all basis orbitals lie on the
$z=0$ plane and all off-diagonal matrix elements
$\me{\varphi_{\R i}}{z}{\varphi_{\R' j}}$ vanish, the matrix $Z_\kk$
that is diagonalized to obtain the HW centers [see \eqs{Z}{Z-solve}]
is the null matrix.

To apply our formalism to flat monolayers, any flat Wannier bands that
may be present must be robust and satisfy the uniform parity
assumption, while all other bands must be dispersive. To ensure that
this is so, one should retain some off-diagonal $z$ matrix elements.
For models based on {\it ab initio} Wannier functions this occurs
naturally, since the position matrix elements between the Wannier
functions are explicitly calculated, and they are generally nonzero
for nearby Wannier functions.  In the case of toy models, one needs to
assign nonzero values to some of the off-diagonal $z$ matrix elements
under reasonable assumptions.

The material chosen for the {\it ab initio} calculations is SnTe,
which we study as a flat monolayer in \sref{2DSnTe} and as a bulk
phase in \sref{3DSnTe}. We first calculate the electronic structure
from density-functional theory (DFT) using the \code{GPAW}
code~\cite{enkovaara-jpcm10}, and then use the \code{Wannier90}
code~\cite{MOSTOFI20142309} to construct well-localized Wannier
functions.  Lastly, TB models are generated by tabulating the matrix
elements of the Kohn-Sham Hamiltonian and of the position operator
between those Wannier functions.
  
The self-consistent DFT calculations are performed without including
spin-orbit coupling, which is added afterwards
non-selfconsistently~\cite{olsen-prb16}.  We use the
Perdew-Burke-Ernzerhof exchange-correlation
functional~\cite{perdew-prl96,perdew-prl97}, and describe the
valence-core interaction via the projector augmented wave
method~\cite{blochl-prb94}. The valence states are expanded in a
plane-wave basis with an energy cutoff of 600 eV, and the BZ is
sampled on $\Gamma$-centered uniform grids containing
$6\times6\times1$ and $6\times6\times6$ points for monolayer and bulk
SnTe, respectively. The projector augmented wave setup includes the
4$d$ semicore states of Sn in addition to the 5$s$ and 5$p$ states of
Sn and Te, yielding a total of 20 valence electrons for each SnTe
formula unit (one per cell for the monolayer, and two for the bulk).

For each formula unit, we construct 16 spinor Wannier functions of $s$
and $p$ character spanning the upper-valence and low-lying conduction
band states.  The Sn 4$d$ states, which give rise to flat bands lying
22~eV below the Fermi level, are excluded from the Wannier
construction.

As a first step towards obtaining well-localized Wannier functions, we
extract from the space of {\it ab initio} Bloch eigenstates at each
grid point $\k$ an $N$-dimensional subspace with the desired orbital
character ($N=16$ for the monolayer, and $N=32$ for the bulk). This is
achieved via the ``band disentanglement'' procedure of
Ref.~\cite{souza-prb01}, which involves specifying two energy windows,
known as the inner and the outer window, and a set of trial orbitals.
The outer window encloses all the valence bands except for the 4$d$
semicore states, as well as all the low-lying conduction states of
$5s$ and $5p$ character. To ensure that the valence states are exactly
preserved in the disentangled subspace, we ``freeze'' them inside an
inner window. An initial guess for the target subspace is obtained by
projecting atom-centered $s$ and $p$ trial orbitals onto the
outer-window states. This is followed by an iterative procedure that
yields an optimally-smooth disentangled subspace across the
BZ~\cite{souza-prb01}.

Having extracted a suitable Bloch subspace, we proceed to construct
well-localized $s$- and $p$-like Wannier functions spanning that
subspace. This is done by projecting onto it the same $s$ and $p$
trial orbitals that were used in the disentanglement step, and then
orthogonalizing the resulting orbitals via the L\"owdin
scheme~\cite{marzari-prb97}. This one-shot procedure, without
additional maximal-localization steps~\cite{marzari-prb97}, ensures
that the Wannier functions retain the orbital character of the trial
orbitals.

To assess the quality of the Wannier basis we calculate the energy
bands from the Hamiltonian matrix elements in that
basis~\cite{souza-prb01}, and find that they are in excellent
agreement with the {\it ab initio} bands obtained using the
\code{GPAW} code~\cite{olsen-prm19}.

In addition to the Hamiltonian and position matrix elements, we also
require the matrix elements of the mirror operator $M_z$ in the
Wannier basis. These are needed to determine the winding numbers of
the nodal touchings between Wannier bands on the mirror planes (see
\sref{winding-number}), as well as the mirror parities $\pA$ and $\pB$
of the flat-band states. To set up the matrix representation of $M_z$,
we assume that the Wannier functions transform under $M_z$ in the same
way as pure $s$ and $p$ orbitals.  We find that the eigenstates of the
Wannier Hamiltonian on the mirror-invariant BZ planes are, to a good
approximation, eigenstates of this approximate $M_z$ operator, which
validates that assumption.

%----------------------------------------------------------------
\subsection{Construction of hybrid Wannier functions and Wannier
  bands}
\secl{HWF-wannier-bands}
%----------------------------------------------------------------

Formally, maximally-localized HW functions satisfy the eigenvalue
equation~\eqref{eq:PzP}.  For a 2D or quasi-2D system extended along
$x$ and $y$, the matrix elements of the $z$ operator appearing in that
equation are well defined.  It is therefore straightforward to set up
the matrix
\beq
Z_{mn\k}=\me{\psi\bmk}{z}{\psi\bnk}\,,
\eql{Z}
\eeq
where $\k=(k_x,k_y)$ and $m$ and $n$ run over the
$J$ occupied energy bands, and to diagonalize it,
\beq
\left[ U^\dagger_\k Z_\k U_\k\right]_{mn}=z_{m\k}\delta_{mn}\,.
\eql{Z-solve}
\eeq
The eigenvalues are the HW centers, and from the eigenvectors (the
columns of the $U_\k$ matrix) we can construct the maximally-localized
HW functions according to
\beq
\ket{h\bnk}=\sum_m\,e^{-i\k\cdot\rr}\ket{\psi\bmk}U_{mn\k}\,,
\eql{hwf-solve}
\eeq
where the phase factor has been included to render them in-plane
periodic.

For bulk systems, which are extended in all directions including the
wannierization direction~$z$, the above procedure fails because the
matrix elements in \eq{Z} become ill defined. In such cases, it is
still possible to construct maximally-localized HW functions by
working in reciprocal space.  We now write $\k=(\kk,k_z)$, and choose
a uniform grid; for each point $\kk$ in the projected 2D BZ, the
problem reduces to the construction of 1D maximally-localized Wannier
functions along~$z$.  The procedure is detailed in
Refs.~\cite{vanderbilt-book18,marzari-prb97}.  Briefly, the first step
is to establish a ``twisted parallel transport gauge'' for the valence
Bloch states along the string of $k_z$ points at each~$\kk$, obtaining
as a byproduct the HW centers $z_{ln\kk}$. The maximally-localized HW
functions $\ket{h_{ln\kk}}$ are then constructed in this gauge using
\eq{hwf}, with the integral over $k_z$ replaced by a summation over
the string of $k_z$ points.

%------------------------------------------------------------
\subsection{Winding number of a point node of order $N$
  % on a mirror plane
}
\secl{winding-number}
%------------------------------------------------------------
%

\subsubsection{Definition}
\secl{winding-number-def}

Earlier, we defined the winding number of a point node where two
  Wannier bands meet on a mirror plane. Since there are situations
  where $N>1$ pairs of bands meet at a node, we need to generalize
  that definition to handle such ``higher-order'' nodes.

Given a point node $\kk_j$ of order
  $N\geq 1$, we introduce
the $2N\times 2N$ matrix
representation of $M_z$
at a
nearby point $\kk$,
\beq
{\cal M}^z_{mn\kk}=\me{h_{m\kk}}{M_z}{h_{n\kk}}\,.
\eeq
Here, $m$ and $n$ run over the $2N$ Wannier bands that meet at
$\kk_j$. By diagonalizing ${\cal M}^z_\kk$ and then transforming the
$\ket{h_{n\kk}}$ states accordingly [see \eqs{Z-solve}{hwf-solve}], we
obtain a new set of $2N$ states $\ket{\tilde h_{n\kk}}$.  Like the
original ones they are cell-periodic in plane and localized along $z$,
but they have definite mirror parities. We choose the first $N$ to be
even under $M_z$, and denote them as $\ket{\tilde h^+_{l\kk}}$; the
remaining $N$ are odd under $M_z$, and we denote them as
$\ket{\tilde h^-_{l\kk}}$.  In both cases, $l$ goes from 1 to $N$.
The matrix representation of $z$ in
the new basis takes the form of \eq{PzP-2bands}, where $f_\kk$ is the
$N\times N$ matrix with elements
\beq
f_{ll'\kk}=\me{\tilde h^+_{l\kk}}{z}{\tilde h^-_{l'\kk}}\,.
\eql{f-k}
\eeq
Letting
\beq
\gamma_\kk=\arg(\det f_\kk)\,,
\eql{gamma-k}
\eeq
the winding number can be evaluated from \eq{W-j} irrespective of
  the order $N$ of the node.
%
% defined as~\cite{asboth-book16}
% %
% \beq
% W_j=\frac{1}{2\pi}\oint_{c_j}
% \partial_\kk\gamma_\kk\cdot d\kk\,,
% \eql{W-j}
% \eeq
% %
% where the integral is over a small circle around the node.

%..................................
\subsubsection{Numerical evaluation}
\secl{winding-number-num}
%..................................

Suppose a single pair of Wannier bands meet at a point node $\kk_j$.
To evaluate the winding number~\eqref{eq:W-j}, the phase $\gamma_\kk$
must be smooth on $c_j$. In practice, we establish a smooth gauge for
the states $\ket{\tilde h^\pm_\kk}$ as follows. We pick a
representation of the two states at a reference point $\kk'_j$ in the
vicinity of the node. Then at any point $\kk'_j+\Delta\kk$ on the
circle $c_j$ we choose the gauge by enforcing maximal phase alignment
with the states at $\kk'_j$, i.e., by requiring that the overlaps
$\ip{\tilde h^+_{\kk'_j}}{\tilde h^+_{\kk'_j+\Delta\kk}}$ and
$\ip{\tilde h^-_{\kk'_j}}{\tilde h^-_{\kk'_j+\Delta\kk}}$ are real and
positive. In other words, we carry out a one-step parallel transport
from $\kk'_j$ to each circumference point.

If several pairs of bands meet at a node, the strategy is basically
the same. The only difference is that one must now use the multiband
version of the parallel-transport
procedure~\cite{vanderbilt-book18,marzari-prb97}.

%===========================
\section{Numerical results}
\secl{results}
%===========================

In this section, we use our formalism to calculate the MCNs of three
different systems. The first is an unbuckled monolayer of SnTe, a
topological crystalline insulator protected by reflection symmetry
about its plane.  The second is rocksalt SnTe, a 3D topological
crystalline insulator protected by a type-2 mirror.  Our last example
is a 3D toy model based on a modified Dirac equation. It is both a
strong topological insulator protected by time-reversal symmetry, and
a topological crystalline insulator with a type-1 mirror. In the first
example the Wannier spectrum is trivially gapped, while in the other
two it is gapless.

%-----------------------------------------
\subsection{Unbuckled monolayer of SnTe}
\secl{2DSnTe}
%-----------------------------------------
 
\begin{figure}
\begin{center}
\includegraphics[width=0.99\columnwidth]{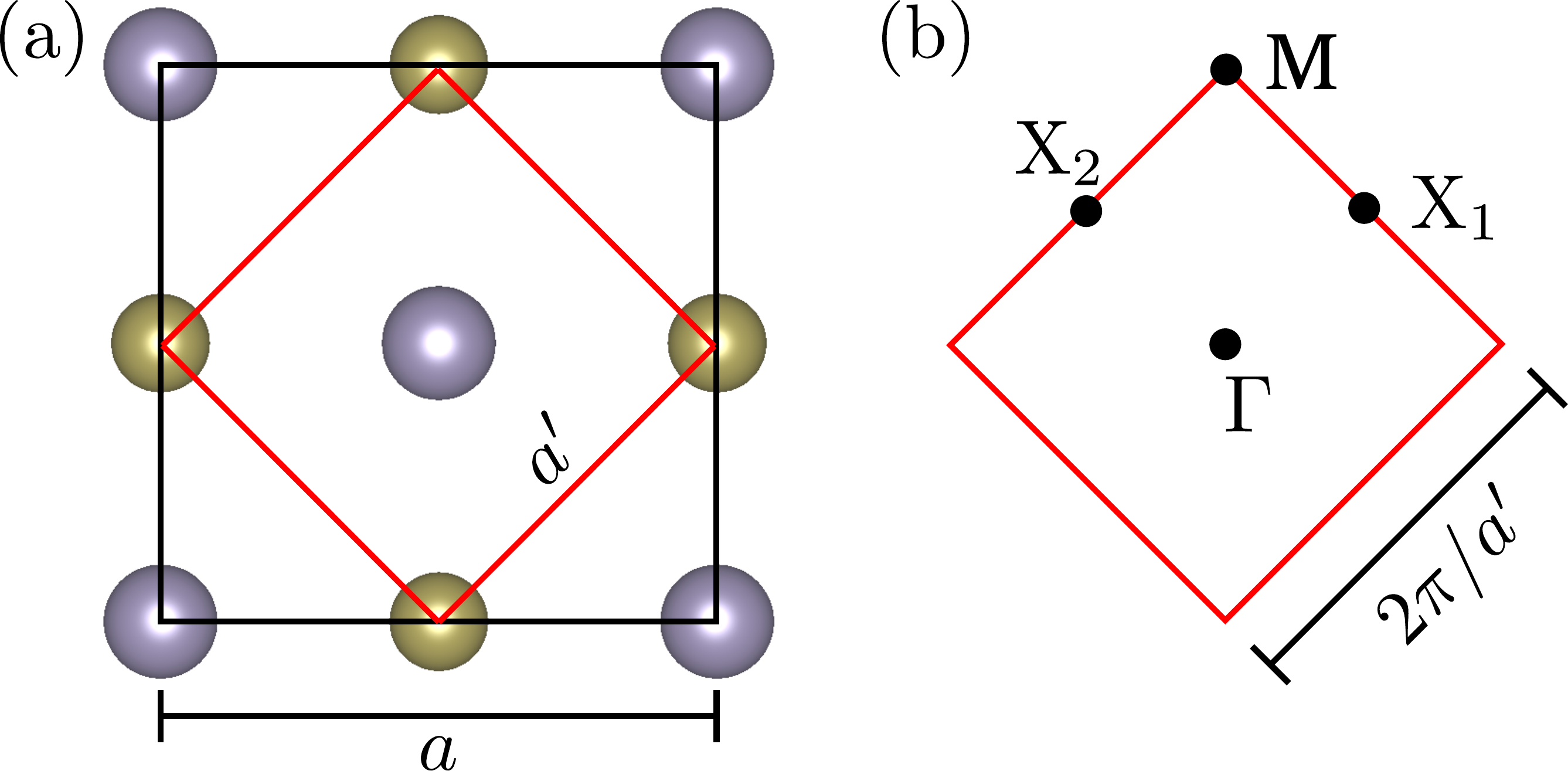}
\caption{(a) Atomic structure of monolayer SnTe. The black square is
  the conventional unit cell with lattice constant $a$, and the red
  square is the primitive cell with lattice constant
  $a^\prime = a/\sqrt{2}$. (b) Brillouin zone and high-symmetry
  points.}
\figl{2dSnTe-struc}
\end{center}
\end{figure}

\begin{figure*}
\begin{center}
\includegraphics[width=0.66\columnwidth]{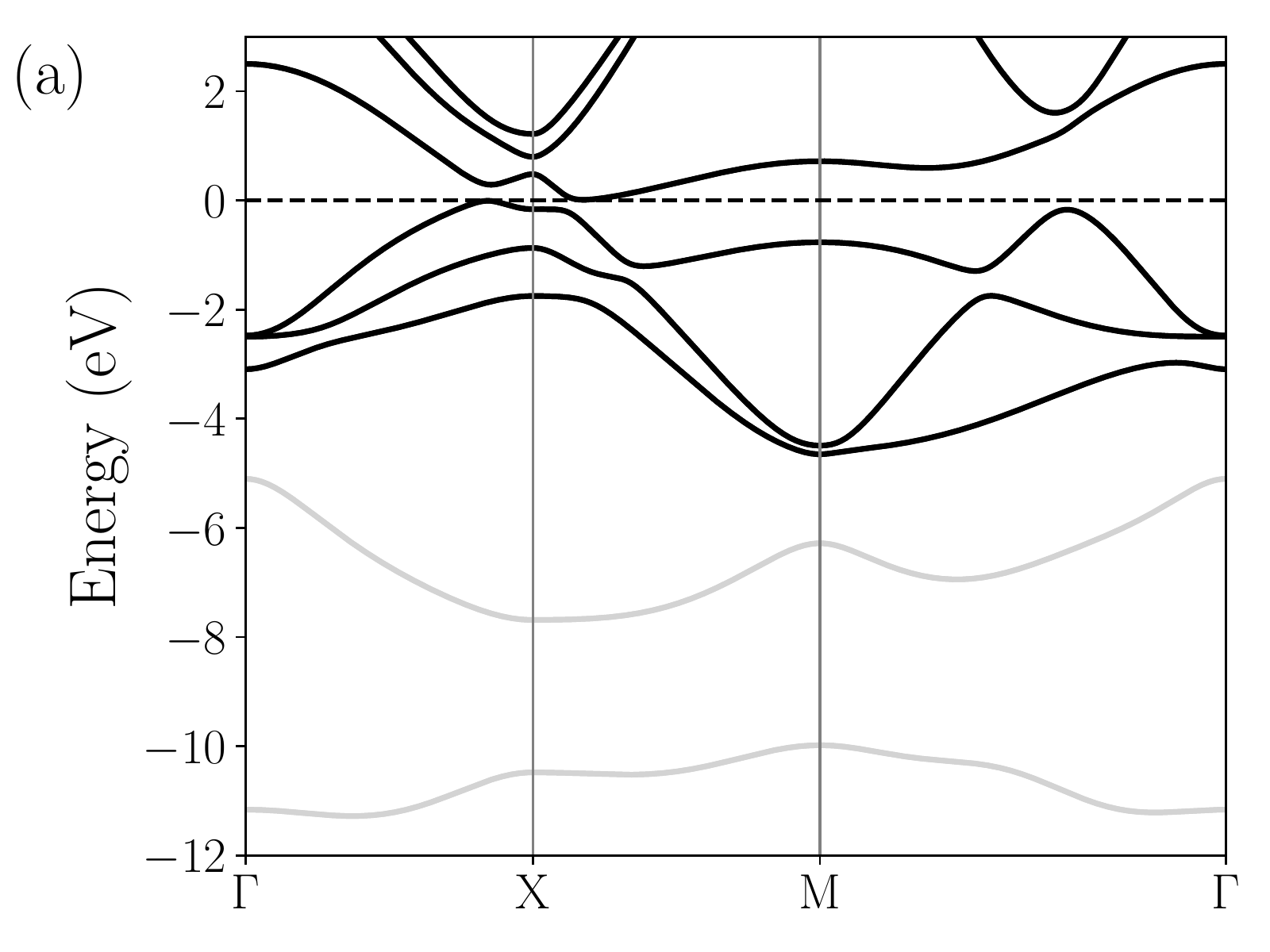}
\includegraphics[width=0.66\columnwidth]{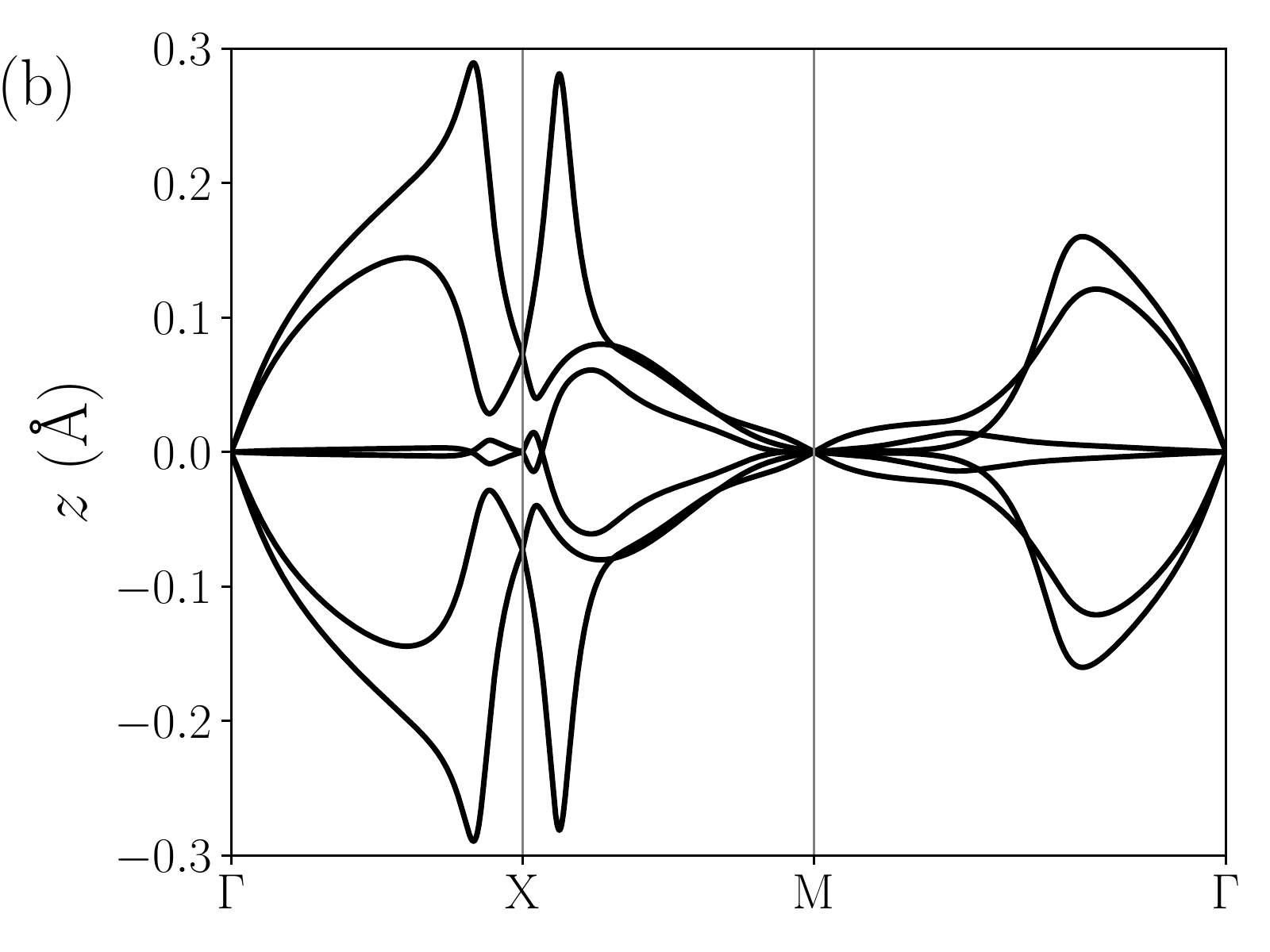}
\includegraphics[trim=10 8 30 20,clip,width=0.66\columnwidth]{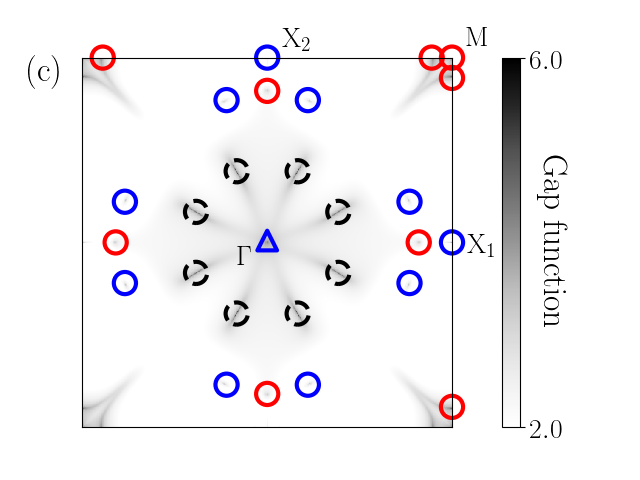}
\caption{(a) Energy bands of monolayer SnTe, with the $s$-type lower
  valence bands that are exluded from the Wannierization shown in
  grey.  All bands are doubly degenerate, and the Fermi level is
  indicated by the dashed line.  (b) Wannier bands obtained from the
  Bloch states in the six $p$-type upper valence bands.  (c)~Heatmap
  plot of the gap function of \eq{gap-fct} for the central pair of
  Wannier bands, where zero-gap points (nodal points) appear as dark
  spots.  Those with winding numbers $W_j=\pm 1$ are indicated by red
  or blue circles, while the one with $W_j=-3$ at the $\Gamma$ point
  is indicated by a blue triangle. Dashed circles denote pairs of
  nearby nodes with equal and opposite winding numbers. When a node
  falls on the BZ boundary, only one of the periodic images is shown.}
\figl{2dSnTe}
\end{center}
\end{figure*}

The structure we consider is shown in \fref{2dSnTe-struc}(a). It
consist of a single unbuckled layer of Sn and Te atoms arranged in a
checkerboard pattern, which can be viewed as a single (001) layer of
the bulk rocksalt structure.

DFT calculations reveal that the system with an optimized lattice
constant of $a=6.16$~\AA\, is situated 0.4 eV above the convex hull
and is dynamically unstable~\cite{haastrup-2dmat18}, and that a
buckled structure that breaks mirror symmetry is energetically
favored~\cite{Kobayashi2015}.  These results imply that a flat SnTe
monolayer is not likely to be experimentally relevant. This system is
nevertheless ideally suited for illustrating our methodology, since it
has reflection symmetry about its own plane and the associated MCN is
nonzero~\cite{liu-nanolett15}.

We carry out calculations using the primitive cell containing one
formula unit.  The Wannier-interpolated energy bands are shown in
\fref{2dSnTe}(a), where all bands are doubly degenerate due to
time-reversal and inversion symmetry.  There is a robust inverted gap
($0.3$~eV) at the X point, and a tiny indirect gap ($0.17$~meV) around
the X point; when the lattice expands the indirect gap increases, and
when it shrinks the system turns into a band overlap
semimetal~\cite{liu-nanolett15,Kobayashi2015}.  The lowest four
valence bands are predominantly $s$-type, and the remaining six
(plotted in red) are predominantly $p$-type.

\Fref{2dSnTe}(b) shows the Wannier bands calculated from the Bloch
states in the $p$-type upper valence bands. The spectrum consists of
three mirror-symmetric band pairs that touch on the A plane $z=0$ at
isolated points in the 2D BZ. There are no flat bands on that plane,
as expected from the presence of time-reversal symmetry
(\sref{no-flat-bands}). \Eq{muG-2D} therefore reduces to
\beq
\mu_{\rm 2D}=\frac{1}{2}\Wa\,,
\eql{mu-SnTe}
\eeq
and the MCN can be determined by evaluating the winding numbers of the
nodal points on the A plane.

To locate those nodal points, we plot in \fref{2dSnTe}(c) the ``gap
function''
\beq
g_\k = -\log(\Delta z_\k/c)\,,
\eql{gap-fct}
\eeq
where $\Delta z(\k)$ is the separation between the central pair of
bands. Regions with a small gap appear in dark gray, and nodal points
as dark spots. The positions and winding numbers of all the nodal
points are indicated in the figure, where we have included only one of
the periodic images when a node falls on the BZ boundary. At $\Gamma$
and M there are nodes where three pairs of Wannier bands touch, with
winding numbers $W_j=-3$ and $W_j=+1$, respectively. All other nodes
on the $z=0$ plane are simple Dirac nodes where only the two central
bands meet, and they have $W_j=\pm 1$. Adding up the winding numbers
of the 36 nodal points in the BZ we obtain $\Wa=-4$, and from
\eq{mu-SnTe} we conclude that the group of six $p$-type valence bands
has a MCN of $-2$.

We repeat the calculation for the four $s$-type lower valence bands,
and find that their net winding number vanishes. The net MCN of the
occupied states is therefore $\mu_{\rm 2D}=-2$, with the nontrivial
topology coming from the $p$ states. This result agrees with the value
$|\mu_{\rm 2D}|=2$ inferred from a $k\cdot p$ analysis of the
simultaneous band inversions at the two X points in the
BZ~\cite{liu-nmat14,liu-nanolett15}.

%----------------------
\subsection{Bulk SnTe}
\secl{3DSnTe}
%----------------------

Bulk SnTe, which crystallizes in the rocksalt structure, is known both
from theory~\cite{hsieh-nc12} and experiment~\cite{tanaka-natphys12}
to be a topological crystalline insulator. The symmetry protecting its
nontrivial band topology is reflection about the $\{110\}$ family of
planes. (Instead, the (001) mirror symmetry responsible for the
topological state of the monolayer is topologically trivial in the
bulk crystal.)

The lattice is face-centered cubic lattice, so that the shortest
lattice vector perpendicular to the (110) planes is
$\a_3=a\xhat/2+a\yhat/2$.  Since its length is twice the separation
between adjacent planes, the (110) mirror operation is of type 2, as
is typical of centered lattices (see \fref{1}).

For our simulations we pick a tetragonal cell subtended by
$\a_1=-a\xhat/2+a\yhat/2$, $\a_2=a\zhat$, and $\a_3$, and reorient the
axes such that those vectors point along $\xhat$, $\yhat$, and
$\zhat$, respectively. In this new frame, the (110) mirror operation
of interest becomes $M_z$. The simulation cell with two formula units
is shown in \fref{3DSnTe-struc}(a), and the associated BZ in
\fref{3DSnTe-struc}(b).

\begin{figure}
\begin{center}
\includegraphics[trim = 0 0 0 190,clip,width=0.99\columnwidth]{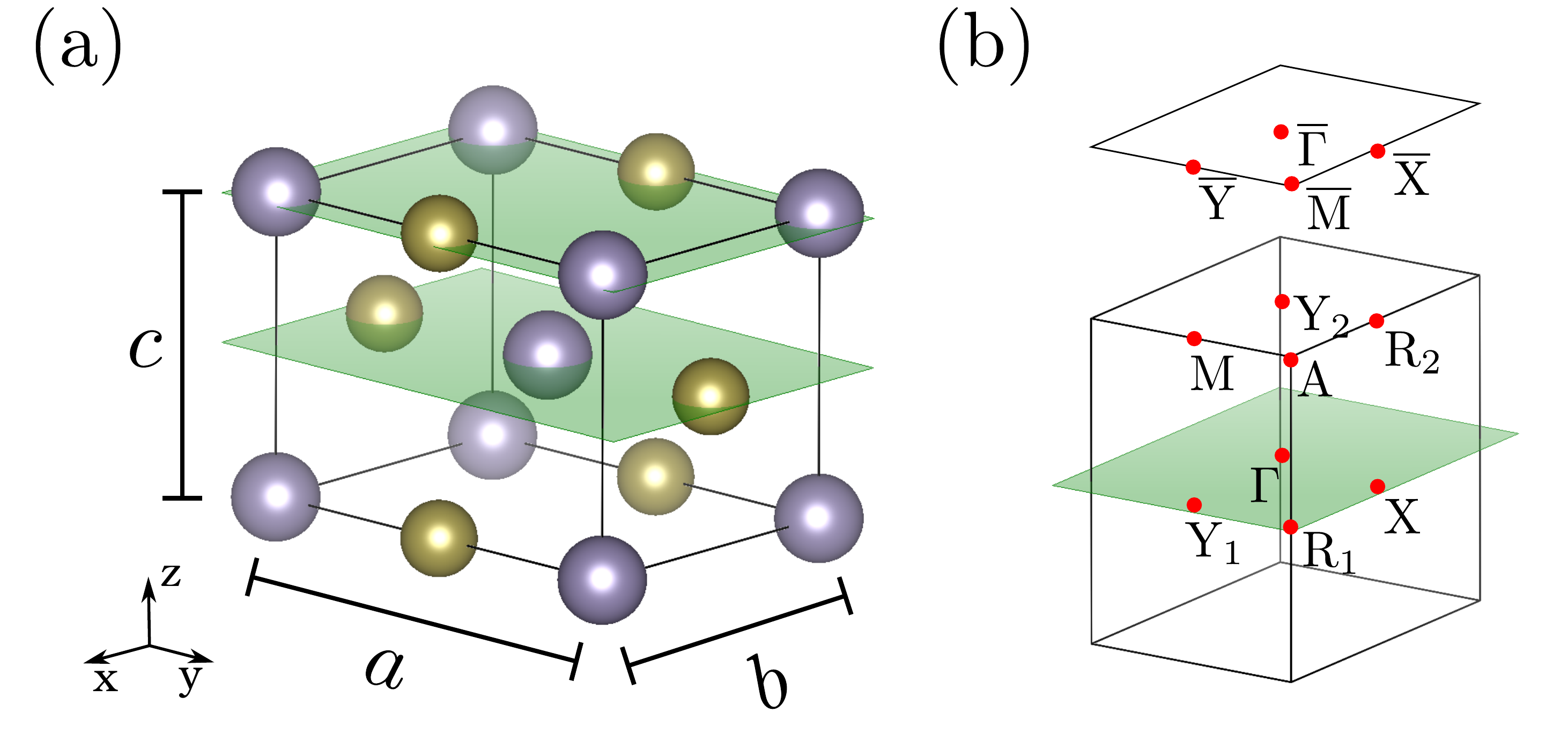}
\caption{(a) Rocksalt structure of bulk SnTe in a tetragonal
  conventional cell. $a$ is the lattice constant of the conventional
  cubic cell, and $b = c = a/\sqrt{2}$. Green planes are equivalent
  mirror planes. (b) Brillouin zone associated with the tetragonal
  cell, with its high-symmetry points indicated in red and the unique
  $M_z$-invariant plane in green. The projected 2D Brillouin zone with
  its high-symmetry points is shown on top.}
\figl{3DSnTe-struc}
\end{center}
\end{figure}

\begin{figure*}
\begin{center}
\includegraphics[width=0.49\columnwidth]{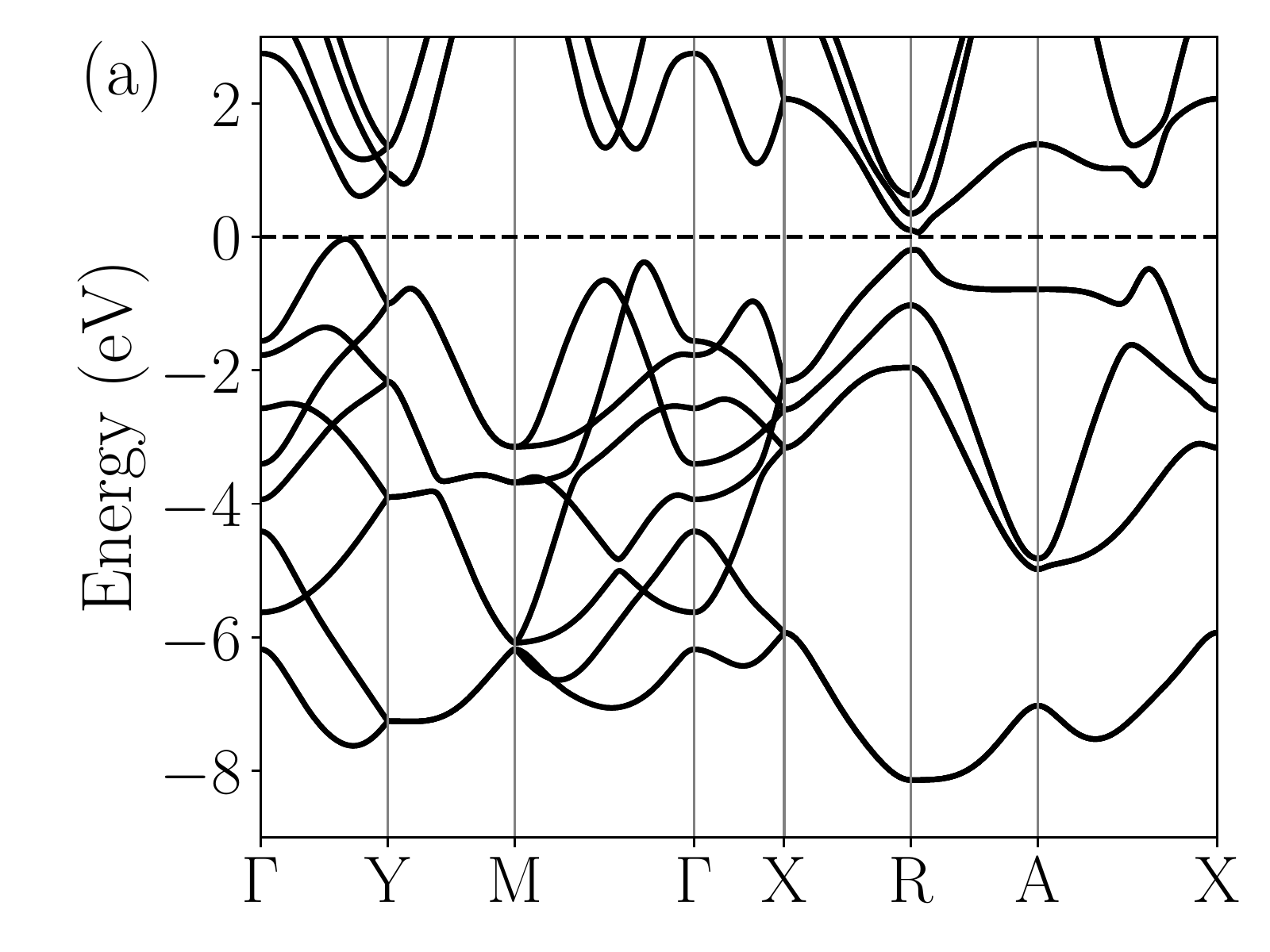}
\includegraphics[width=0.49\columnwidth]{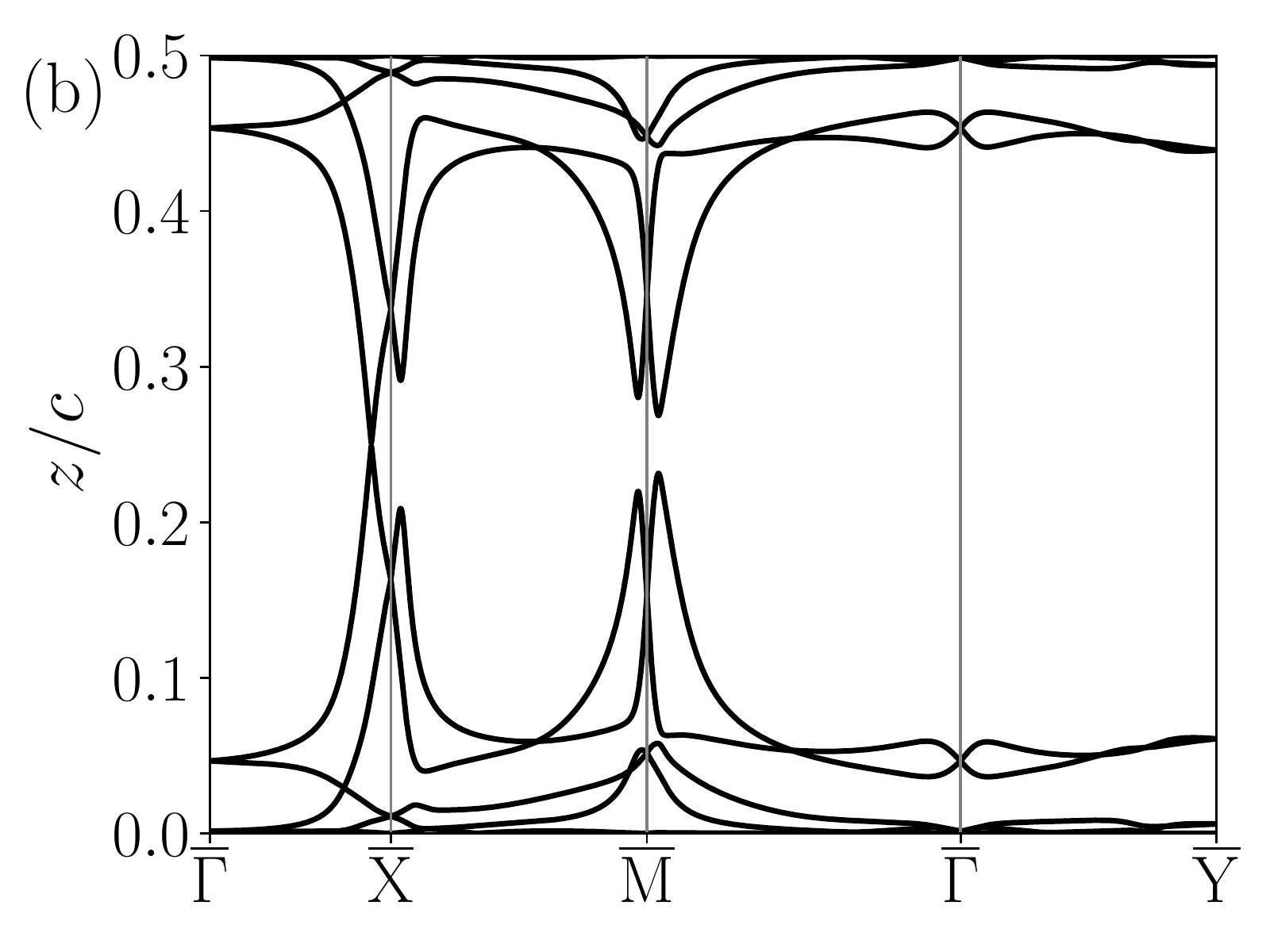}
\includegraphics[width=0.49\columnwidth]{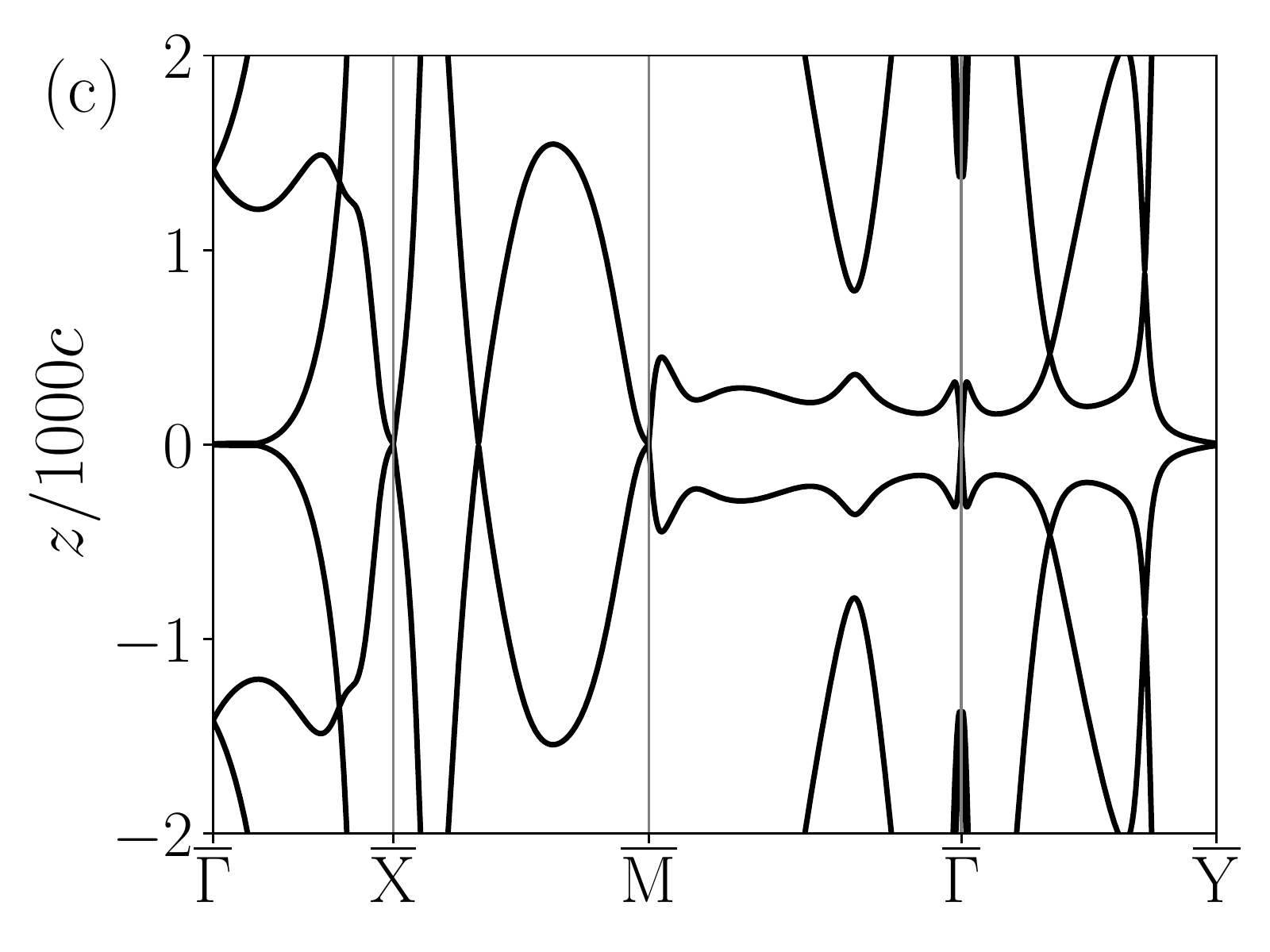}
\includegraphics[trim = 30 20 20 20,clip,width=0.49\columnwidth]{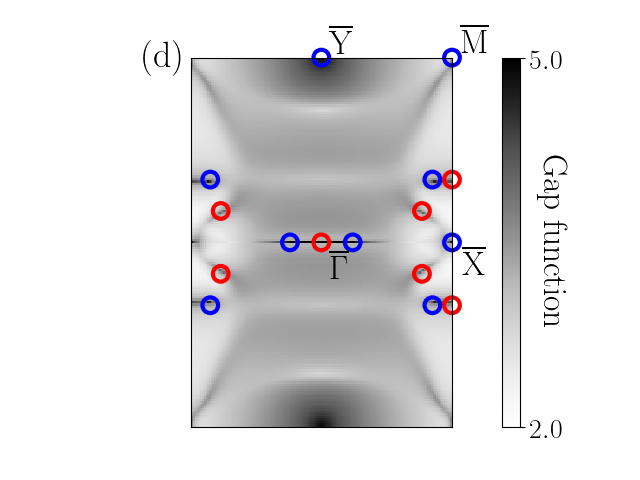}
\caption{(a) Energy bands of bulk SnTe along high-symmetry lines of
  the folded tetragonal BZ.  The Fermi level is indicated by the
  dashed line. (b) Wannier band structure obtained from the full set
  of valence states.  (c) Detail of the Wannier bands around the $z=0$
  mirror plane. (d) Heatmap plot of the gap function of \eq{gap-fct}
  for the central pair of Wannier bands around $z=0$, with the nodal
  points color-coded as in \fref{2dSnTe}(c).}
\figl{3DSnTe}
\end{center}
\end{figure*}

In \fref{3DSnTe}(a) we present the energy bands calculated along the
high-symmetry lines of the folded BZ.  The nontrivial topology arises
from simultaneous band inversions at the two L points in the unfolded
BZ~\cite{hsieh-nc12}, which map onto the two R points in
\fref{3DSnTe-struc}(b).  The inverted band gap at R and the global
indirect band gap amount to $0.3$ and $0.1$~eV, respectively.
  
From the full set of valence band states, we construct HW functions
localized along $z$. The Wannier spectrum is shown in
\fref{3DSnTe}(b). Its periodicity is $c/2$ because the cell is doubled
along $z$, and only one period is shown.  The spectrum is gapless,
with two pairs of bands crossing in opposite directions, between
$\overline{\rm X}$ and $\overline{\Gamma}$, the gap centered at
$z=c/4$ (only one of the two crossings is shown).  This spectral flow
arises from the nonzero MCN associated with $M_y$ symmetry (equivalent
to $M_z$), which leaves invariant the BZ plane containing the
$\Gamma$, X, ${\rm R}_2$, and ${\rm Y}_2$ points.  For a discussion of
such ``in-plane'' Wannier flow associated with a nonzero MCN, see
Ref.~\cite{gresch-prb17}.

Since $M_z$ is a type-2 mirror, we evaluate its unique MCN using
\eq{muG-type2}. And since the Wannier spectrum is gapless, and hence
devoid of flat bands, we set $\Caf=0$ in that equation to obtain
\beq
\mu\G=\Wa\,,
\eeq
which says that the MCN equals the sum of the winding numbers of all
the point nodes on the $z=0$ plane.

As indicated in \fref{3DSnTe}(d), there are 16 independent point nodes
in total on that plane, all of them simple nodes where only two bands
meet.  Seven have winding numbers $+1$ and the other nine have winding
numbers $-1$, yielding $\mu\G = -2$ for the MCN. This value is in
agreement with that originally obtained in Ref.~\cite{hsieh-nc12} from
a $k\cdot p$ analysis of the band inversions. Using \eq{theta-type2},
we confirm that the system is axion-trivial.

%--------------------------------------------------
\subsection{ Modified Dirac model on a cubic lattice}
\secl{dirac}
%--------------------------------------------------

In this section we study a 3D toy model constructed by first modifying
the free Dirac equation to enable topological phases for certain
parameter values, and then placing it on a cubic lattice.  The
4$\times$4 Hamiltonian matrix in reciprocal space
reads~\cite{doi:10.1142/S2010324711000057,Rauch2017}
\begin{widetext}
\begin{equation}
H(\k)=\left(
\begin{matrix}
m - 2 M K(\mathbf{k}) % + B_z
& 0 & c \sin k_z & c(\sin k_x-i \sin k_y) \\
0 & m - 2 M K(\mathbf{k}) % - B_z
& c(\sin k_x+i \sin k_y) & -c \sin k_z \\
c \sin k_z & c(\sin k_x-i \sin k_y) & -m + 2 M K(\mathbf{k}) % - B_z
& 0 \\
c(\sin k_x+i \sin k_y) & -c \sin k_z & 0 & -m + 2 M K(\mathbf{k}) % + B_z
\end{matrix}
\right)\,,
\eql{dirac-model}
\end{equation}
\end{widetext}
where $K(\mathbf{k}) = 3-\cos k_x-\cos k_y-\cos k_z$, and $c$, $m$,
and $M$ are dimensionless parameters inherited from the original
isotropic modified Dirac equation~\cite{doi:10.1142/S2010324711000057}
by setting the rest mass $m_0 c^{2}$ to be the energy scale of the
model~\cite{Rauch2017}.

\begin{figure}
\begin{center}
\includegraphics[width=0.99\columnwidth]{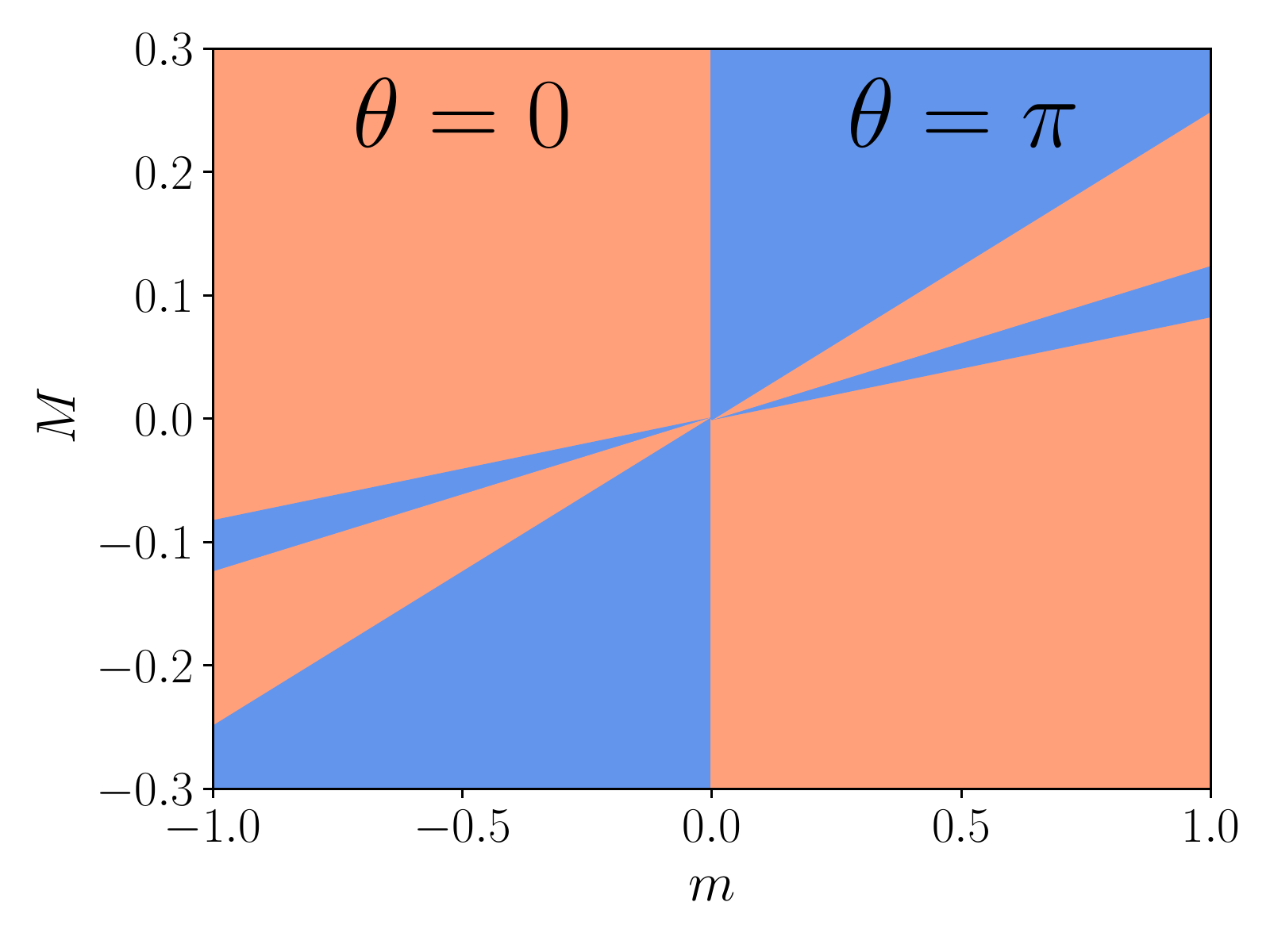}
\caption{Topological phase diagram of the model of \eq{dirac-model}
  for $c=1.0$. Orange and blue regions denote axion-even
  ($\theta = 0$) and axion-odd ($\theta = \pi$) phases, respectively.}
\figl{phase-Dirac-model}
\end{center}
\end{figure}

The topological phase diagram of the half-filled model is shown in
\fref{phase-Dirac-model} for $c=1.0$.  The system is gapped except on
the $m=0,4M,8M,12M$ lines, where the gap closes at $\Gamma=(0,0,0)$,
${\rm X}=(\pi,0,0)$, ${\rm M}=(\pi,\pi,0)$, and
${\rm A}=(\pi,\pi,\pi)$, respectively. As shown in
\aref{phase-diagram}, those metallic lines separate axion-trivial from
axion-odd insulating phases.

The axion angle is quantized by several axion-odd symmetries.  Some
are $z$-reversing (inversion and horizontal mirror $M_z$), and others
are $z$-preserving (spinful time reversal and vertical mirrrors).  As
$M_z$ is a type-1 mirror, it protects two MCNs that are related to the
axion angle by \eq{theta}.

%.........................................................
\subsubsection{Axion-odd phase with protected Wannier flow}
\secl{Dirac-connected}
% .........................................................

For our numerical tests we set $c=m=1.0$ and $M=0.5$ to put the model
in the axion-odd phase. The energy band structure is shown in
\fref{Dirac-model}(a). The bands are pairwise degenerate due to the
presence of time-reversal and inversion symmetry, with a finite gap
between the two pairs over the entire BZ.  The Fermi level is placed
at midgap.

\begin{figure}
\begin{center}
\includegraphics[width=0.49\columnwidth]{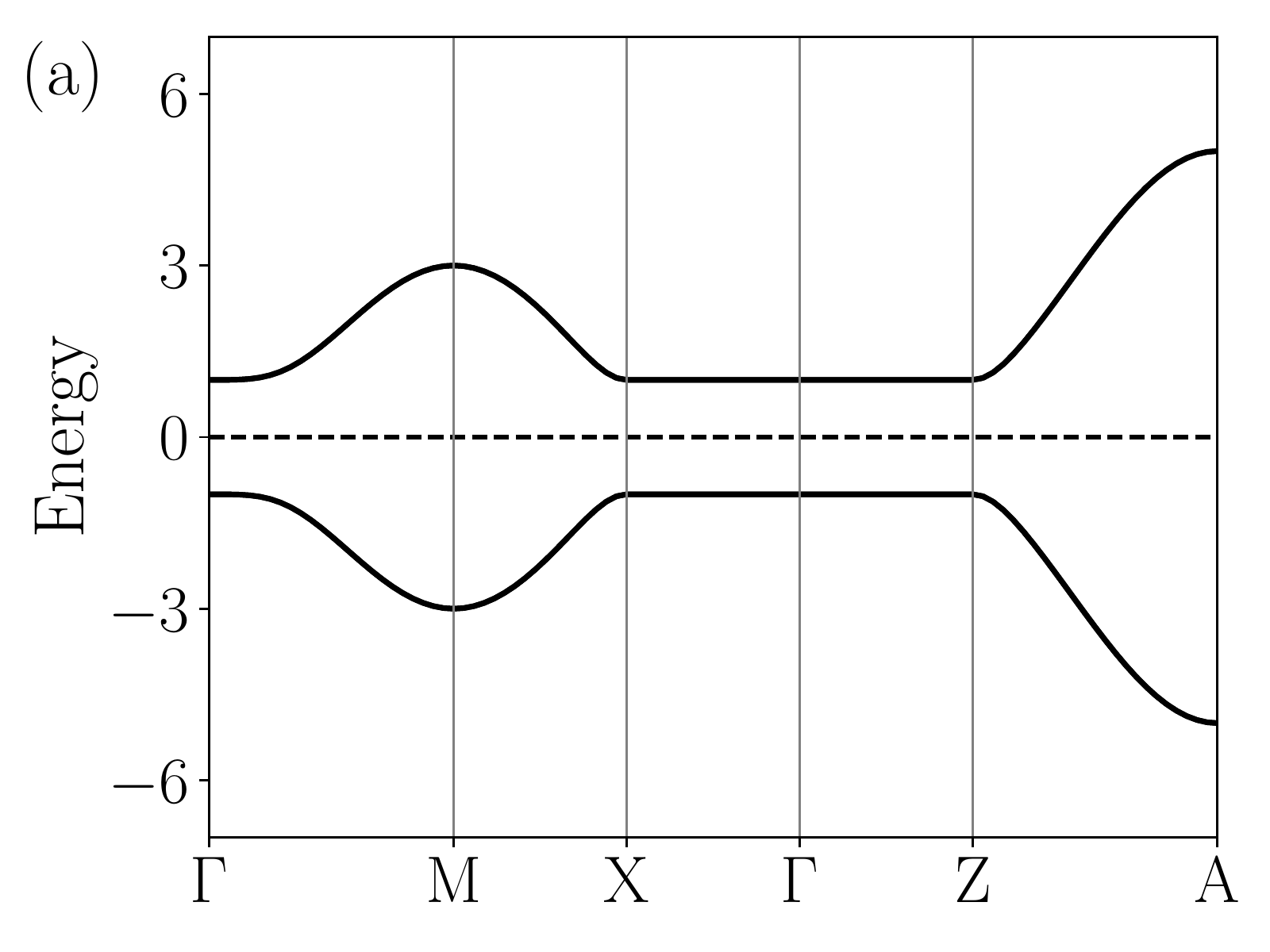}
\includegraphics[width=0.49\columnwidth]{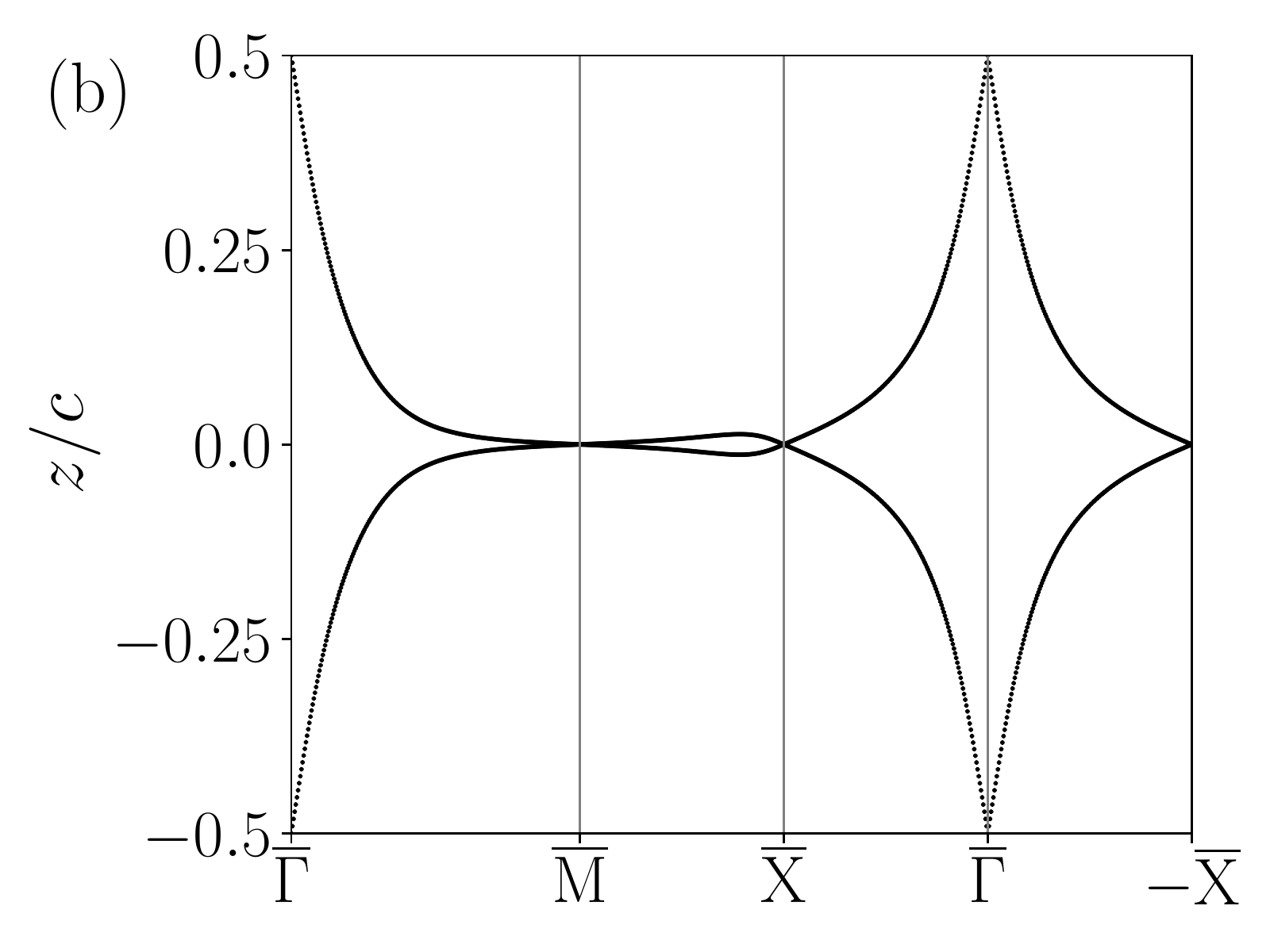}\\
\includegraphics[trim = 15 20 30 15,clip,width=0.49\columnwidth]{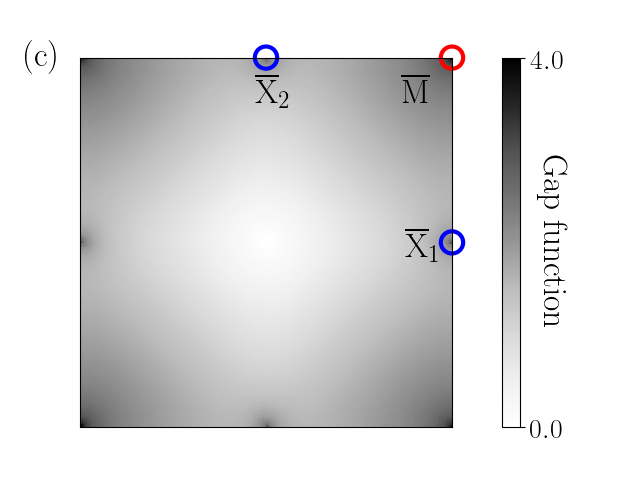}
\includegraphics[trim = 15 20 30 15,clip,width=0.49\columnwidth]{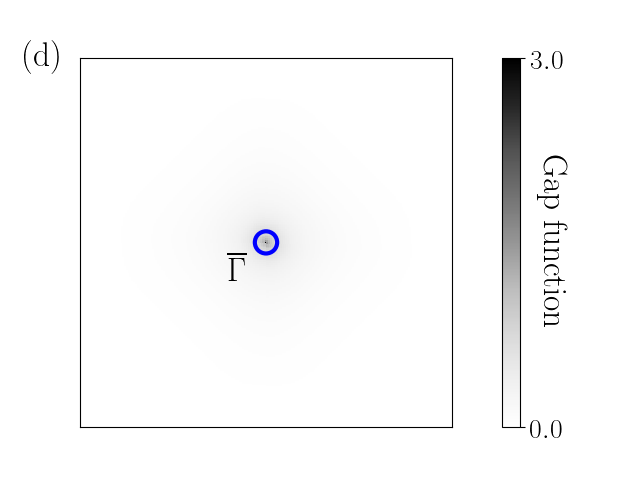}
\caption{(a) Energy bands of the model described by \eq{dirac-model}
  with $c=m=1.0$ and $M=0.5$.  The bands are doubly degenerate, and
  the Fermi level (dashed line) has been placed at midgap. (b)~Wannier
  band structure obtained from the valence states.  (c) and (d)
  Heatmap plots of the gap function of \eq{gap-fct} about the $z=0$
  and $z=c/2$ planes, respectively, with the nodal points color-coded
  as in \fref{2dSnTe}(c).}
\figl{Dirac-model}
\end{center}
\end{figure}

Since the system is axion-odd and has $z$-preserving axion-odd
symmetries, the connectivity (or ``flow'') of the Wannier bands is
topologically protected~\cite{varnava-prb20}. In particular, spinful
time reversal symmetry requires that the two bands per vertical cell
are glued together as follows: one band touches the band above at one
of the four time-reversal invariant momenta (TRIM), and it touches the
periodic image below at the other three.  As for the $z$-reversing
axion-odd symmetries, the effect of $M_z$ is to pin the up-touching to
one of the mirror planes and the three down-touchings to the other,
while inversion further constrains the four touchings to occur at TRIM
on those planes, as already mandated by time reversal.

The pattern of band touchings described above is confirmed by
\fref{Dirac-model}(b), where we plot the Wannier bands. They were
obtained by placing at the origin the four basis orbitals that belong
to the home unit cell, and making the diagonal approximation of
\eq{diag-r} for the position matrix.  There is one band touching at
$\overline{\Gamma}$ on the B plane, and three more on the A plane: one
at $\overline{\rm M}$, and the others at the two $\overline{\rm X}$
points.

Since the Wannier spectrum is gapless, the MCNs $\mu\G$ and $\mu\X$
are given respectively by the half-sum and the half-difference of the
net winding numbers on the A and B planes
[\eqs{muG-gapless}{muX-gapless}]. As indicated in the gap-function
plots of \fsref{Dirac-model}(c,d), the three nodes at A give $\Wa=-1$
and the single node at B gives $\Wb=-1$, so that $\mu\G=-1$ and
$\mu\X=0$. Note that $\mu\G+\mu\X$ is an odd number, as required by
\eq{theta} for an axion-odd system.

%......................................................
\subsubsection{Axion-odd phase with fragile Wannier flow}
\secl{Dirac-disconnected}
%......................................................

If the $z$-preserving axion-odd symmetries of the model (time reversal
and vertical mirrors) are weakly broken, the system will remain in an
axion-odd phase protected by $M_z$ and inversion.  But since these are
$z$-reversing operations, the Wannier spectrum is no longer
topologically required to be gapless.  The Wannier flow is only
protected in a ``fragile'' sense, and it can be destroyed, while
preserving $M_z$, by adding some weakly-coupled trivial bands to the
valence manifold~\cite{varnava-prb20,wieder-arxiv18}.  Below we carry
out this procedure in two different ways, and confirm that the MCNs
remain the same as in the original model.

%- - - - - - - - - - - - - - - - - - - - - - - - - -
\paragraph{Insertion of a symmetric pair of occupied orbitals}
%- - - - - - - - - - - - - - - - - - - - - - - - - -

\begin{figure}
\begin{center}
\includegraphics[width=0.49\columnwidth]{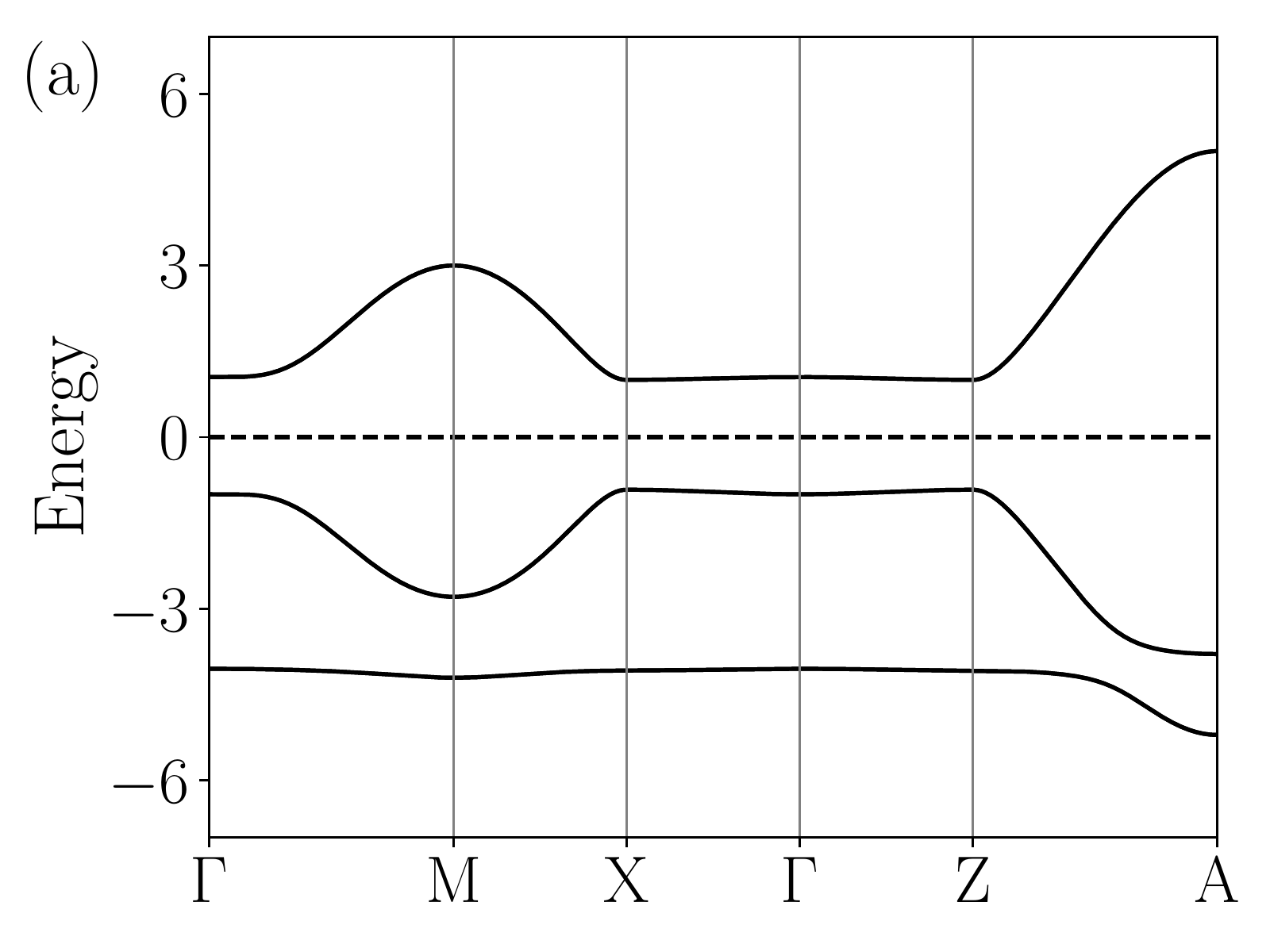}
\includegraphics[width=0.49\columnwidth]{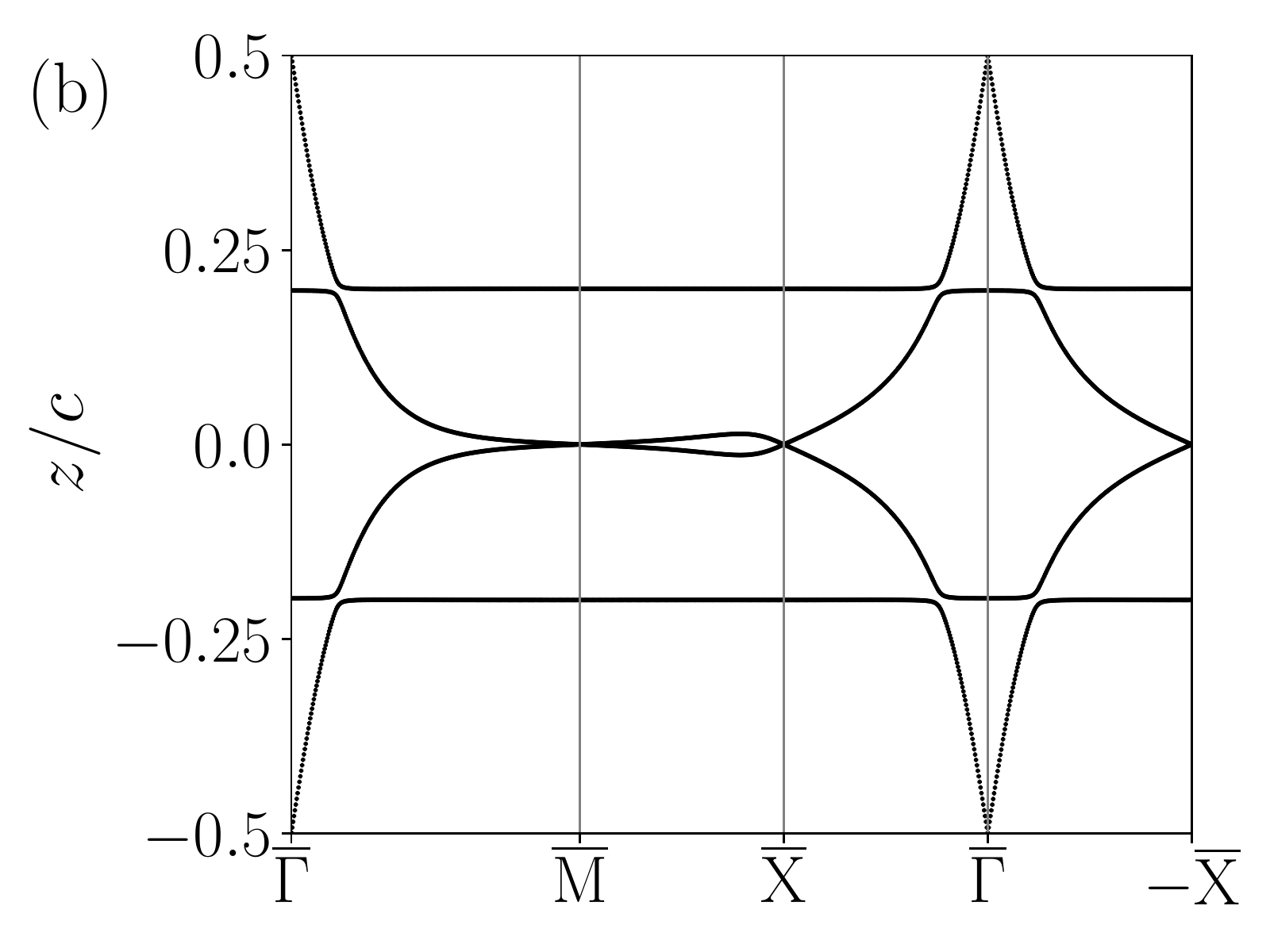}
\caption{(a) Energy bands of the same model as in \fref{Dirac-model},
  after adding an extra pair of occupied orbitals with
    $E=-4.0$ at $z=\pm 0.2c$ and coupling them to the other
  orbitals. The bands are doubly degenerate, and the Fermi level
  (dashed line) has been placed at midgap. (b) Wannier band structure
  obtained from the valence states, with small gaps around
  $z=\pm 0.2c$ due to the added orbitals.}
\figl{Dirac-model-2-pairs}
\end{center}
\end{figure}

Here we implement the strategy outlined in \sref{MCN-gapless}.  We
insert in the unit cell two more orbitals, denoted as $\ket{5}$ and
$\ket{6}$, that have opposite spins and the same on-site energy
$E=-4.0$. To break time reversal and the vertical mirrors while
preserving $M_z$ and inversion, we place the spin-up orbital $\ket{5}$
at $(x,y,z)=(0.0,0.0,0.2c)$, and the spin-down orbital $\ket{6}$ at
$(x,y,z)=(0.0,0.0,-0.2c)$, keeping the original orbitals $\ket{1}$ to
$\ket{4}$ at the origin.  Finally, we couple the new orbitals to the
old via the matrix elements $\me{5}{H}{1}=\me{6}{H}{2}=0.5$. The
resulting model retains the $M_z$ and inversion symmetries of the
original model, and it breaks the time-reversal and vertical mirror
symmetries in the $Z$ matrix of \eq{Z} (but not in the Hamiltonian).

The energy and Wannier band structures are plotted in
\fsref{Dirac-model-2-pairs}(a,b). Because the Hamiltonian has both
inversion and time-reveral symmetry, the energy bands remain doubly
degenerate as in \fref{Dirac-model}(a).  The breaking of the
$z$-preserving symmetries in the $Z$ matrix is reflected in the
Wannier spectrum which is no longer connected as in
\fref{Dirac-model}(b), with small gaps opening up near
$z = \pm 0.2 c$. The node at $\overline{\Gamma}$ on the B plane and
those at $\overline{\rm X}_1$, $\overline{\rm X}_2$, and
$\overline{\rm M}$ on the A plane remain intact, protected by $M_z$
and inversion. Their winding numbers are also unchanged, leading to
the same MCNs as in the original model.

%- - - - - - - - - - - - - - - - - - - - - - - - -
\paragraph{Insertion of a single occupied orbital at $z=0$.}
%- - - - - - - - - - - - - - - - - - - - - - - - -

\begin{figure}
\begin{center}
\includegraphics[width=0.54\columnwidth]{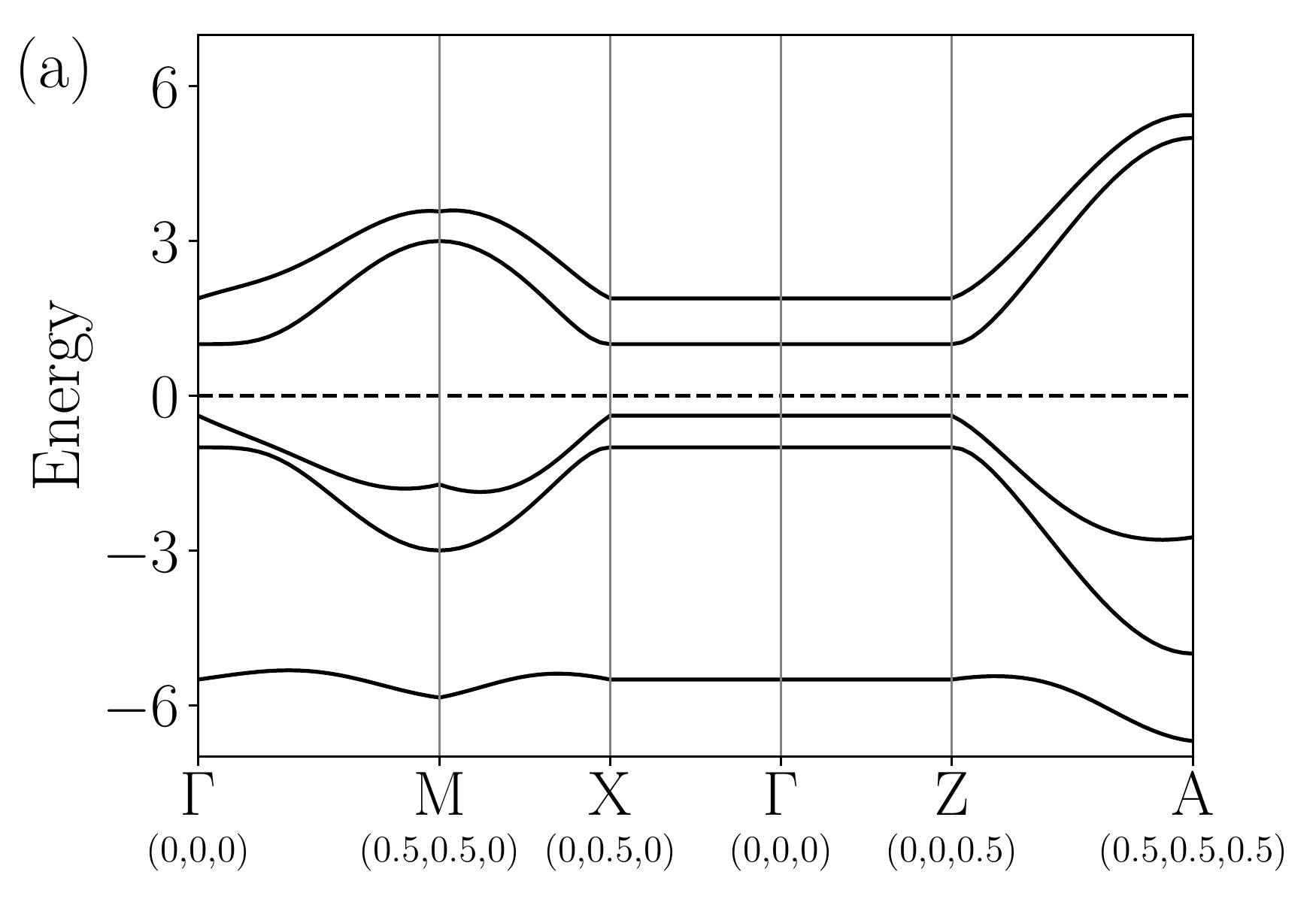}
\includegraphics[trim=80 20 0 10,clip,width=0.44\columnwidth]{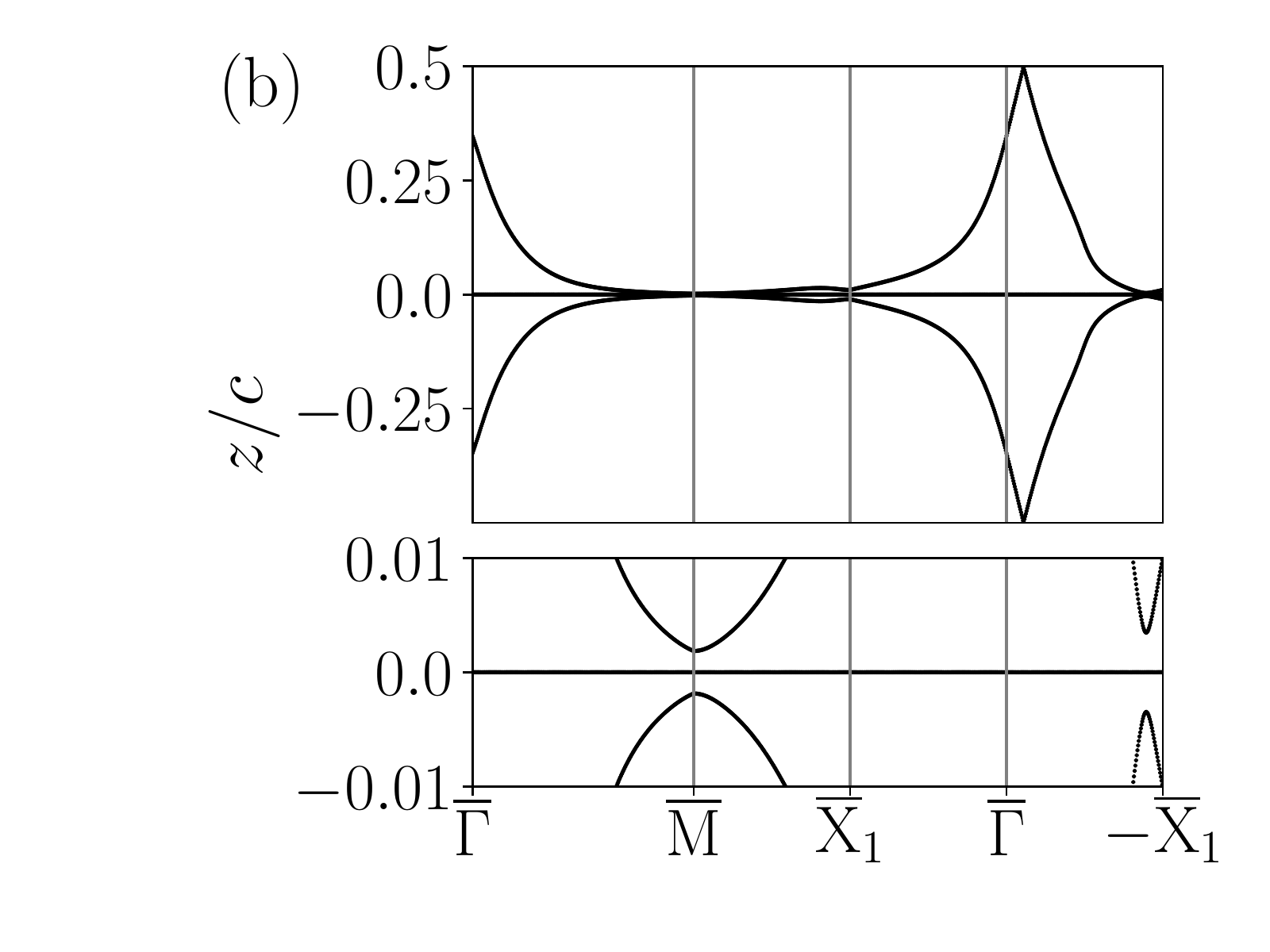}
\caption{(a) Energy bands of the same model as in \fref{Dirac-model},
  after adding an extra occupied orbital at $z=0$ and coupling it to
  the other orbitals. The Fermi level (dashed line) has been placed in
  the gap. (b) Wannier band structure obtained from the valence
  states. The added orbital generates a flat band at $z=0$, which
  repels the nodal points on that plane (lower panel).}
\figl{Dirac-model-1-pair}
\end{center}
\end{figure}

An alternative way of opening up a gap in the Wannier spectrum is to
insert a flat band on a mirror plane.  To illustrate this procedure,
we add at the origin a single spin-up orbital $\ket{5}$ with on-site
energy $E=-4.0$ and odd parity about that plane, and couple it to the
model via $\me{5}{H}{1} = \me{5}{H}{4} = 2.0$. Because the orbital is
spin-polarized, it breaks time reversal; and because the spin points
in the vertical direction, it also breaks all vertical mirrors while
preserving $M_z$.  In addition, the coupling terms break inversion
symmetry, leaving $M_z$ as the only axion-odd symmetry.  The energy
bands of the modified model are shown in \fref{Dirac-model-1-pair}(a).
A new band has appeared below the other four, so that there are now
three valence bands in total, leading to three Wannier bands.

The added orbital, which belongs to the ${\rm A}^+$ class in
Table~\ref{tab:parity}, generates an extra even-parity state at both G
and X. This creates an imbalance $\Delta N_{\rm G}=\Delta N_{\rm X}=1$
between even- and odd-parity states on the two mirror-invariant BZ
planes, which according to \eq{num-flat-1-A} results in a flat band at
A.  We emphasize that this extra band remains flat even after the
added orbital is coupled to the model, as long as the coupling terms
respect $M_z$ symmetry.  As already mentioned, those terms are chosen
to break inversion symmetry. This is needed to ensure that the three
point nodes on the A plane are repelled by the flat band in the manner
described in \sref{repel}, since inversion symmetry would otherwise
protect them.

The resulting Wannier bands are displayed in the upper panel of
\fref{Dirac-model-1-pair}(b); because of the lowered symmetry, the
node at $z=c/2$ is no longer pinned to $\overline{\Gamma}$ as in
\fref{Dirac-model}(b). The lower panel reveals a perfectly flat band
at $z=0$, well separated from a pair of dispersive bands whose three
touchings on the $z=0$ plane in \fref{Dirac-model}(c) have been gapped
out.  Under these circumstances, \eqs{muG-w}{muX-w} for the MCNs
reduce to
\beq
\mu\G = \half(\pA \Caf + \Wb)
\eql{muG-Dirac}
\eeq
and 
\beq
\mu\X = \half(\pA \Caf - \Wb)\,.
\eql{muX-Dirac}
\eeq
The single node at B has the same winding number $\Wb = -1$ as in the
original model, while the net winding number $\Wa = -1$ of the
gapped-out nodes at A has been transferred to the index $\pA\Caf$ of
the flat band ($\pA=-1$, and $\Caf = +1$). Overall, the MCNs remain
unchanged.

%===================
\section{Summary}
\secl{summary}
%===================

In summary, we have investigated the topological properties of
mirror-symmetric insulating crystals from the viewpoint of HW
functions localized along the direction orthogonal to the mirror
plane.  We first clarified the generic behaviors of the associated
Wannier bands, and then derived a set of rules for deducing the MCNs.
To validate and illustrate the formalism, we applied it to SnTe in the
monolayer and bulk forms, and to a toy model of an axion-odd
insulator.

In the HW representation, the MCNs are expressed in terms of a set of
integer-valued properties of the Wannier bands on the mirror planes:
the Chern numbers and mirror parities of flat bands lying on those
planes, and the winding numbers of the touching points on those planes
between symmetric pairs of dispersive bands. One advantage of this
representation is that it reveals the relation between the MCNs and
the axion $\zt$ index from purely bulk considerations.  That relation
is far from obvious in the standard Bloch representation, and
previously it had only been obtained via an indirect argument
involving surface states.

In some cases the axion $\zt$ index can be determined by visual
inspection of the Wannier band structure, e.g., by counting the number
of nodal points between certain bands~\cite{varnava-prb20}. We have
found that mere visual inspection does not suffice for obtaining the
MCNs since it does not reveal, for example, the relative signs of the
winding numbers of different nodes.

Interestingly, in certain cases where flat Wannier bands are present
the magnitudes of the MCN can be determined without having to divide
the occupied manifold into two mirror sectors. This follows from the
uniform-parity assumption for the flat bands, which has no counterpart
in the Bloch representation.  Since the determination of the mirror
parities is the most cumbersome step in the calculation of MCNs, this
feature of the HW formalism could lead to a more automated algorithm
for computing MCNs. Even without such further developments, the
formalism has already proven useful for discussing the topological
classification of mirror-symmetric insulators.

%--------------------------------------------------------------------
\acknowledgments Work by T.R. was supported by the Deutsche
Forschungsgemeinschaft Grant No. Ra 3025/1-1 from the Deutsche
Forschungsgemeinschaft.  Work by D.V. was supported by National
Science Foundation Grant DMR-1954856.  Work by I.S. was supported by
Grant No.~FIS2016-77188-P from the Spanish Ministerio de Econom\'ia y
Competitividad.
%--------------------------------------------------------------------

\appendix

%=====================================================
\section{Derivation of \eqr{num-flat-1-A}{num-flat-2}}
\secl{num-flat}
% ====================================================

According to \tref{parity}, the numbers of occupied states with each
mirror parity at G and X are
\begin{subequations}
\begin{align}
N_{{\rm G}^\pm} &=  \Nafpm + \Nbfpm + \frac{1}{2}\wt N\,,\eql{N-Gpm}\\
N_{{\rm X}^\pm} &= \Nafpm + \Nbfmp + \frac{1}{2}\wt N\,,
\end{align}
\end{subequations}
where $\wt N=\Nad+\Nbd+\Nuc$ is the total number of dispersive Wannier
bands per cell. Letting
$\Delta N_{{\rm G}}=N_{{\rm G}^+}-N_{{\rm G}^-}$ and
$\Delta\overline N_{{\rm A}}=\Nafp-\Nafm$, and defining
$\Delta N_{{\rm X}}$ and $\Delta N_{{\rm B}}$ in the same way, we find
\begin{subequations}
\begin{align}
\Delta \overline N_{{\rm A}}&=\frac{1}{2}
\left( \Delta N_{{\rm G}}+\Delta N_{{\rm X}} \right)\,,\eql{DeltaNA}\\
\Delta \overline N_{{\rm B}}&=\frac{1}{2}
\left( \Delta N_{{\rm G}}-\Delta N_{{\rm X}} \right)\,.\eql{DeltaNB}
\end{align}
\end{subequations}
Under the uniform parity assumption
$\vert\Delta \overline N_{{\rm A}}\vert=\Naf$ and
$\vert\Delta \overline N_{{\rm B}}\vert=\Nbf$, resulting in
\eqs{num-flat-1-A}{num-flat-1-B}.  In the case of a type-2 mirror A
and B are equivalent, and from \eq{N-Gpm}
$\Delta \overline N_{\rm A}+\Delta \overline N_{\rm B}=\Delta N_{\rm
  G}$.  Hence
$\Delta \overline N_{\rm A}=\Delta \overline N_{\rm B}=\Delta N_{\rm
  G}/2$, yielding \eq{num-flat-2} under the same assumption.

%===============================
\section{Derivation of \eq{Nw}}
\secl{winding}
%===============================

Let us prove \eq{Nw} for the case of a single pair of dispersive
Wannier bands connected by point nodes on the A plane.  In this case
the matrix $f_\kk$ of \eq{f-k} reduces to the scalar
\beq
f_\kk\equiv\me{\wt h_\kk^+}{z}{\wt h_\kk^-}=|f_\kk|e^{i\gamma_\kk}\,,
\eql{f}
\eeq
where $\ket{\wt h_\kk^\pm}$ are states of even or odd mirror parity
constructed from the pair of HW functions as described in
\sref{winding-number-def}. These states are cell-periodic in plane and
localized along $z$, and we also define new states
$\ket{\psi_\kk^\pm}=e^{i\kk\cdot\rr}\ket{\wt h_\kk^\pm}$ that are
Wannier-like along $z$ and Bloch-like in plane.

When the Chern numbers $\wt C_{{\rm A}^\pm}$ are nonzero, it becomes
impossible to choose a gauge for the states $\ket{\psi_\kk^\pm}$ that
is both smooth and periodic in the projected 2D
BZ~\cite{vanderbilt-book18}.  We assume a square BZ with
$k_x,k_y\in [0,2\pi]$, and choose a smooth but nonperiodic gauge for
the $\ket{\psi_\kk^-}$ states.  To characterize the lack of
periodicity, let the phase relations between the edges of the BZ be
\beq
\ket{\psi^-_{\rm R}}=e^{-i\mu}\ket{\psi^-_{\rm L}}\,,\quad
\ket{\psi^-_{\rm T}}=e^{-i\nu}\ket{\psi^-_{\rm B}}\,,
\eql{gauge}
\eeq
where $\{\text{L,R,T,B}\}=\{\text{left,right,top,bottom}\}$,
$\mu=\mu(k_y)$, and $\nu=\nu(k_x)$. Also let
\beq
\Delta\mu=\mu(2\pi)-\mu(0)\,,\quad
\Delta\nu=\nu(2\pi)-\nu(0)\,.
\eeq
When computing the Berry phase around the BZ boundary as an integral
of the connection
$\A_\kk^-=i\ip{\wt h_\kk^-}{\partial_\kk \wt h_\kk^-}$,
\beq
\phi_-=\oint_{\partial\text{BZ}} \A_\kk^-\cdot d\kk\,,
\eeq
the contribution from the L and R segments cancel except for terms
coming from $\mu$, and similarly for the top and bottom segments. It
follows that
\beq
\phi_-=\Delta\mu-\Delta\nu\,.
\eeq

We assume a smooth but nonperiodic gauge for the $\ket{\psi^+_ \kk}$
states as well, so that the phase $\gamma_\kk$ in \eq{f} becomes a
smooth function of $\kk$ (except at the nodes, where $f_\kk$ vanishes
and $\gamma_\kk$ becomes ill defined). Now we phase-align
$\ket{\psi^+_\kk}$ with $\ket{\psi^-_\kk}$ by re-gauging as follows,
\beq
\ket{\psi^+_\kk}^\prime=e^{i\gamma_\kk}\ket{\psi^+_\kk}\,.
\eql{phase-align}
\eeq
(In this new gauge $f^\prime_\kk$ is real, and $\gamma^\prime_\kk$ is
zero everywhere.) This will make a gauge for $\ket{\psi^+_\kk}^\prime$
that is also nonperiodic. For the moment we only assume that this
gauge is smooth in a neighborhood extending some small distance inside
the boundary; we ignore what is going on deeper inside. It is not hard
to see that the same relations as in \eq{gauge}, with the same
functions $\mu$ and $\nu$, apply to the $\ket{\psi^+_\kk}^\prime$
states, and it follows that
\beq
\phi^\prime_+=\phi_-\quad\text{(call it $\phi$)\,.}
\eeq
Now, in the case of the $\ket{\psi^-_\kk}$ states the interior was
smooth, so by applying Stokes' theorem to
\beq 
2\pi\Cadm=\int_{\rm BZ}\Omega_\kk^-\,d^2k
\eql{C-minus}
\eeq
where
$\Omega^-_\kk=\partial_{k_x} A^-_{\kk,y}-\partial_{k_y} A^-_{\kk,x}$
is the Berry curvature of state $\ket{u^-_\kk}$, we get
\beq
2\pi\Cadm=\phi\,.
\eeq
If the interior of $\ket{\psi^+_\kk}^\prime$ were also smooth, we
would conclude that $\Cadp=\Cadm$. Conversely, when the MCN is nonzero
there must exist nonanalytic points where the phase of
$\ket{u^+_\kk}^\prime$ changes discontinuously. Those points are
precisely the nodes of $f_\kk$, which we label by $j$; they act as
vortex singularities of the Berry connection
\beq
\left(\A_\kk^+\right)^\prime=\A_\kk^+-\partial_\kk\gamma_\kk\,,
\eql{A-vortex}
\eeq
and we extract their winding numbers $W_j$ using \eq{W-j}.
  % [\eq{W-j}],
%
Let $S$
be the interior of the projected BZ with a small circle $c_j$ cut
around each node, and apply Stokes' theorem over the region~$S$ to
find
\beq
\label{eq:stokes}
\int_S\Omega_\kk^+\,d^2k=
\int_{\partial\text{BZ}}\left(\A_\kk^+\right)^\prime\cdot d\kk-
\sum_j\oint_{c_j}\left(\A_\kk^+\right)^\prime\cdot d\kk\,.
\eeq
The first term on the right-hand side is equal to
$\phi^\prime_+=\phi=2\pi\Cadm$.  In the limit of small circles the
left-hand side becomes $2\pi\Cadp$, and the second term on the
right-hand side reduces to $2\pi\sum_j\,W_j$ (this follows from
\eq{A-vortex} by noting that $\A^+_\kk$ is smooth everywhere). Thus
$\Cadp-\Cadm$ equals $\Wa=\sum_{j\in{\rm A}}\,W_j$, which is what we
set out to prove.
The same result holds if more than one pair of bands meet at some of
the point nodes, in which case $\gamma_\kk$ is given by the more
  general expression in \eq{gamma-k}.
%
% The same result holds if more than one pair of bands meet at some of
% the point nodes.  Their winding number are still given by \eq{W-j},
% but $\gamma_\kk$ is now given by the more general expression in
% \eq{gamma-k} instead of \eq{f}.

%=======================================================================
\section{Phase diagram of the modified Dirac model on a cubic lattice}
\secl{phase-diagram}
%=======================================================================

\begin{figure}
\begin{center}
\includegraphics[width=0.49\columnwidth]{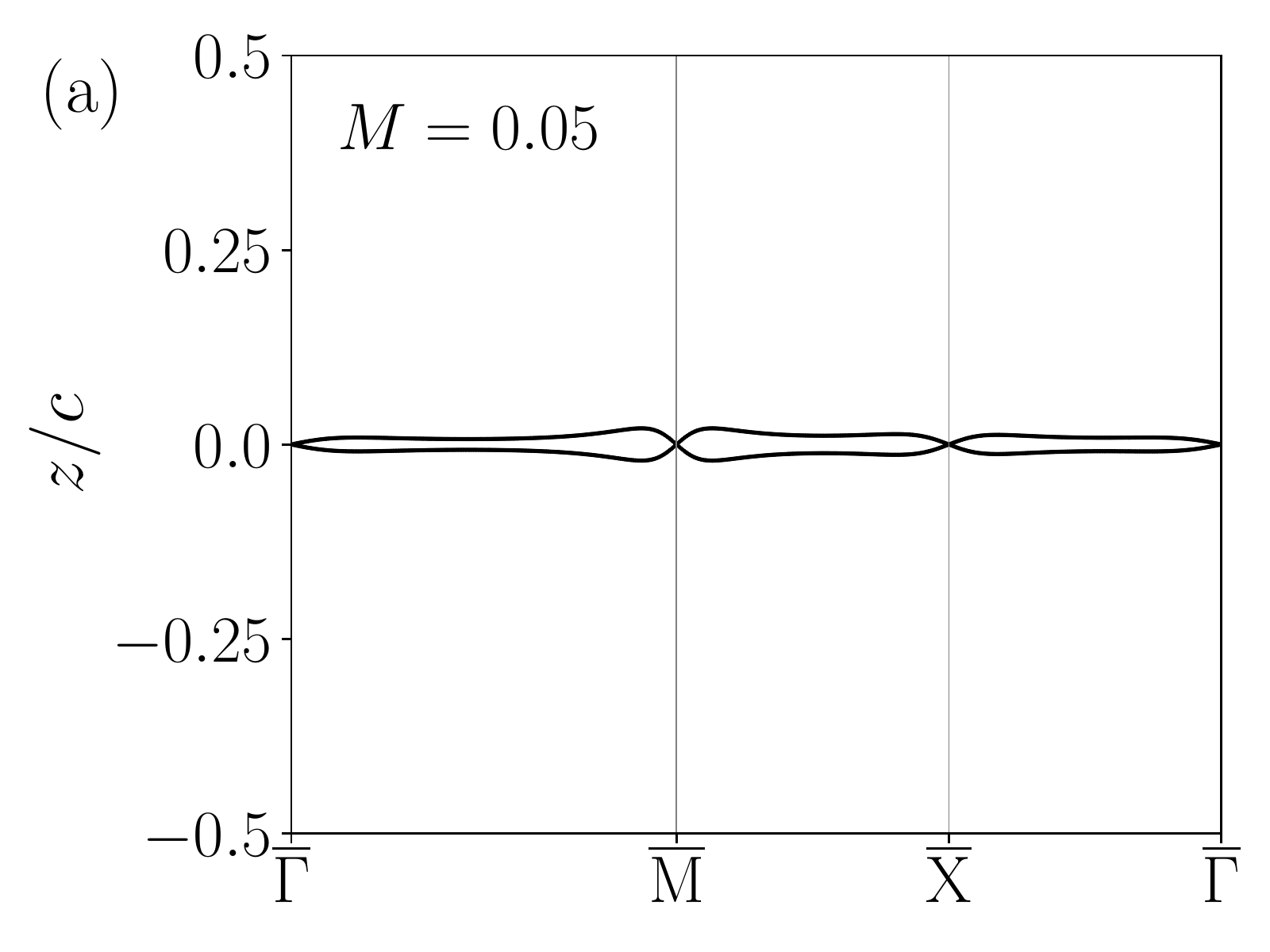}
\includegraphics[width=0.49\columnwidth]{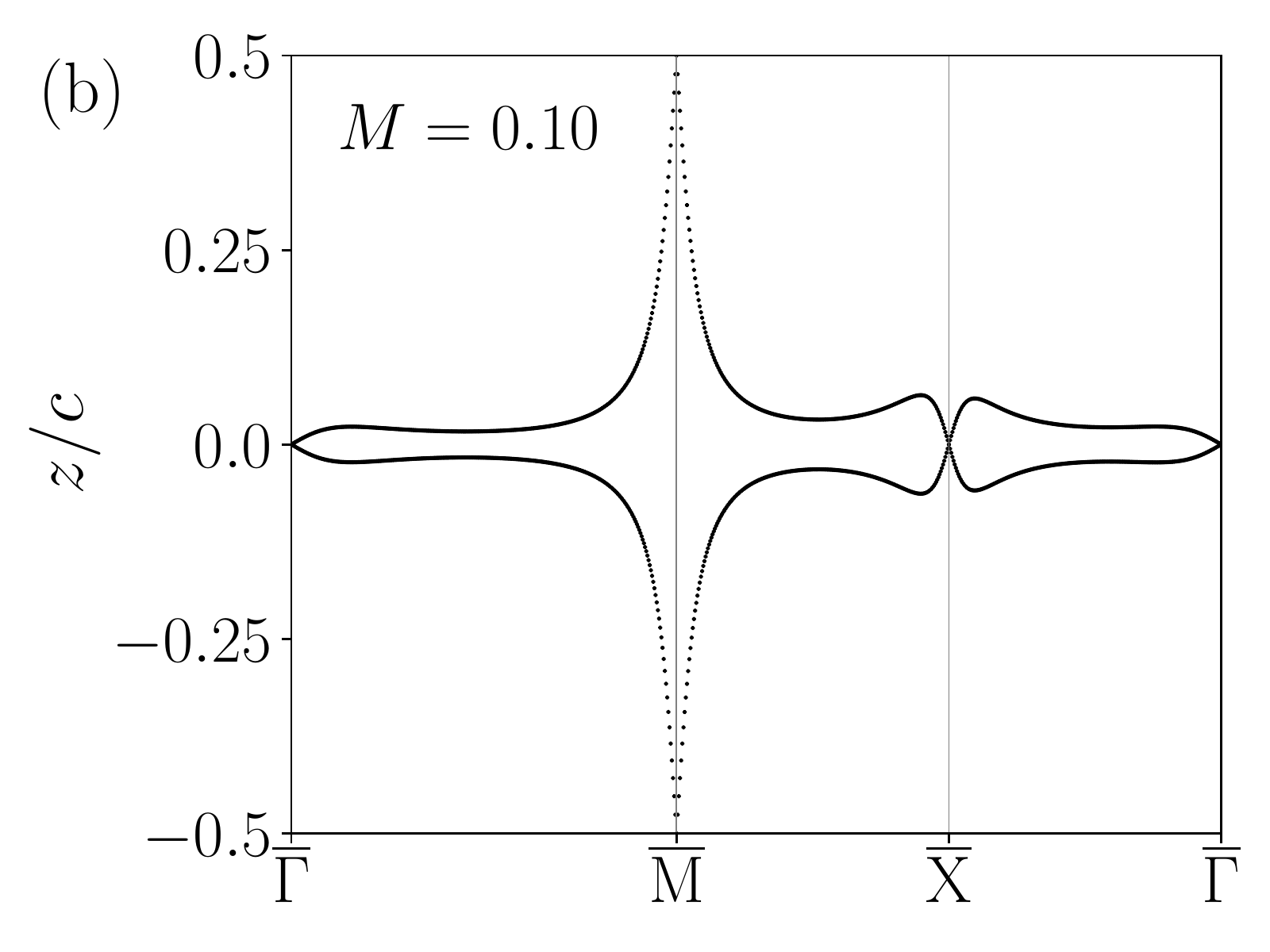}\\
\includegraphics[width=0.49\columnwidth]{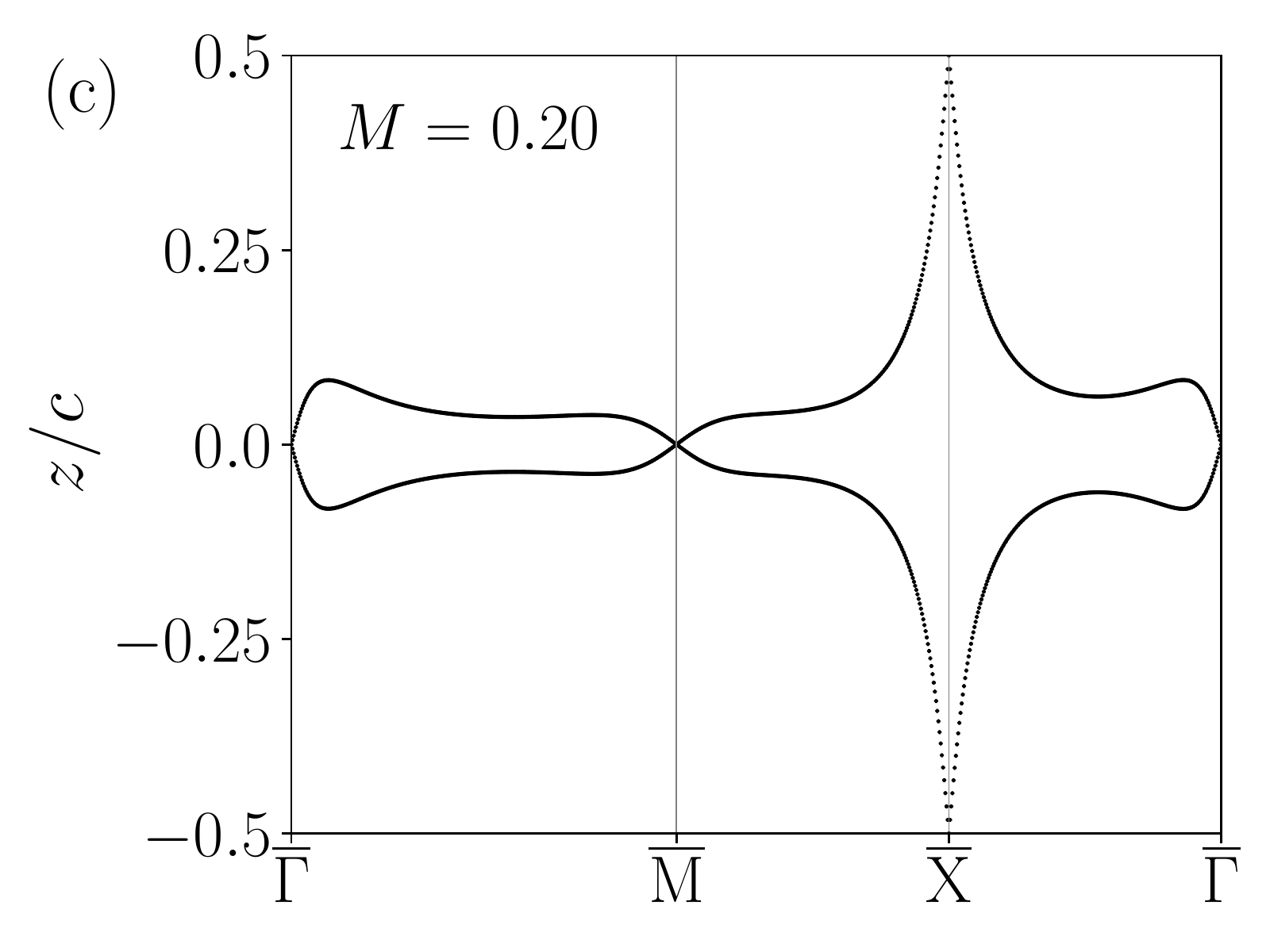}
\includegraphics[width=0.49\columnwidth]{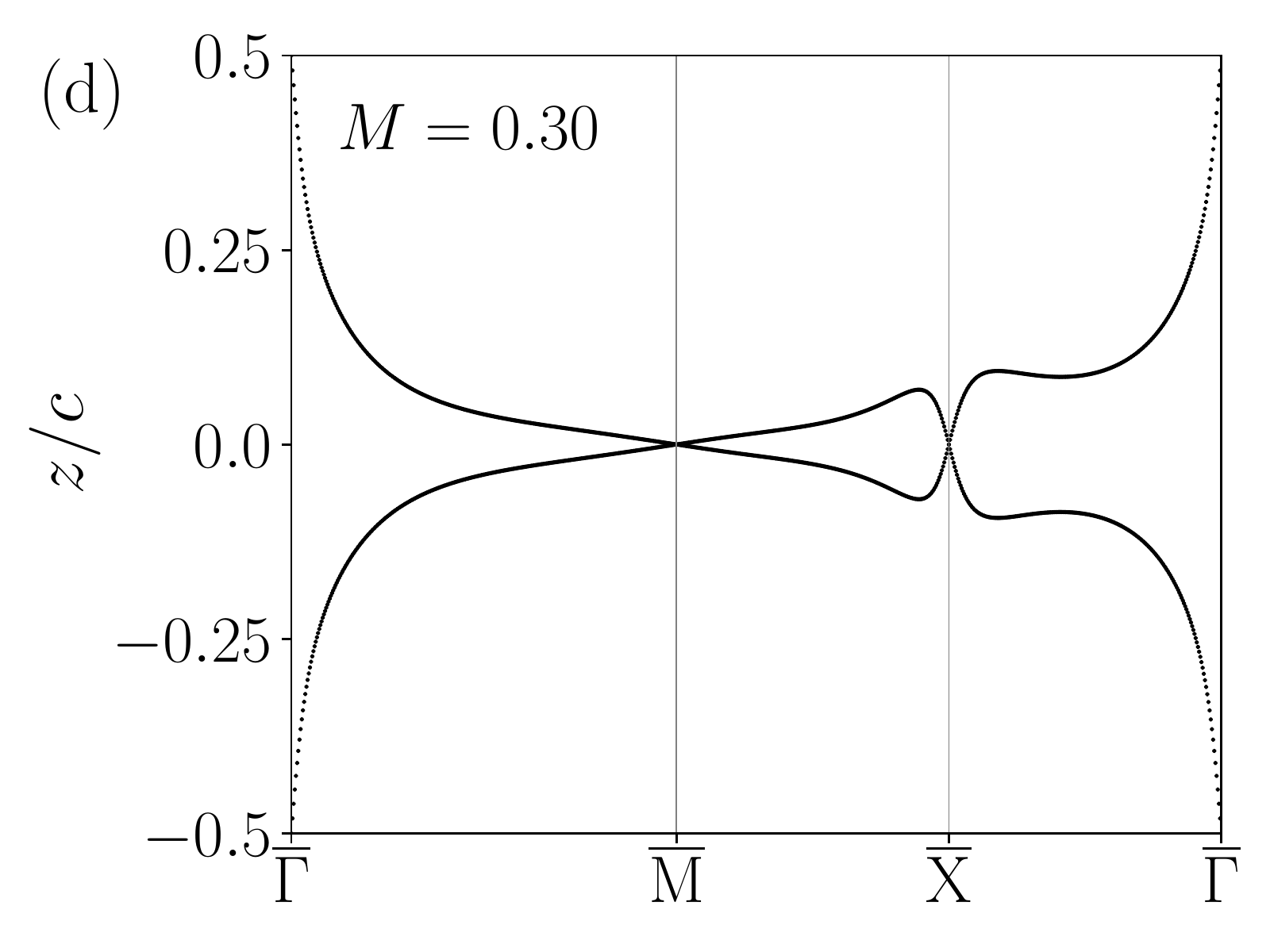}
\caption{Wannier bands of the modified Dirac model on a cubic lattice
  [\eq{dirac-model}], for $m=1.0$ and varying $M$.}
\figl{Dirac-model-phases}
\end{center}
\end{figure}

In this Appendix, we map out the topological phase diagram of the
model of \eq{dirac-model} as a function of the parameters $m$ and $M$,
for $c=1.0$.  The band gap closes for $m=0,4M,8M,12M$ at the points
$\Gamma$, $\rm X$, $\rm M$, and $\rm A$,
respectively~\cite{shen-dirac-book}. Those lines in the phase diagram
mark the topological phase transitions between axion-even and
axion-odd phases.

To decide which phases are trivial and which are topological, it is
sufficient to inspect the Wannier band structures in
\fref{Dirac-model-phases}, obtained for representative states in each
of the four phases along the $m=1.0$ line.  Since the model has
several axion-odd symmetries (time reversal, inversion, and multiple
mirrors), we can base our analysis on either of them, applying in each
case the rules given in Ref.~\cite{varnava-prb20} to determine the
axion $\zt$ index. In the following, we choose to focus on
time-reversal symmetry.

The Wannier spectrum of an axion-odd phase with spinful time-reversal
symmetry must be gapless, with each band touching the band above at
one of the four TRIM and the band below at the other three (or
vice-versa).  From this criterion we conclude that
\fsref{Dirac-model-phases}(a,c) correspond to axion-trivial phases,
and \fsref{Dirac-model-phases}(b,d) to axion-odd topological
phases. Hence the system is topological for $0<m/M<4$ and $8<m/M<12$,
producing the phase diagram in \fref{phase-Dirac-model}.  This is in
agreement with Ref.~\onlinecite{shen-dirac-book}, where the strong
topological index $\nu_0=\theta/\pi$ of each phase was determined from
the parity eigenvalues of the Bloch states at the eight TRIM in the 3D
BZ~\cite{fu-prb07}.

%\bibliography{bib_arXiv}
\bibliography{pap_arXiv.bbl}

\begin{thebibliography}{43}%
\makeatletter
\providecommand \@ifxundefined [1]{%
 \@ifx{#1\undefined}
}%
\providecommand \@ifnum [1]{%
 \ifnum #1\expandafter \@firstoftwo
 \else \expandafter \@secondoftwo
 \fi
}%
\providecommand \@ifx [1]{%
 \ifx #1\expandafter \@firstoftwo
 \else \expandafter \@secondoftwo
 \fi
}%
\providecommand \natexlab [1]{#1}%
\providecommand \enquote  [1]{``#1''}%
\providecommand \bibnamefont  [1]{#1}%
\providecommand \bibfnamefont [1]{#1}%
\providecommand \citenamefont [1]{#1}%
\providecommand \href@noop [0]{\@secondoftwo}%
\providecommand \href [0]{\begingroup \@sanitize@url \@href}%
\providecommand \@href[1]{\@@startlink{#1}\@@href}%
\providecommand \@@href[1]{\endgroup#1\@@endlink}%
\providecommand \@sanitize@url [0]{\catcode `\\12\catcode `\$12\catcode
  `\&12\catcode `\#12\catcode `\^12\catcode `\_12\catcode `\%12\relax}%
\providecommand \@@startlink[1]{}%
\providecommand \@@endlink[0]{}%
\providecommand \url  [0]{\begingroup\@sanitize@url \@url }%
\providecommand \@url [1]{\endgroup\@href {#1}{\urlprefix }}%
\providecommand \urlprefix  [0]{URL }%
\providecommand \Eprint [0]{\href }%
\providecommand \doibase [0]{http://dx.doi.org/}%
\providecommand \selectlanguage [0]{\@gobble}%
\providecommand \bibinfo  [0]{\@secondoftwo}%
\providecommand \bibfield  [0]{\@secondoftwo}%
\providecommand \translation [1]{[#1]}%
\providecommand \BibitemOpen [0]{}%
\providecommand \bibitemStop [0]{}%
\providecommand \bibitemNoStop [0]{.\EOS\space}%
\providecommand \EOS [0]{\spacefactor3000\relax}%
\providecommand \BibitemShut  [1]{\csname bibitem#1\endcsname}%
\let\auto@bib@innerbib\@empty
%</preamble>
\bibitem [{\citenamefont {Teo}\ \emph {et~al.}(2008)\citenamefont {Teo},
  \citenamefont {Fu},\ and\ \citenamefont {Kane}}]{teo-prb08}%
  \BibitemOpen
  \bibfield  {author} {\bibinfo {author} {\bibfnamefont {J.~C.~Y.}\
  \bibnamefont {Teo}}, \bibinfo {author} {\bibfnamefont {L.}~\bibnamefont
  {Fu}}, \ and\ \bibinfo {author} {\bibfnamefont {C.~L.}\ \bibnamefont
  {Kane}},\ }\bibfield  {title} {\enquote {\bibinfo {title} {{Surface states
  and topological invariants in three-dimensional topological insulators:
  Application to ${\text{Bi}}_{1\ensuremath{-}x}{\text{Sb}}_{x}$}},}\ }\href
  {\doibase 10.1103/PhysRevB.78.045426} {\bibfield  {journal} {\bibinfo
  {journal} {Phys. Rev. B}\ }\textbf {\bibinfo {volume} {78}},\ \bibinfo
  {pages} {045426} (\bibinfo {year} {2008})}\BibitemShut {NoStop}%
\bibitem [{\citenamefont {Ando}\ and\ \citenamefont {Fu}(2015)}]{ando-arcmp15}%
  \BibitemOpen
  \bibfield  {author} {\bibinfo {author} {\bibfnamefont {Y.}~\bibnamefont
  {Ando}}\ and\ \bibinfo {author} {\bibfnamefont {L.}~\bibnamefont {Fu}},\
  }\bibfield  {title} {\enquote {\bibinfo {title} {{Topological Crystalline
  Insulators and Topological Superconductors: From Concepts to Materials}},}\
  }\href {\doibase 10.1146/annurev-conmatphys-031214-014501} {\bibfield
  {journal} {\bibinfo  {journal} {Annu. Rev. Condens. Matter Phys.}\ }\textbf
  {\bibinfo {volume} {6}},\ \bibinfo {pages} {361} (\bibinfo {year}
  {2015})}\BibitemShut {NoStop}%
\bibitem [{\citenamefont {Qi}\ \emph {et~al.}(2008)\citenamefont {Qi},
  \citenamefont {Hughes},\ and\ \citenamefont {Zhang}}]{qi-prb08}%
  \BibitemOpen
  \bibfield  {author} {\bibinfo {author} {\bibfnamefont {X.-L.}\ \bibnamefont
  {Qi}}, \bibinfo {author} {\bibfnamefont {T.~L.}\ \bibnamefont {Hughes}}, \
  and\ \bibinfo {author} {\bibfnamefont {S.-C.}\ \bibnamefont {Zhang}},\
  }\bibfield  {title} {\enquote {\bibinfo {title} {{Topological field theory of
  time-reversal invariant insulators}},}\ }\href {\doibase
  10.1103/PhysRevB.78.195424} {\bibfield  {journal} {\bibinfo  {journal} {Phys.
  Rev. B}\ }\textbf {\bibinfo {volume} {78}},\ \bibinfo {pages} {195424}
  (\bibinfo {year} {2008})}\BibitemShut {NoStop}%
\bibitem [{\citenamefont {Essin}\ \emph {et~al.}(2009)\citenamefont {Essin},
  \citenamefont {Moore},\ and\ \citenamefont {Vanderbilt}}]{essin-prl09}%
  \BibitemOpen
  \bibfield  {author} {\bibinfo {author} {\bibfnamefont {A.~M.}\ \bibnamefont
  {Essin}}, \bibinfo {author} {\bibfnamefont {J.~E.}\ \bibnamefont {Moore}}, \
  and\ \bibinfo {author} {\bibfnamefont {D.}~\bibnamefont {Vanderbilt}},\
  }\bibfield  {title} {\enquote {\bibinfo {title} {{Magnetoelectric
  Polarizability and Axion Electrodynamics in Crystalline Insulators}},}\
  }\href {\doibase 10.1103/PhysRevLett.102.146805} {\bibfield  {journal}
  {\bibinfo  {journal} {Phys. Rev. Lett.}\ }\textbf {\bibinfo {volume} {102}},\
  \bibinfo {pages} {146805} (\bibinfo {year} {2009})}\BibitemShut {NoStop}%
\bibitem [{\citenamefont {Vanderbilt}(2018)}]{vanderbilt-book18}%
  \BibitemOpen
  \bibfield  {author} {\bibinfo {author} {\bibfnamefont {D.}~\bibnamefont
  {Vanderbilt}},\ }\href {\doibase 10.1017/9781316662205} {\emph {\bibinfo
  {title} {{Berry Phases in Electronic Structure Theory: Electric Polarization,
  Orbital Magnetization and Topological Insulators}}}}\ (\bibinfo  {publisher}
  {Cambridge University Press},\ \bibinfo {address} {Cambridge (United
  Kingdom)},\ \bibinfo {year} {2018})\BibitemShut {NoStop}%
\bibitem [{\citenamefont {Armitage}\ and\ \citenamefont
  {Wu}(2019)}]{armitage-scipost19}%
  \BibitemOpen
  \bibfield  {author} {\bibinfo {author} {\bibfnamefont {N.~P.}\ \bibnamefont
  {Armitage}}\ and\ \bibinfo {author} {\bibfnamefont {Liang}\ \bibnamefont
  {Wu}},\ }\bibfield  {title} {\enquote {\bibinfo {title} {{On the matter of
  topological insulators as magnetoelectrics}},}\ }\href {\doibase
  10.21468/SciPostPhys.6.4.046} {\bibfield  {journal} {\bibinfo  {journal}
  {SciPost Phys.}\ }\textbf {\bibinfo {volume} {6}},\ \bibinfo {pages} {46}
  (\bibinfo {year} {2019})}\BibitemShut {NoStop}%
\bibitem [{\citenamefont {Nenno}\ \emph {et~al.}(2020)\citenamefont {Nenno},
  \citenamefont {Garcia}, \citenamefont {Gooth}, \citenamefont {Felser},\ and\
  \citenamefont {Narang}}]{nenno-nrp20}%
  \BibitemOpen
  \bibfield  {author} {\bibinfo {author} {\bibfnamefont {D.~M.}\ \bibnamefont
  {Nenno}}, \bibinfo {author} {\bibfnamefont {C.~A.~C.}\ \bibnamefont
  {Garcia}}, \bibinfo {author} {\bibfnamefont {J.}~\bibnamefont {Gooth}},
  \bibinfo {author} {\bibfnamefont {C.}~\bibnamefont {Felser}}, \ and\ \bibinfo
  {author} {\bibfnamefont {P.}~\bibnamefont {Narang}},\ }\bibfield  {title}
  {\enquote {\bibinfo {title} {Axion physics in condensed-matter systems},}\
  }\href {\doibase 10.1038/s42254-020-0240-2} {\bibfield  {journal} {\bibinfo
  {journal} {Nat. Rev. Phys.}\ }\textbf {\bibinfo {volume} {2}},\ \bibinfo
  {pages} {682} (\bibinfo {year} {2020})}\BibitemShut {NoStop}%
\bibitem [{\citenamefont {Sekine}\ and\ \citenamefont
  {Nomura}(2021)}]{sekine-jap21}%
  \BibitemOpen
  \bibfield  {author} {\bibinfo {author} {\bibfnamefont {A.}~\bibnamefont
  {Sekine}}\ and\ \bibinfo {author} {\bibfnamefont {K.}~\bibnamefont
  {Nomura}},\ }\bibfield  {title} {\enquote {\bibinfo {title} {{Axion
  electrodynamics in topological materials}},}\ }\href {\doibase
  10.1063/5.0038804} {\bibfield  {journal} {\bibinfo  {journal} {J. Appl.
  Phys.}\ }\textbf {\bibinfo {volume} {129}},\ \bibinfo {pages} {141101}
  (\bibinfo {year} {2021})}\BibitemShut {NoStop}%
\bibitem [{\citenamefont {Otrokov}\ \emph {et~al.}(2019)\citenamefont {Otrokov}
  \emph {et~al.}}]{otrokov-nat19}%
  \BibitemOpen
  \bibfield  {author} {\bibinfo {author} {\bibfnamefont {M.~M.}\ \bibnamefont
  {Otrokov}} \emph {et~al.},\ }\bibfield  {title} {\enquote {\bibinfo {title}
  {Prediction and observation of an antiferromagnetic topological insulator},}\
  }\href {\doibase 10.1038/s41586-019-1840-9} {\bibfield  {journal} {\bibinfo
  {journal} {Nature}\ }\textbf {\bibinfo {volume} {576}},\ \bibinfo {pages}
  {416} (\bibinfo {year} {2019})}\BibitemShut {NoStop}%
\bibitem [{\citenamefont {Mong}\ \emph {et~al.}(2010)\citenamefont {Mong},
  \citenamefont {Essin},\ and\ \citenamefont {Moore}}]{mong-prb10}%
  \BibitemOpen
  \bibfield  {author} {\bibinfo {author} {\bibfnamefont {R.~S.~K.}\
  \bibnamefont {Mong}}, \bibinfo {author} {\bibfnamefont {A.~M.}\ \bibnamefont
  {Essin}}, \ and\ \bibinfo {author} {\bibfnamefont {J.~E.}\ \bibnamefont
  {Moore}},\ }\bibfield  {title} {\enquote {\bibinfo {title} {Antiferromagnetic
  topological insulators},}\ }\href {\doibase 10.1103/PhysRevB.81.245209}
  {\bibfield  {journal} {\bibinfo  {journal} {Phys. Rev. B}\ }\textbf {\bibinfo
  {volume} {81}},\ \bibinfo {pages} {245209} (\bibinfo {year}
  {2010})}\BibitemShut {NoStop}%
\bibitem [{\citenamefont {Fu}\ and\ \citenamefont {Kane}(2007)}]{fu-prb07}%
  \BibitemOpen
  \bibfield  {author} {\bibinfo {author} {\bibfnamefont {L.}~\bibnamefont
  {Fu}}\ and\ \bibinfo {author} {\bibfnamefont {C.~L.}\ \bibnamefont {Kane}},\
  }\bibfield  {title} {\enquote {\bibinfo {title} {Topological insulators with
  inversion symmetry},}\ }\href {\doibase 10.1103/PhysRevB.76.045302}
  {\bibfield  {journal} {\bibinfo  {journal} {Phys. Rev. B}\ }\textbf {\bibinfo
  {volume} {76}},\ \bibinfo {pages} {045302} (\bibinfo {year}
  {2007})}\BibitemShut {NoStop}%
\bibitem [{\citenamefont {Turner}\ \emph {et~al.}(2012)\citenamefont {Turner},
  \citenamefont {Zhang}, \citenamefont {Mong},\ and\ \citenamefont
  {Vishwanath}}]{turner-prb12}%
  \BibitemOpen
  \bibfield  {author} {\bibinfo {author} {\bibfnamefont {A.~M.}\ \bibnamefont
  {Turner}}, \bibinfo {author} {\bibfnamefont {Y.}~\bibnamefont {Zhang}},
  \bibinfo {author} {\bibfnamefont {R.~S.~K.}\ \bibnamefont {Mong}}, \ and\
  \bibinfo {author} {\bibfnamefont {A.}~\bibnamefont {Vishwanath}},\ }\bibfield
   {title} {\enquote {\bibinfo {title} {Quantized response and topology of
  magnetic insulators with inversion symmetry},}\ }\href {\doibase
  10.1103/PhysRevB.85.165120} {\bibfield  {journal} {\bibinfo  {journal} {Phys.
  Rev. B}\ }\textbf {\bibinfo {volume} {85}},\ \bibinfo {pages} {165120}
  (\bibinfo {year} {2012})}\BibitemShut {NoStop}%
\bibitem [{\citenamefont {Varnava}\ \emph {et~al.}(2020)\citenamefont
  {Varnava}, \citenamefont {Souza},\ and\ \citenamefont
  {Vanderbilt}}]{varnava-prb20}%
  \BibitemOpen
  \bibfield  {author} {\bibinfo {author} {\bibfnamefont {N.}~\bibnamefont
  {Varnava}}, \bibinfo {author} {\bibfnamefont {I.}~\bibnamefont {Souza}}, \
  and\ \bibinfo {author} {\bibfnamefont {D.}~\bibnamefont {Vanderbilt}},\
  }\bibfield  {title} {\enquote {\bibinfo {title} {{Axion coupling in the
  hybrid Wannier representation}},}\ }\href {\doibase
  10.1103/PhysRevB.101.155130} {\bibfield  {journal} {\bibinfo  {journal}
  {Phys. Rev. B}\ }\textbf {\bibinfo {volume} {101}},\ \bibinfo {pages}
  {155130} (\bibinfo {year} {2020})}\BibitemShut {NoStop}%
\bibitem [{\citenamefont {Varjas}\ \emph {et~al.}(2015)\citenamefont {Varjas},
  \citenamefont {de~Juan},\ and\ \citenamefont {Lu}}]{varjas-prb15}%
  \BibitemOpen
  \bibfield  {author} {\bibinfo {author} {\bibfnamefont {D.}~\bibnamefont
  {Varjas}}, \bibinfo {author} {\bibfnamefont {F.}~\bibnamefont {de~Juan}}, \
  and\ \bibinfo {author} {\bibfnamefont {Y.-M.}\ \bibnamefont {Lu}},\
  }\bibfield  {title} {\enquote {\bibinfo {title} {Bulk invariants and
  topological response in insulators and superconductors with nonsymmorphic
  symmetries},}\ }\href {\doibase 10.1103/PhysRevB.92.195116} {\bibfield
  {journal} {\bibinfo  {journal} {Phys. Rev. B}\ }\textbf {\bibinfo {volume}
  {92}},\ \bibinfo {pages} {195116} (\bibinfo {year} {2015})}\BibitemShut
  {NoStop}%
\bibitem [{\citenamefont {Fulga}\ \emph {et~al.}(2016)\citenamefont {Fulga},
  \citenamefont {Avraham}, \citenamefont {Beidenkopf},\ and\ \citenamefont
  {Stern}}]{fulga-prb16}%
  \BibitemOpen
  \bibfield  {author} {\bibinfo {author} {\bibfnamefont {I.~C.}\ \bibnamefont
  {Fulga}}, \bibinfo {author} {\bibfnamefont {N.}~\bibnamefont {Avraham}},
  \bibinfo {author} {\bibfnamefont {H.}~\bibnamefont {Beidenkopf}}, \ and\
  \bibinfo {author} {\bibfnamefont {A.}~\bibnamefont {Stern}},\ }\bibfield
  {title} {\enquote {\bibinfo {title} {Coupled-layer description of topological
  crystalline insulators},}\ }\href {\doibase 10.1103/PhysRevB.94.125405}
  {\bibfield  {journal} {\bibinfo  {journal} {Phys. Rev. B}\ }\textbf {\bibinfo
  {volume} {94}},\ \bibinfo {pages} {125405} (\bibinfo {year}
  {2016})}\BibitemShut {NoStop}%
\bibitem [{\citenamefont {Hsieh}\ \emph {et~al.}(2012)\citenamefont {Hsieh},
  \citenamefont {Lin}, \citenamefont {Liu}, \citenamefont {Duan}, \citenamefont
  {Bansil},\ and\ \citenamefont {Fu}}]{hsieh-nc12}%
  \BibitemOpen
  \bibfield  {author} {\bibinfo {author} {\bibfnamefont {T.~H.}\ \bibnamefont
  {Hsieh}}, \bibinfo {author} {\bibfnamefont {H.}~\bibnamefont {Lin}}, \bibinfo
  {author} {\bibfnamefont {J.}~\bibnamefont {Liu}}, \bibinfo {author}
  {\bibfnamefont {W.}~\bibnamefont {Duan}}, \bibinfo {author} {\bibfnamefont
  {A.}~\bibnamefont {Bansil}}, \ and\ \bibinfo {author} {\bibfnamefont
  {L.}~\bibnamefont {Fu}},\ }\bibfield  {title} {\enquote {\bibinfo {title}
  {{Topological crystalline insulators in the SnTe material class}},}\ }\href
  {\doibase 10.1038/ncomms1969} {\bibfield  {journal} {\bibinfo  {journal}
  {Nat. Commun.}\ }\textbf {\bibinfo {volume} {3}},\ \bibinfo {pages} {982}
  (\bibinfo {year} {2012})}\BibitemShut {NoStop}%
\bibitem [{\citenamefont {Liu}\ \emph {et~al.}(2014)\citenamefont {Liu},
  \citenamefont {Hsieh}, \citenamefont {Wei}, \citenamefont {Duan},
  \citenamefont {Moodera},\ and\ \citenamefont {Fu}}]{liu-nmat14}%
  \BibitemOpen
  \bibfield  {author} {\bibinfo {author} {\bibfnamefont {J.}~\bibnamefont
  {Liu}}, \bibinfo {author} {\bibfnamefont {T.~H.}\ \bibnamefont {Hsieh}},
  \bibinfo {author} {\bibfnamefont {P.}~\bibnamefont {Wei}}, \bibinfo {author}
  {\bibfnamefont {W.}~\bibnamefont {Duan}}, \bibinfo {author} {\bibfnamefont
  {J.}~\bibnamefont {Moodera}}, \ and\ \bibinfo {author} {\bibfnamefont
  {L.}~\bibnamefont {Fu}},\ }\bibfield  {title} {\enquote {\bibinfo {title}
  {Spin-filtered edge states with an electrically tunable gap in a
  two-dimensional topological crystalline insulator},}\ }\href {\doibase
  10.1038/nmat3828} {\bibfield  {journal} {\bibinfo  {journal} {Nature Mater.}\
  }\textbf {\bibinfo {volume} {13}},\ \bibinfo {pages} {178} (\bibinfo {year}
  {2014})}\BibitemShut {NoStop}%
\bibitem [{\citenamefont {Marzari}\ and\ \citenamefont
  {Vanderbilt}(1997)}]{marzari-prb97}%
  \BibitemOpen
  \bibfield  {author} {\bibinfo {author} {\bibfnamefont {N.}~\bibnamefont
  {Marzari}}\ and\ \bibinfo {author} {\bibfnamefont {D.}~\bibnamefont
  {Vanderbilt}},\ }\bibfield  {title} {\enquote {\bibinfo {title} {{Maximally
  localized generalized Wannier functions for composite energy bands}},}\
  }\href {\doibase 10.1103/PhysRevB.56.12847} {\bibfield  {journal} {\bibinfo
  {journal} {Phys. Rev. B}\ }\textbf {\bibinfo {volume} {56}},\ \bibinfo
  {pages} {12847} (\bibinfo {year} {1997})}\BibitemShut {NoStop}%
\bibitem [{\citenamefont {Taherinejad}\ and\ \citenamefont
  {Vanderbilt}(2015)}]{taherinejad-prl15}%
  \BibitemOpen
  \bibfield  {author} {\bibinfo {author} {\bibfnamefont {M.}~\bibnamefont
  {Taherinejad}}\ and\ \bibinfo {author} {\bibfnamefont {D.}~\bibnamefont
  {Vanderbilt}},\ }\bibfield  {title} {\enquote {\bibinfo {title} {{Adiabatic
  Pumping of Chern-Simons Axion Coupling}},}\ }\href {\doibase
  10.1103/PhysRevLett.114.096401} {\bibfield  {journal} {\bibinfo  {journal}
  {Phys. Rev. Lett.}\ }\textbf {\bibinfo {volume} {114}},\ \bibinfo {pages}
  {096401} (\bibinfo {year} {2015})}\BibitemShut {NoStop}%
\bibitem [{\citenamefont {Sutherland}(1986)}]{sutherland-prb86}%
  \BibitemOpen
  \bibfield  {author} {\bibinfo {author} {\bibfnamefont {B.}~\bibnamefont
  {Sutherland}},\ }\bibfield  {title} {\enquote {\bibinfo {title} {Localization
  of electronic wave functions due to local topology},}\ }\href {\doibase
  10.1103/PhysRevB.34.5208} {\bibfield  {journal} {\bibinfo  {journal} {Phys.
  Rev. B}\ }\textbf {\bibinfo {volume} {34}},\ \bibinfo {pages} {5208}
  (\bibinfo {year} {1986})}\BibitemShut {NoStop}%
\bibitem [{\citenamefont {Lieb}(1989)}]{lieb-prl89}%
  \BibitemOpen
  \bibfield  {author} {\bibinfo {author} {\bibfnamefont {E.~H.}\ \bibnamefont
  {Lieb}},\ }\bibfield  {title} {\enquote {\bibinfo {title} {{Two theorems on
  the Hubbard model}},}\ }\href {\doibase 10.1103/PhysRevLett.62.1201}
  {\bibfield  {journal} {\bibinfo  {journal} {Phys. Rev. Lett.}\ }\textbf
  {\bibinfo {volume} {62}},\ \bibinfo {pages} {1201} (\bibinfo {year}
  {1989})}\BibitemShut {NoStop}%
\bibitem [{\citenamefont {Ramachandran}\ \emph {et~al.}(2017)\citenamefont
  {Ramachandran}, \citenamefont {Andreanov},\ and\ \citenamefont
  {Flach}}]{ramachandran-prb17}%
  \BibitemOpen
  \bibfield  {author} {\bibinfo {author} {\bibfnamefont {A.}~\bibnamefont
  {Ramachandran}}, \bibinfo {author} {\bibfnamefont {A.}~\bibnamefont
  {Andreanov}}, \ and\ \bibinfo {author} {\bibfnamefont {S.}~\bibnamefont
  {Flach}},\ }\bibfield  {title} {\enquote {\bibinfo {title} {{Chiral flat
  bands: Existence, engineering, and stability}},}\ }\href {\doibase
  10.1103/PhysRevB.96.161104} {\bibfield  {journal} {\bibinfo  {journal} {Phys.
  Rev. B}\ }\textbf {\bibinfo {volume} {96}},\ \bibinfo {pages} {161104(R)}
  (\bibinfo {year} {2017})}\BibitemShut {NoStop}%
\bibitem [{\citenamefont {Asb{\'o}th}\ \emph {et~al.}(2016)\citenamefont
  {Asb{\'o}th}, \citenamefont {P{\'a}lyi},\ and\ \citenamefont
  {Oroszl{\'a}ny}}]{asboth-book16}%
  \BibitemOpen
  \bibfield  {author} {\bibinfo {author} {\bibfnamefont {J.~A.}\ \bibnamefont
  {Asb{\'o}th}}, \bibinfo {author} {\bibfnamefont {A.}~\bibnamefont
  {P{\'a}lyi}}, \ and\ \bibinfo {author} {\bibfnamefont {L.}~\bibnamefont
  {Oroszl{\'a}ny}},\ }\href {\doibase 10.1007/978-3-319-25607-8} {\emph
  {\bibinfo {title} {{A Short Course on Topological Insulators}}}}\ (\bibinfo
  {publisher} {Springer},\ \bibinfo {address} {Cham},\ \bibinfo {year}
  {2016})\BibitemShut {NoStop}%
\bibitem [{\citenamefont {Kim}\ \emph {et~al.}(2015)\citenamefont {Kim},
  \citenamefont {Kane}, \citenamefont {Mele},\ and\ \citenamefont
  {Rappe}}]{kim-prl15}%
  \BibitemOpen
  \bibfield  {author} {\bibinfo {author} {\bibfnamefont {Y.}~\bibnamefont
  {Kim}}, \bibinfo {author} {\bibfnamefont {C.~L.}\ \bibnamefont {Kane}},
  \bibinfo {author} {\bibfnamefont {E.~J.}\ \bibnamefont {Mele}}, \ and\
  \bibinfo {author} {\bibfnamefont {A.~M.}\ \bibnamefont {Rappe}},\ }\bibfield
  {title} {\enquote {\bibinfo {title} {{Layered Topological Crystalline
  Insulators}},}\ }\href {\doibase 10.1103/PhysRevLett.115.086802} {\bibfield
  {journal} {\bibinfo  {journal} {Phys. Rev. Lett.}\ }\textbf {\bibinfo
  {volume} {115}},\ \bibinfo {pages} {086802} (\bibinfo {year}
  {2015})}\BibitemShut {NoStop}%
\bibitem [{\citenamefont {Wieder}\ and\ \citenamefont
  {Bernevig}(2018)}]{wieder-arxiv18}%
  \BibitemOpen
  \bibfield  {author} {\bibinfo {author} {\bibfnamefont {B.~J.}\ \bibnamefont
  {Wieder}}\ and\ \bibinfo {author} {\bibfnamefont {B.~A.}\ \bibnamefont
  {Bernevig}},\ }\href {https://arxiv.org/abs/1810.02373} {\enquote {\bibinfo
  {title} {{The Axion Insulator as a Pump of Fragile Topology}},}\ } (\bibinfo
  {year} {2018}),\ \Eprint {http://arxiv.org/abs/1810.02373} {arXiv:1810.02373}
  \BibitemShut {NoStop}%
\bibitem [{\citenamefont {Park}\ and\ \citenamefont
  {Marzari}(2011)}]{park-prb11}%
  \BibitemOpen
  \bibfield  {author} {\bibinfo {author} {\bibfnamefont {C.-H.}\ \bibnamefont
  {Park}}\ and\ \bibinfo {author} {\bibfnamefont {N.}~\bibnamefont {Marzari}},\
  }\bibfield  {title} {\enquote {\bibinfo {title} {Berry phase and pseudospin
  winding number in bilayer graphene},}\ }\href {\doibase
  10.1103/PhysRevB.84.205440} {\bibfield  {journal} {\bibinfo  {journal} {Phys.
  Rev. B}\ }\textbf {\bibinfo {volume} {84}},\ \bibinfo {pages} {205440}
  (\bibinfo {year} {2011})}\BibitemShut {NoStop}%
\bibitem [{pyt()}]{pythtb}%
  \BibitemOpen
  \href@noop {} {}\bibinfo {note} {The \textsc{PythTB} code package is
  available at
  \url{http://www.physics.rutgers.edu/pythtb/about.html}}\BibitemShut {NoStop}%
\bibitem [{\citenamefont {Enkovaara}\ \emph {et~al.}(2010)\citenamefont
  {Enkovaara}, \citenamefont {Rostgaard}, \citenamefont {Mortensen},
  \citenamefont {Chen}, \citenamefont {Du{\l}ak}, \citenamefont {Ferrighi},
  \citenamefont {Gavnholt}, \citenamefont {Glinsvad}, \citenamefont {Haikola},
  \citenamefont {Hansen}, \citenamefont {Kristoffersen}, \citenamefont
  {Kuisma}, \citenamefont {Larsen}, \citenamefont {Lehtovaara}, \citenamefont
  {Ljungberg}, \citenamefont {Lopez-Acevedo}, \citenamefont {Moses},
  \citenamefont {Ojanen}, \citenamefont {Olsen}, \citenamefont {Petzold},
  \citenamefont {Romero}, \citenamefont {Stausholm-M{\o}ller}, \citenamefont
  {Strange}, \citenamefont {Tritsaris}, \citenamefont {Vanin}, \citenamefont
  {Walter}, \citenamefont {Hammer}, \citenamefont {H{\"{a}}kkinen},
  \citenamefont {Madsen}, \citenamefont {Nieminen}, \citenamefont {N{\o}rskov},
  \citenamefont {Puska}, \citenamefont {Rantala}, \citenamefont {Schi{\o}tz},
  \citenamefont {Thygesen},\ and\ \citenamefont {Jacobsen}}]{enkovaara-jpcm10}%
  \BibitemOpen
  \bibfield  {author} {\bibinfo {author} {\bibfnamefont {J.}~\bibnamefont
  {Enkovaara}}, \bibinfo {author} {\bibfnamefont {C.}~\bibnamefont
  {Rostgaard}}, \bibinfo {author} {\bibfnamefont {J.~J.}\ \bibnamefont
  {Mortensen}}, \bibinfo {author} {\bibfnamefont {J.}~\bibnamefont {Chen}},
  \bibinfo {author} {\bibfnamefont {M.}~\bibnamefont {Du{\l}ak}}, \bibinfo
  {author} {\bibfnamefont {L.}~\bibnamefont {Ferrighi}}, \bibinfo {author}
  {\bibfnamefont {J.}~\bibnamefont {Gavnholt}}, \bibinfo {author}
  {\bibfnamefont {C.}~\bibnamefont {Glinsvad}}, \bibinfo {author}
  {\bibfnamefont {V.}~\bibnamefont {Haikola}}, \bibinfo {author} {\bibfnamefont
  {H.~A.}\ \bibnamefont {Hansen}}, \bibinfo {author} {\bibfnamefont {H.~H.}\
  \bibnamefont {Kristoffersen}}, \bibinfo {author} {\bibfnamefont
  {M.}~\bibnamefont {Kuisma}}, \bibinfo {author} {\bibfnamefont {A.~H.}\
  \bibnamefont {Larsen}}, \bibinfo {author} {\bibfnamefont {L.}~\bibnamefont
  {Lehtovaara}}, \bibinfo {author} {\bibfnamefont {M.}~\bibnamefont
  {Ljungberg}}, \bibinfo {author} {\bibfnamefont {O.}~\bibnamefont
  {Lopez-Acevedo}}, \bibinfo {author} {\bibfnamefont {P.~G.}\ \bibnamefont
  {Moses}}, \bibinfo {author} {\bibfnamefont {J.}~\bibnamefont {Ojanen}},
  \bibinfo {author} {\bibfnamefont {T.}~\bibnamefont {Olsen}}, \bibinfo
  {author} {\bibfnamefont {V.}~\bibnamefont {Petzold}}, \bibinfo {author}
  {\bibfnamefont {N.~A.}\ \bibnamefont {Romero}}, \bibinfo {author}
  {\bibfnamefont {J.}~\bibnamefont {Stausholm-M{\o}ller}}, \bibinfo {author}
  {\bibfnamefont {M.}~\bibnamefont {Strange}}, \bibinfo {author} {\bibfnamefont
  {G.~A.}\ \bibnamefont {Tritsaris}}, \bibinfo {author} {\bibfnamefont
  {M.}~\bibnamefont {Vanin}}, \bibinfo {author} {\bibfnamefont
  {M.}~\bibnamefont {Walter}}, \bibinfo {author} {\bibfnamefont
  {B.}~\bibnamefont {Hammer}}, \bibinfo {author} {\bibfnamefont
  {H.}~\bibnamefont {H{\"{a}}kkinen}}, \bibinfo {author} {\bibfnamefont
  {G.~K.~H.}\ \bibnamefont {Madsen}}, \bibinfo {author} {\bibfnamefont {R.~M.}\
  \bibnamefont {Nieminen}}, \bibinfo {author} {\bibfnamefont {J.~K.}\
  \bibnamefont {N{\o}rskov}}, \bibinfo {author} {\bibfnamefont
  {M.}~\bibnamefont {Puska}}, \bibinfo {author} {\bibfnamefont {T.~T.}\
  \bibnamefont {Rantala}}, \bibinfo {author} {\bibfnamefont {J.}~\bibnamefont
  {Schi{\o}tz}}, \bibinfo {author} {\bibfnamefont {K.~S.}\ \bibnamefont
  {Thygesen}}, \ and\ \bibinfo {author} {\bibfnamefont {K.~W.}\ \bibnamefont
  {Jacobsen}},\ }\bibfield  {title} {\enquote {\bibinfo {title} {{Electronic
  structure calculations with GPAW: a real-space implementation of the
  projector augmented-wave method}},}\ }\href {\doibase
  10.1088/0953-8984/22/25/253202} {\bibfield  {journal} {\bibinfo  {journal}
  {J. Phys. Condens. Matter}\ }\textbf {\bibinfo {volume} {22}},\ \bibinfo
  {pages} {253202} (\bibinfo {year} {2010})}\BibitemShut {NoStop}%
\bibitem [{\citenamefont {Mostofi}\ \emph {et~al.}(2014)\citenamefont
  {Mostofi}, \citenamefont {Yates}, \citenamefont {Pizzi}, \citenamefont {Lee},
  \citenamefont {Souza}, \citenamefont {Vanderbilt},\ and\ \citenamefont
  {Marzari}}]{MOSTOFI20142309}%
  \BibitemOpen
  \bibfield  {author} {\bibinfo {author} {\bibfnamefont {A.~A.}\ \bibnamefont
  {Mostofi}}, \bibinfo {author} {\bibfnamefont {J.~R.}\ \bibnamefont {Yates}},
  \bibinfo {author} {\bibfnamefont {G.}~\bibnamefont {Pizzi}}, \bibinfo
  {author} {\bibfnamefont {Y.-S.}\ \bibnamefont {Lee}}, \bibinfo {author}
  {\bibfnamefont {I.}~\bibnamefont {Souza}}, \bibinfo {author} {\bibfnamefont
  {D.}~\bibnamefont {Vanderbilt}}, \ and\ \bibinfo {author} {\bibfnamefont
  {N.}~\bibnamefont {Marzari}},\ }\bibfield  {title} {\enquote {\bibinfo
  {title} {{An updated version of wannier90: A tool for obtaining
  maximally-localised Wannier functions}},}\ }\href {\doibase
  https://doi.org/10.1016/j.cpc.2014.05.003} {\bibfield  {journal} {\bibinfo
  {journal} {Comput. Phys. Commun.}\ }\textbf {\bibinfo {volume} {185}},\
  \bibinfo {pages} {2309} (\bibinfo {year} {2014})}\BibitemShut {NoStop}%
\bibitem [{\citenamefont {Olsen}(2016)}]{olsen-prb16}%
  \BibitemOpen
  \bibfield  {author} {\bibinfo {author} {\bibfnamefont {T.}~\bibnamefont
  {Olsen}},\ }\bibfield  {title} {\enquote {\bibinfo {title} {{Designing
  in-plane heterostructures of quantum spin Hall insulators from first
  principles: $1\text{T}^\prime$-MoS$_2$ with adsorbates}},}\ }\href {\doibase
  10.1103/PhysRevB.94.235106} {\bibfield  {journal} {\bibinfo  {journal} {Phys.
  Rev. B}\ }\textbf {\bibinfo {volume} {94}},\ \bibinfo {pages} {235106}
  (\bibinfo {year} {2016})}\BibitemShut {NoStop}%
\bibitem [{\citenamefont {Perdew}\ \emph {et~al.}(1996)\citenamefont {Perdew},
  \citenamefont {Burke},\ and\ \citenamefont {Ernzerhof}}]{perdew-prl96}%
  \BibitemOpen
  \bibfield  {author} {\bibinfo {author} {\bibfnamefont {J.~P.}\ \bibnamefont
  {Perdew}}, \bibinfo {author} {\bibfnamefont {K.}~\bibnamefont {Burke}}, \
  and\ \bibinfo {author} {\bibfnamefont {M.}~\bibnamefont {Ernzerhof}},\
  }\bibfield  {title} {\enquote {\bibinfo {title} {{Generalized Gradient
  Approximation Made Simple}},}\ }\href {\doibase 10.1103/PhysRevLett.77.3865}
  {\bibfield  {journal} {\bibinfo  {journal} {Phys. Rev. Lett.}\ }\textbf
  {\bibinfo {volume} {77}},\ \bibinfo {pages} {3865} (\bibinfo {year}
  {1996})}\BibitemShut {NoStop}%
\bibitem [{\citenamefont {Perdew}\ \emph {et~al.}(1997)\citenamefont {Perdew},
  \citenamefont {Burke},\ and\ \citenamefont {Ernzerhof}}]{perdew-prl97}%
  \BibitemOpen
  \bibfield  {author} {\bibinfo {author} {\bibfnamefont {J.~P.}\ \bibnamefont
  {Perdew}}, \bibinfo {author} {\bibfnamefont {K.}~\bibnamefont {Burke}}, \
  and\ \bibinfo {author} {\bibfnamefont {M.}~\bibnamefont {Ernzerhof}},\
  }\bibfield  {title} {\enquote {\bibinfo {title} {{Generalized Gradient
  Approximation Made Simple [Phys. Rev. Lett. 77, 3865 (1996)]}},}\ }\href
  {\doibase 10.1103/PhysRevLett.78.1396} {\bibfield  {journal} {\bibinfo
  {journal} {Phys. Rev. Lett.}\ }\textbf {\bibinfo {volume} {78}},\ \bibinfo
  {pages} {1396(E)} (\bibinfo {year} {1997})}\BibitemShut {NoStop}%
\bibitem [{\citenamefont {Bl\"ochl}(1994)}]{blochl-prb94}%
  \BibitemOpen
  \bibfield  {author} {\bibinfo {author} {\bibfnamefont {P.~E.}\ \bibnamefont
  {Bl\"ochl}},\ }\bibfield  {title} {\enquote {\bibinfo {title} {Projector
  augmented-wave method},}\ }\href {\doibase 10.1103/PhysRevB.50.17953}
  {\bibfield  {journal} {\bibinfo  {journal} {Phys. Rev. B}\ }\textbf {\bibinfo
  {volume} {50}},\ \bibinfo {pages} {17953} (\bibinfo {year}
  {1994})}\BibitemShut {NoStop}%
\bibitem [{\citenamefont {Souza}\ \emph {et~al.}(2001)\citenamefont {Souza},
  \citenamefont {Marzari},\ and\ \citenamefont {Vanderbilt}}]{souza-prb01}%
  \BibitemOpen
  \bibfield  {author} {\bibinfo {author} {\bibfnamefont {I.}~\bibnamefont
  {Souza}}, \bibinfo {author} {\bibfnamefont {N.}~\bibnamefont {Marzari}}, \
  and\ \bibinfo {author} {\bibfnamefont {D.}~\bibnamefont {Vanderbilt}},\
  }\bibfield  {title} {\enquote {\bibinfo {title} {{Maximally localized Wannier
  functions for entangled energy bands}},}\ }\href {\doibase
  10.1103/PhysRevB.65.035109} {\bibfield  {journal} {\bibinfo  {journal} {Phys.
  Rev. B}\ }\textbf {\bibinfo {volume} {65}},\ \bibinfo {pages} {035109}
  (\bibinfo {year} {2001})}\BibitemShut {NoStop}%
\bibitem [{\citenamefont {Olsen}\ \emph {et~al.}(2019)\citenamefont {Olsen},
  \citenamefont {Andersen}, \citenamefont {Okugawa}, \citenamefont {Torelli},
  \citenamefont {Deilmann},\ and\ \citenamefont {Thygesen}}]{olsen-prm19}%
  \BibitemOpen
  \bibfield  {author} {\bibinfo {author} {\bibfnamefont {T.}~\bibnamefont
  {Olsen}}, \bibinfo {author} {\bibfnamefont {E.}~\bibnamefont {Andersen}},
  \bibinfo {author} {\bibfnamefont {T.}~\bibnamefont {Okugawa}}, \bibinfo
  {author} {\bibfnamefont {D.}~\bibnamefont {Torelli}}, \bibinfo {author}
  {\bibfnamefont {T.}~\bibnamefont {Deilmann}}, \ and\ \bibinfo {author}
  {\bibfnamefont {K.~S.}\ \bibnamefont {Thygesen}},\ }\bibfield  {title}
  {\enquote {\bibinfo {title} {{Discovering two-dimensional topological
  insulators from high-throughput computations}},}\ }\href {\doibase
  10.1103/PhysRevMaterials.3.024005} {\bibfield  {journal} {\bibinfo  {journal}
  {Phys. Rev. Mater.}\ }\textbf {\bibinfo {volume} {3}},\ \bibinfo {pages}
  {024005} (\bibinfo {year} {2019})}\BibitemShut {NoStop}%
\bibitem [{\citenamefont {Haastrup}\ \emph {et~al.}(2018)\citenamefont
  {Haastrup}, \citenamefont {Strange}, \citenamefont {Pandey}, \citenamefont
  {Deilmann}, \citenamefont {Schmidt}, \citenamefont {Hinsche}, \citenamefont
  {Gjerding}, \citenamefont {Torelli}, \citenamefont {Larsen}, \citenamefont
  {Riis-Jensen}, \citenamefont {Gath}, \citenamefont {Jacobsen}, \citenamefont
  {Mortensen}, \citenamefont {Olsen},\ and\ \citenamefont
  {Thygesen}}]{haastrup-2dmat18}%
  \BibitemOpen
  \bibfield  {author} {\bibinfo {author} {\bibfnamefont {Sten}\ \bibnamefont
  {Haastrup}}, \bibinfo {author} {\bibfnamefont {Mikkel}\ \bibnamefont
  {Strange}}, \bibinfo {author} {\bibfnamefont {Mohnish}\ \bibnamefont
  {Pandey}}, \bibinfo {author} {\bibfnamefont {Thorsten}\ \bibnamefont
  {Deilmann}}, \bibinfo {author} {\bibfnamefont {Per~S.}\ \bibnamefont
  {Schmidt}}, \bibinfo {author} {\bibfnamefont {Nicki~F.}\ \bibnamefont
  {Hinsche}}, \bibinfo {author} {\bibfnamefont {Morten~N.}\ \bibnamefont
  {Gjerding}}, \bibinfo {author} {\bibfnamefont {Daniele}\ \bibnamefont
  {Torelli}}, \bibinfo {author} {\bibfnamefont {Peter~M.}\ \bibnamefont
  {Larsen}}, \bibinfo {author} {\bibfnamefont {Anders~C.}\ \bibnamefont
  {Riis-Jensen}}, \bibinfo {author} {\bibfnamefont {Jakob}\ \bibnamefont
  {Gath}}, \bibinfo {author} {\bibfnamefont {Karsten~W.}\ \bibnamefont
  {Jacobsen}}, \bibinfo {author} {\bibfnamefont {Jens~J{\o}rgen}\ \bibnamefont
  {Mortensen}}, \bibinfo {author} {\bibfnamefont {Thomas}\ \bibnamefont
  {Olsen}}, \ and\ \bibinfo {author} {\bibfnamefont {Kristian~S.}\ \bibnamefont
  {Thygesen}},\ }\bibfield  {title} {\enquote {\bibinfo {title} {{The
  Computational 2D Materials Database: High-throughput modeling and discovery
  of atomically thin crystals}},}\ }\href {\doibase 10.1088/2053-1583/aacfc1}
  {\bibfield  {journal} {\bibinfo  {journal} {2D Mater.}\ }\textbf {\bibinfo
  {volume} {5}},\ \bibinfo {pages} {042002} (\bibinfo {year}
  {2018})}\BibitemShut {NoStop}%
\bibitem [{\citenamefont {Kobayashi}(2015)}]{Kobayashi2015}%
  \BibitemOpen
  \bibfield  {author} {\bibinfo {author} {\bibfnamefont {K.}~\bibnamefont
  {Kobayashi}},\ }\bibfield  {title} {\enquote {\bibinfo {title} {{Electronic
  states of SnTe and PbTe (001) monolayers with supports}},}\ }\href {\doibase
  https://doi.org/10.1016/j.susc.2015.04.009} {\bibfield  {journal} {\bibinfo
  {journal} {Surf. Sci.}\ }\textbf {\bibinfo {volume} {639}},\ \bibinfo {pages}
  {54} (\bibinfo {year} {2015})}\BibitemShut {NoStop}%
\bibitem [{\citenamefont {Liu}\ \emph {et~al.}(2015)\citenamefont {Liu},
  \citenamefont {Qian},\ and\ \citenamefont {Fu}}]{liu-nanolett15}%
  \BibitemOpen
  \bibfield  {author} {\bibinfo {author} {\bibfnamefont {J.}~\bibnamefont
  {Liu}}, \bibinfo {author} {\bibfnamefont {X.}~\bibnamefont {Qian}}, \ and\
  \bibinfo {author} {\bibfnamefont {L.}~\bibnamefont {Fu}},\ }\bibfield
  {title} {\enquote {\bibinfo {title} {{Crystal Field Effect Induced
  Topological Crystalline Insulators In Monolayer IV--VI Semiconductors}},}\
  }\href {\doibase 10.1021/acs.nanolett.5b00308} {\bibfield  {journal}
  {\bibinfo  {journal} {Nano Lett.}\ }\textbf {\bibinfo {volume} {15}},\
  \bibinfo {pages} {2657} (\bibinfo {year} {2015})}\BibitemShut {NoStop}%
\bibitem [{\citenamefont {Tanaka}\ \emph {et~al.}(2012)\citenamefont {Tanaka},
  \citenamefont {Ren}, \citenamefont {Sato}, \citenamefont {Nakayama},
  \citenamefont {Souma}, \citenamefont {Takahashi}, \citenamefont {Segawa},\
  and\ \citenamefont {Ando}}]{tanaka-natphys12}%
  \BibitemOpen
  \bibfield  {author} {\bibinfo {author} {\bibfnamefont {Y.}~\bibnamefont
  {Tanaka}}, \bibinfo {author} {\bibfnamefont {Z.}~\bibnamefont {Ren}},
  \bibinfo {author} {\bibfnamefont {T.}~\bibnamefont {Sato}}, \bibinfo {author}
  {\bibfnamefont {K.}~\bibnamefont {Nakayama}}, \bibinfo {author}
  {\bibfnamefont {S.}~\bibnamefont {Souma}}, \bibinfo {author} {\bibfnamefont
  {T.}~\bibnamefont {Takahashi}}, \bibinfo {author} {\bibfnamefont
  {K.}~\bibnamefont {Segawa}}, \ and\ \bibinfo {author} {\bibfnamefont
  {Y.}~\bibnamefont {Ando}},\ }\bibfield  {title} {\enquote {\bibinfo {title}
  {{Experimental realization of a topological crystalline insulator in
  SnTe}},}\ }\href {\doibase 10.1038/NPHYS2442} {\bibfield  {journal} {\bibinfo
   {journal} {Nat. Phys.}\ }\textbf {\bibinfo {volume} {8}},\ \bibinfo {pages}
  {800} (\bibinfo {year} {2012})}\BibitemShut {NoStop}%
\bibitem [{\citenamefont {Gresch}\ \emph {et~al.}(2017)\citenamefont {Gresch},
  \citenamefont {Aut\`es}, \citenamefont {Yazyev}, \citenamefont {Troyer},
  \citenamefont {Vanderbilt}, \citenamefont {Bernevig},\ and\ \citenamefont
  {Soluyanov}}]{gresch-prb17}%
  \BibitemOpen
  \bibfield  {author} {\bibinfo {author} {\bibfnamefont {D.}~\bibnamefont
  {Gresch}}, \bibinfo {author} {\bibfnamefont {G.}~\bibnamefont {Aut\`es}},
  \bibinfo {author} {\bibfnamefont {O.~V.}\ \bibnamefont {Yazyev}}, \bibinfo
  {author} {\bibfnamefont {M.}~\bibnamefont {Troyer}}, \bibinfo {author}
  {\bibfnamefont {D.}~\bibnamefont {Vanderbilt}}, \bibinfo {author}
  {\bibfnamefont {B.~A.}\ \bibnamefont {Bernevig}}, \ and\ \bibinfo {author}
  {\bibfnamefont {A.~A.}\ \bibnamefont {Soluyanov}},\ }\bibfield  {title}
  {\enquote {\bibinfo {title} {{Z2Pack: Numerical implementation of hybrid
  Wannier centers for identifying topological materials}},}\ }\href {\doibase
  10.1103/PhysRevB.95.075146} {\bibfield  {journal} {\bibinfo  {journal} {Phys.
  Rev. B}\ }\textbf {\bibinfo {volume} {95}},\ \bibinfo {pages} {075146}
  (\bibinfo {year} {2017})}\BibitemShut {NoStop}%
\bibitem [{\citenamefont {Shen}\ \emph {et~al.}(2011)\citenamefont {Shen},
  \citenamefont {Shan},\ and\ \citenamefont
  {Lu}}]{doi:10.1142/S2010324711000057}%
  \BibitemOpen
  \bibfield  {author} {\bibinfo {author} {\bibfnamefont {S.-Q.}\ \bibnamefont
  {Shen}}, \bibinfo {author} {\bibfnamefont {W.-Y.}\ \bibnamefont {Shan}}, \
  and\ \bibinfo {author} {\bibfnamefont {H.-Z.}\ \bibnamefont {Lu}},\
  }\bibfield  {title} {\enquote {\bibinfo {title} {{Topological Insulator and
  the Dirac Equation}},}\ }\href {\doibase 10.1142/S2010324711000057}
  {\bibfield  {journal} {\bibinfo  {journal} {SPIN}\ }\textbf {\bibinfo
  {volume} {01}},\ \bibinfo {pages} {33} (\bibinfo {year} {2011})}\BibitemShut
  {NoStop}%
\bibitem [{\citenamefont {Rauch}\ \emph {et~al.}(2017)\citenamefont {Rauch},
  \citenamefont {Nguyen~Minh}, \citenamefont {Henk},\ and\ \citenamefont
  {Mertig}}]{Rauch2017}%
  \BibitemOpen
  \bibfield  {author} {\bibinfo {author} {\bibfnamefont {T.}~\bibnamefont
  {Rauch}}, \bibinfo {author} {\bibfnamefont {H.}~\bibnamefont {Nguyen~Minh}},
  \bibinfo {author} {\bibfnamefont {J.}~\bibnamefont {Henk}}, \ and\ \bibinfo
  {author} {\bibfnamefont {I.}~\bibnamefont {Mertig}},\ }\bibfield  {title}
  {\enquote {\bibinfo {title} {{Model for ferromagnetic Weyl and nodal line
  semimetals: Topological invariants, surface states, anomalous and spin Hall
  effect}},}\ }\href {\doibase 10.1103/PhysRevB.96.235103} {\bibfield
  {journal} {\bibinfo  {journal} {Phys. Rev. B}\ }\textbf {\bibinfo {volume}
  {96}},\ \bibinfo {pages} {235103} (\bibinfo {year} {2017})}\BibitemShut
  {NoStop}%
\bibitem [{\citenamefont {Shen}(2012)}]{shen-dirac-book}%
  \BibitemOpen
  \bibfield  {author} {\bibinfo {author} {\bibfnamefont {S.-Q.}\ \bibnamefont
  {Shen}},\ }\href {\doibase https://doi.org/10.1007/978-3-642-32858-9} {\emph
  {\bibinfo {title} {Topological Insulators -- Dirac Equation in Condensed
  Matters}}}\ (\bibinfo  {publisher} {Springer, Berlin, Heidelberg},\ \bibinfo
  {year} {2012})\BibitemShut {NoStop}%
\end{thebibliography}%

\end{document}